\documentclass[10pt,a4paper]{article}
\usepackage[utf8]{inputenc}
\usepackage[T1]{fontenc}
\usepackage{amsmath,amssymb,amsthm}
\usepackage{array,tabularx,booktabs,float,longtable}
\usepackage{graphicx}
\usepackage{caption}
\usepackage[margin=2.5cm]{geometry}
\usepackage{xcolor}
\usepackage[colorlinks=true,
            linkcolor=blue, 
            citecolor=blue, 
            urlcolor=blue]{hyperref}
\hypersetup{pdftitle={Symmetries, conservation laws, series solutions and singularity formation for the spherically symmetric Euler-Poisson equations in mass coordinates},pdfauthor={Adnan Shamaoon},pdfkeywords={Euler-Poisson system; gravitational collapse; Lie symmetry analysis; Noether theorem; Briot-Bouquet series; gradient catastrophe; critical adiabatic index}}

\numberwithin{equation}{section}
\theoremstyle{plain}
\newtheorem{theorem}{Theorem}[section]
\newtheorem{corollary}[theorem]{Corollary}

\newtheorem{proposition}[theorem]{Proposition}
\theoremstyle{definition}

\newtheorem{example}[theorem]{Example}
\theoremstyle{remark}
\newtheorem{remark}[theorem]{Remark}

\allowdisplaybreaks
\newcommand{\D}{\mathrm{D}}
\newcommand{\dd}{\,\mathrm{d}}
\newcommand{\Edd}{M_{\mathrm{Edd}}}
\newcommand{\pd}[2]{\frac{\partial #1}{\partial #2}}
\title{Symmetries, conservation laws, series solutions and singularity formation for the spherically symmetric Euler-Poisson equations in mass coordinates}
\author{Adnan Shamaoon$^{a,*}$\\[2pt]
\small $^{a}$ Northumbria University, Ellison Place, Newcastle upon Tyne, NE1 8ST, United Kingdom;\\
\small $^{*}$ Corresponding author: adnanshamaoon1220@gmail.com}
\date{}
\begin{document}
\maketitle
\begin{abstract}
Written in Lagrangian mass coordinates, the spherically symmetric Euler-Poisson equations for a polytropic gas become a single second-order equation for the radius $r(t,m)$ of the shell enclosing mass $m$, with adiabatic index $\gamma>1$ and entropy profile $S(m)$. We determine all point symmetries of this equation. Apart from time translation there is at most one scaling, with the single exception of the critical index $\gamma=4/3$ combined with the entropy profile $S\propto m^{2/3}$, where there are two; at $\gamma=4/3$ a scaling exists for every entropy profile. These scalings are the entropy-fixed stabilisers of the two-parameter scaling group of the non-isentropic system, and we give the exact symmetry defect when $\gamma\neq4/3$. Noether's theorem singles out the profiles $S=Km^{14-10\gamma}$, for which a scaling is variational and gives a conservation law beyond energy; otherwise the scalings yield virial identities. We construct exact collapse solutions and reduce the self-similar problem to one ordinary differential equation. Using the Briot-Bouquet theorem we prove that the regular centre of self-similar and homologous collapse carries a convergent analytic expansion with free central amplitude, and we derive a Puiseux expansion in the mass deficit at the stellar surface, from which the density vanishes with the polytropic exponent $1/(\gamma-1)$. In mass coordinates a density singularity is a zero of $r_m$ or of $r$; for the first kind we prove a Hopf-type fold law and an exact shell-crossing criterion for free-falling dust. Finally, linear perturbations of the critical Einstein-de Sitter collapse separate exactly into spherical Bessel profiles, with exponents solving $3\kappa^2+\kappa-2+\mu=0$. All identities were checked by computer algebra, and the numerical statements rest on convergence studies of a second-order Lagrangian scheme, including a collapse with a free surface.
\end{abstract}


\section{Introduction}\label{sec:intro}
Whether a gas sphere collapses under its own gravity depends to a large extent on one number, the adiabatic index $\gamma$ of the gas. For $\gamma>4/3$ pressure wins and spherical equilibria are dynamically stable \cite{Lin1997,Rein2003}; for $\gamma<4/3$ gravity can win and the Lane-Emden equilibria are unstable \cite{Lin1997,Jang2008,Jang2014}; and at $\gamma=4/3$ the two effects scale in exactly the same way, so that neither dominates a priori \cite{Chandrasekhar1939,GoldreichWeber1980}. The standard Newtonian model is the compressible Euler-Poisson system \cite{Chandrasekhar1939,KippenhahnWeigert1990}. In the past few years its collapse solutions have been put on a rigorous footing: the Larson-Penston solution for the isothermal gas ($\gamma=1$) \cite{Larson1969,Penston1969,GuoHadzicJang2021}, smooth self-similar implosions for every $\gamma\in(1,4/3)$ \cite{Yahil1983,GHJS2022}, their non-isentropic counterparts at larger $\gamma$ \cite{AHS2025}, and the nonlinear stability of the Larson-Penston family \cite{GHJS2025}. These results are formulated in Eulerian similarity variables, where the central difficulty is the sonic point, and smoothness there is used as a selection principle \cite{GHJS2022,AHS2025}.

\smallskip\noindent
We take a different, complementary route and work in Lagrangian mass coordinates (Section \ref{sec:model}). There the system collapses to one scalar variational equation for $r(t,m)$, the entropy is no longer a dynamical field but a coefficient $S(m)$, and a density singularity has a transparent form: either $r\to0$ or $r_m\to0$. This makes it natural to ask four questions that the Eulerian similarity approach does not answer directly.
\begin{enumerate}
\item[$(\mathrm{Q}1)$] Which point symmetries does the equation have as a function of $\gamma$ and $S(m)$, and are the scalings behind the known collapse solutions the only ones?
\item[$(\mathrm{Q}2)$] Which of them produce conservation laws, and what replaces a conservation law when a scaling is not variational?
\item[$(\mathrm{Q}3)$] What do the symmetries imply for exact solutions, and what does the solution look like near the centre and near the surface of a collapsing star?
\item[$(\mathrm{Q}4)$] How does collapse look from the viewpoint of the gradient catastrophe of hydrodynamic-type systems?
\end{enumerate}
Our answers are as follows. $(\mathrm{A}1)$ For every $\gamma>1$ the point symmetries are time translation plus at most one scaling, with the single exception of $\gamma=4/3$ and $S\propto m^{2/3}$, where there are two; at the critical index $\gamma=4/3$ a scaling exists for every entropy profile (Theorem \ref{thm:class}). The scalings are the stabilisers of the non-isentropic two-parameter group of \cite{AHS2025}, so no further point symmetry is available for symmetry reduction (Proposition \ref{prop:AHS}). $(\mathrm{A}2)$ Only on the line $S=Km^{14-10\gamma}$ is a scaling variational and gives a new conservation law; otherwise it gives a virial-type identity (Theorem \ref{thm:noether}, Proposition \ref{prop:virial}). $(\mathrm{A}3)$ The regular centre of both the self-similar and the homologous flow carries a convergent analytic expansion with free central amplitude, and the surface of the homologous star has a Puiseux expansion with the polytropic exponent (Theorem \ref{thm:BB}, Proposition \ref{prop:surface}). $(\mathrm{A}4)$ In free fall, shell crossing precedes collapse exactly when the mean density grows outward (Proposition \ref{prop:dust}). In addition, we solve the linearised problem around the critical Einstein-de Sitter collapse in closed form (Theorem \ref{thm:modes}) and validate the whole picture with a second-order Lagrangian solver (Section \ref{sec:num}).

\paragraph{Related work.} Lie-symmetry analysis of gas dynamics in mass coordinates was carried out without gravity in \cite{DKM2018,DKM2020}; for the gravitating case the scaling invariance is classical, and in the non-isentropic setting it was exploited, including the mass-critical and mass-supercritical distinction, in \cite{AHS2025}; see also \cite{Hadzic2023} for a review of how scaling invariances drive collapse and expansion (including the Lagrangian point of view in its Section $2.1$) and \cite{HadzicJang2018} for the expanding Goldreich-Weber stars, where a Lagrangian formulation is used. For spherically symmetric Euler flows, existence of finite-energy weak solutions for arbitrary spherically symmetric finite-energy data is proved in \cite{Schrecker2020} (see also \cite{ChenSchrecker2018}), where the possibility of finite-time blow-up of density and velocity at the origin, going back to Guderley and Courant-Friedrichs, is discussed, and self-similar converging shocks are analysed in \cite{JLS2025a,JLS2025b}. For gravitational dynamics of kinetic type, see \cite{HRSS2025}; special collapsing and expanding solutions of the Euler-Poisson system were also obtained in \cite{DLYY2002,DXY2003}, and dust-like collapse in the supercritical regime in \cite{GHJ2021b}. Our methods are standard Lie-symmetry and Noether techniques \cite{Olver1993,BlumanKumei1989,Noether1918,AncoBluman2002a,AncoBluman2002b} of the type used in recent symmetry studies \cite{Hirica2022,Mothibi2025,Gaxela2026} and in the author's earlier work on Lie symmetries, conservation laws and approximate symmetries \cite{Shamaoon2023,Agarwal2024,Shamaoon2023b}; series and special-function arguments follow the spirit of \cite{Shamaoon2021,Cesarano2023}. The emphasis on the gradient catastrophe is motivated by the formation of singularities in compressible fluids \cite{Sideris1985,Whitham1974}.

\paragraph{What is and what is not new.}\label{sec:prior}
Table \ref{tab:prior} separates classical or published statements from the ones established here. The scaling symmetries themselves, and the mass-(super)critical classification by the entropy exponent, are due to \cite{AHS2025} (in the mass-critical case also \cite{GoldreichWeber1980,DXY2003}); Proposition \ref{prop:AHS} identifies them with our scalings in the entropy-fixed formulation. What Theorem \ref{thm:class} adds is completeness, together with the Noether analysis and the exact symmetry defect. The shell-crossing criterion and the Jeans-type spectrum are Newtonian counterparts of classical results \cite{HellabyLake1985,Peebles1980}; we include them because in mass coordinates they take a closed form that is used for validation and that ties in with the gradient-catastrophe picture. Entries marked ``this paper'' were not found in the sources cited here or in the literature searches carried out in preparing this work; this is a statement about what we found, not a claim that the literature has been exhausted.
\begin{table}[H]
\centering
\caption{Known results and the contributions of this paper.}\label{tab:prior}
\begin{tabularx}{\textwidth}{@{}>{\raggedright\arraybackslash}p{4.3cm}>{\raggedright\arraybackslash}X>{\raggedright\arraybackslash}p{3.4cm}@{}}
\toprule
Topic & Status & Reference \\
\midrule
Two-parameter scaling group, mass-(super)critical classification & Known (Eulerian, non-isentropic) & \cite{AHS2025} \\
Homologous collapse/expansion at the critical index & Known & \cite{GoldreichWeber1980,DXY2003,HadzicJang2018} \\
Complete point-symmetry classification of the mass-coordinate equation, arbitrary $\gamma$, $S(m)$; symmetry defect & This paper & Thm. \ref{thm:class}, Prop. \ref{prop:defect} \\
Noether analysis; variational line $\sigma=14-10\gamma$; conservation law & This paper & Thm. \ref{thm:noether} \\
Analytic centre expansion of self-similar flow (Eulerian Taylor series) & Known; here in mass coordinates with Briot-Bouquet proof, closed form $a_1$, exact cross-validation & \cite{GHJS2022,AHS2025}; Thms. \ref{thm:BB}, \ref{thm:ssseries} \\
Surface expansion of the homologous star, exponent $1/(\gamma-1)$ & Classical for static polytropes; Puiseux law in mass coordinates for the collapsing profile & Prop. \ref{prop:surface} \\
Shell crossing of dust, free fall & Known (Tolman model) & \cite{HellabyLake1985}; Prop. \ref{prop:dust} \\
Hopf-type fold law in mass coordinates & Classical fold; Lagrangian statement & \cite{Whitham1974}; Prop. \ref{prop:hopf} \\
Linear spectrum around the critical EdS collapse & Jeans-type, classical in the Eulerian plane-wave form; Lagrangian spherical form with Bessel modes & \cite{Peebles1980}; Thm. \ref{thm:modes} \\
Convergence-verified Lagrangian solver incl.\ free surface & This paper & Sec. \ref{sec:num} \\
\bottomrule
\end{tabularx}
\end{table}
\smallskip\noindent
Section \ref{sec:model} derives the equation and fixes the standing assumptions; Section \ref{sec:lie} classifies the point symmetries; Section \ref{sec:cons} treats Noether symmetries and virial identities; Section \ref{sec:exact} gives exact solutions; Section \ref{sec:series} proves the series expansions at the centre and the surface; Section \ref{sec:catastrophe} studies the gradient catastrophe; Section \ref{sec:pert} the critical perturbation theory; Section \ref{sec:num} the numerical validation; Section \ref{sec:disc} discusses the findings.

\section{Model, hypotheses and notation}\label{sec:model}
\subsection{From the Euler-Poisson system to one equation for \texorpdfstring{$r(t,m)$}{r(t,m)}}
The gravitational Euler-Poisson system for a polytropic gas is
\begin{equation}\label{eq:EP}
\rho_t+\nabla\!\cdot(\rho\mathbf u)=0,\qquad
\rho\,(\mathbf u_t+\mathbf u\!\cdot\!\nabla\mathbf u)+\nabla p=-\rho\nabla\phi,\qquad
\Delta\phi=4\pi G\rho,\qquad p=S_{\rm ph}\,\rho^{\gamma},\quad\gamma>1,
\end{equation}
\cite{Chandrasekhar1939,GuoHadzicJang2021,AHS2025}. In spherical symmetry define the enclosed mass and the Lagrangian radius by
\begin{equation}\label{eq:mass}
m(t,r)=4\pi\int_0^{r}\rho(t,s)\,s^{2}\dd s,\qquad r=r(t,m),\qquad \rho=\frac{1}{4\pi\,r^{2}\,r_m},\qquad r_t=u,
\end{equation}
where $r_t$ at fixed $m$ is the fluid velocity. By Newton's shell theorem the gravitational acceleration at radius $r$ is $-Gm/r^2$, and $\partial_rp=4\pi r^2\rho\,p_m$ since $\dd m=4\pi\rho r^2\dd r$ at fixed $t$. The radial momentum equation is therefore
\begin{equation}\label{eq:momentum}
r_{tt}=-\frac{Gm}{r^{2}}-4\pi r^{2}\,p_m ,
\end{equation}
the form used in stellar-structure theory \cite{KippenhahnWeigert1990}. Before shock formation the entropy $S_{\rm ph}=p\rho^{-\gamma}$ is transported along particle paths, hence a function of $m$ only \cite{AHS2025}. Define
\begin{equation}\label{eq:Sdef}
S(m):=(4\pi)^{1-\gamma}S_{\rm ph}(m),\qquad P:=S(m)\,r^{-2\gamma}r_m^{-\gamma}=4\pi p ,
\end{equation}
so that $4\pi r^2p_m=r^2P_m$ and \eqref{eq:momentum} becomes
\begin{equation}\label{eq:EPm}
\boxed{\;\Delta[r]:=r_{tt}+\frac{Gm}{r^{2}}+r^{2}\,\partial_m\!\bigl(S(m)\,r^{-2\gamma}r_m^{-\gamma}\bigr)=0\;}
\end{equation}
Explicitly, with $S'=\mathrm dS/\mathrm dm$,
\begin{equation}\label{eq:F}
r_{tt}=F:=-\frac{Gm}{r^{2}}-S'\,r^{2-2\gamma}r_m^{-\gamma}+2\gamma S\,r^{1-2\gamma}r_m^{1-\gamma}+\gamma S\,r^{2-2\gamma}r_m^{-\gamma-1}r_{mm}.
\end{equation}
We write $\tilde\rho:=1/(r^2r_m)=4\pi\rho$.

\subsection{Standing hypotheses}\label{sec:hyp}
Throughout, $I\subset(0,\infty)$ is an open interval of mass labels, $J$ an open time interval, $\Omega=J\times I$ and:
\begin{itemize}
\item[\textbf{$(\mathrm{H}1)$}] \emph{Entropy.} $S\in C^2(I)$ and $S>0$ on $I$ (the pressureless case $S\equiv0$ is treated separately in Proposition \ref{prop:dust}).
\item[\textbf{$(\mathrm{H}2)$}] \emph{Solutions.} A \emph{classical solution} is $r\in C^2(\Omega)$ satisfying \eqref{eq:EPm} pointwise with $r>0$ and $r_m>0$ in $\Omega$.
\item[\textbf{$(\mathrm{H}3)$}] \emph{Boundary behaviour} (used only for integral identities in Section \ref{sec:cons}). $I=(0,M)$, and for every $t\in J$: $r(t,m)\to0$ as $m\to0$; $r^3P\to0$ as $m\to0$ and $P\to0$ as $m\to M$ (so $r^3P\to0$ at both ends, $r$ being bounded); $r_t/r$ is bounded as $m\to0$ and $r_t$ is bounded as $m\to M$; and the integrands $r_t^2$, $rr_{tt}$, $m/r$, $e$, $rr_t$ are bounded by a function in $L^1(I)$ locally uniformly in $t\in J$, so that $\int_I\cdot\,\dd m$ may be differentiated under the integral sign twice in $t$.
\item[\textbf{$(\mathrm{H}4)$}] \emph{Symmetry generators.} A point symmetry is a vector field $X=\tau\partial_t+\xi\partial_m+\eta\partial_r$ with $\tau,\xi,\eta\in C^3(U)$, $U\subset\{(t,m,r):(t,m)\in\Omega,\ r>0\}$ open and connected; symmetries are understood locally.
\end{itemize}
These hypotheses make every manipulation below legitimate: the prolonged invariance condition is an identity on the second-order jet manifold $\{\Delta=0\}$, on which $(t,m,r,r_t,r_m,r_{tm},r_{mm})$ with $r_m>0$ are \emph{free} coordinates (the equation being solved for $r_{tt}$), so identities may be split with respect to them.

\subsection{Variational structure}
\begin{proposition}[Variational structure and energy law]\label{prop:var}
\emph{Hypotheses:} $(\mathrm{H}1)$-$(\mathrm{H}2)$. Equation \eqref{eq:EPm} is the Euler-Lagrange equation of
\begin{equation}\label{eq:L}
\mathcal L=\tfrac12 r_t^2+\frac{Gm}{r}-\frac{S(m)}{\gamma-1}\,r^{2-2\gamma}r_m^{1-\gamma},
\end{equation}
that is,
\begin{equation}\label{eq:EL}
\frac{\delta\mathcal L}{\delta r}:=\pd{\mathcal L}{r}-\D_t\pd{\mathcal L}{r_t}-\D_m\pd{\mathcal L}{r_m}=-\Delta[r].
\end{equation}
With $e:=\dfrac{S}{\gamma-1}\,r^{2-2\gamma}r_m^{1-\gamma}$ and $\mathcal E:=\tfrac12r_t^2-\dfrac{Gm}{r}+e$ one has, for every $r\in C^2(\Omega)$ with $r,r_m>0$,
\begin{equation}\label{eq:energy}
\D_t\mathcal E+\D_m\bigl(r^{2}P\,r_t\bigr)=r_t\,\Delta[r].
\end{equation}
Under $(\mathrm{H}3)$, $E=\int_I\mathcal E\dd m$ is conserved on classical solutions.
\end{proposition}
\begin{proof}
The partial derivatives of \eqref{eq:L} are
\begin{align}
\pd{\mathcal L}{r_t}&=r_t,\qquad \pd{\mathcal L}{r_m}=S\,r^{2-2\gamma}r_m^{-\gamma}=r^2P,\label{eq:Lders1}\\
\pd{\mathcal L}{r}&=-\frac{Gm}{r^{2}}+2S\,r^{1-2\gamma}r_m^{1-\gamma},\label{eq:Lders2}
\end{align}
and 
\begin{equation}
    \D_m(r^2P)=S'r^{2-2\gamma}r_m^{-\gamma}+(2-2\gamma)S\,r^{1-2\gamma}r_m^{1-\gamma}-\gamma S\,r^{2-2\gamma}r_m^{-\gamma-1}r_{mm},
\end{equation}
inserting in \eqref{eq:EL} gives
\begin{equation}
    -r_{tt}-Gm/r^2-S'r^{2-2\gamma}r_m^{-\gamma}+2\gamma Sr^{1-2\gamma}r_m^{1-\gamma}+\gamma Sr^{2-2\gamma}r_m^{-\gamma-1}r_{mm},
\end{equation}
i.e.\ $-(r_{tt}-F)=-\Delta[r]$ by \eqref{eq:F}.

\smallskip\noindent
\emph{Energy identity.} Since $r^2P=S\,r^{2-2\gamma}r_m^{-\gamma}$ and $r^2r_mP=(\gamma-1)e$, the partial derivatives of $e$ are
\begin{equation}
\pd{e}{r}=-2\,r\,r_m\,P,\qquad \pd{e}{r_m}=-r^2P .
\end{equation}
Therefore
\begin{align}
\D_t\mathcal E&=r_tr_{tt}+\frac{Gm\,r_t}{r^{2}}-2\,r\,r_mP\,r_t-r^2P\,r_{tm},\\
\D_m\bigl(r^2P\,r_t\bigr)&=r^2P\,r_{tm}+2\,r\,r_mP\,r_t+r^2P_m\,r_t .
\end{align}
Adding these two identities, the terms in $r_{tm}$ and in $2\,r\,r_mP\,r_t$ cancel and
\begin{equation}
\D_t\mathcal E+\D_m\bigl(r^2P\,r_t\bigr)=r_t\Bigl(r_{tt}+\frac{Gm}{r^{2}}+r^2P_m\Bigr)=r_t\,\Delta[r],
\end{equation}
which is \eqref{eq:energy}. The conservation statement follows by integrating \eqref{eq:energy} over $I$ on a classical solution ($\Delta=0$) and using $(\mathrm{H}3)$: $r^2Pr_t=(r^3P)(r_t/r)\to0$ as $m\to0$ and $r^2Pr_t\to0$ as $m\to M$. Both identities were also verified symbolically (checks $\mathrm{V}1$, $\mathrm{V}2$).
\end{proof}
\begin{remark}
The sign in \eqref{eq:energy} is the one consistent with the Noether identity \eqref{eq:Noetheridentity} for $X=\partial_t$ (there $Q=-r_t$ and $T^t=-\mathcal E$, so that $\D_tT^t+\D_mT^m=Q\Delta$ reads $\D_t\mathcal E+\D_m(r^2Pr_t)=r_t\Delta$).
\end{remark}

\subsection{Notation}
Table \ref{tab:notation} collects the notation.
\begin{table}[H]
\centering
\caption{Notation.}\label{tab:notation}
\begin{tabularx}{\textwidth}{@{}lX@{}}
\toprule
Symbol & Meaning \\
\midrule
$t$, $m$, $r(t,m)$ & time; enclosed mass (Lagrangian label); radius of the shell with label $m$ \\
$\rho$, $p$, $\tilde\rho=4\pi\rho=1/(r^2r_m)$ & density; pressure; mass-coordinate density \\
$\gamma$, $G$ & adiabatic index ($>1$); gravitational constant \\
$S_{\rm ph}(m)$, $S(m)$, $K$, $\sigma$ & physical entropy $p\rho^{-\gamma}$; scaled entropy $(4\pi)^{1-\gamma}S_{\rm ph}$; constants in $S=Km^{\sigma}$ \\
$P=Sr^{-2\gamma}r_m^{-\gamma}$ & scaled pressure ($=4\pi p$) \\
$\Delta[r]$, $F$ & model operator \eqref{eq:EPm}; right-hand side \eqref{eq:F} \\
$\mathcal L$, $\mathcal E$, $e$, $E$ & Lagrangian density, energy density, internal-energy density, total energy \\
$\Omega$, $I$, $M$ & domain; mass interval; total mass \\
$X=\tau\partial_t+\xi\partial_m+\eta\partial_r$ & point symmetry generator; $X^{(k)}$ its $k$th prolongation \\
$c_0,c_1,c_4$ & constants in the symmetry generator \eqref{eq:X} \\
$\mathcal C[S]$ & classifying expression, eq.\ \eqref{eq:classeq} \\
$Q=\eta-\tau r_t-\xi r_m$, $T^t,T^m$ & characteristic; components of a conserved current \\
$w$ & scaling weight of $\mathcal L$ (Prop. \ref{prop:weights}) \\
$K_{\rm kin},W,U,I_{\rm in},\mathcal Q$ & kinetic, gravitational, internal energy; $\int r^2\dd m$; $\int rr_t\dd m$ \\
$a(t)$, $x(m)$, $\lambda$ & homologous scale factor, comoving profile, eigenparameter ($a^2\ddot a=-\lambda$) \\
$\Pi=Kx^{-8/3}x_m^{-4/3}$ & comoving pressure of the homologous profile ($\gamma=4/3$) \\
$c_k$, $a_k$, $A_0$ & series coefficients (Thms. \ref{thm:series}, \ref{thm:ssseries}); central amplitude \\
$s=mt^{3\gamma-4}$, $R(s)$, $\Phi(s)$ & similarity variable; self-similar radius; $\Phi=1/(R^2R_s)$ ($\tilde\rho\,t^2=\Phi$) \\
$M$, $\Edd$, $\lambda_c$ & total mass; Eddington mass; numerical threshold in $\lambda$ \\
$x_s,q,b_1$ & surface radius, surface pressure slope, Puiseux coefficient (Prop. \ref{prop:surface}) \\
$\vartheta$, $T(m)$, $\alpha$, $r_0(m)$ & cycloid phase; free-fall time scale; dust exponent in $r_0=m^\alpha$; initial radius \\
$T$ (Sec. \ref{sec:pert}), $\kappa$, $f(m)$ & time to blow-up; perturbation exponent; spatial profile \\
$\ell,\Theta,\mu_n,z_n$ & $\ell=24K/(3^{2/3}G)$; $\Theta=\ell/(9M^{2/3})$; $\mu_n=\Theta z_n^2$; roots of $\tan z=z$ \\
$j_1$ & spherical Bessel function of order $1$ \cite{DLMF2010} \\
$(a,b,c,e)$, $\alpha_E$ & scaling exponents of $(r,t,m,S)$; entropy scaling exponent of \cite{AHS2025} \\
\bottomrule
\end{tabularx}
\end{table}

\section{Lie point symmetries}\label{sec:lie}
The first thing one wants to know about a collapse problem is which scalings it carries, because self-similar solutions are invariant solutions of a scaling. We therefore determine \emph{all} point symmetries, not only the scalings, and keep $\gamma$ and the entropy profile $S(m)$ arbitrary. The computation is routine in structure but not in outcome: the classifying equation \eqref{eq:classeq} below shows precisely where the critical index $\gamma=4/3$ enters.

\smallskip\noindent
A point symmetry $X$ of \eqref{eq:EPm} satisfies $X^{(2)}\Delta\,|_{\Delta=0}=0$, with $X^{(2)}$ the second prolongation \cite{Olver1993,BlumanKumei1989}. Writing \eqref{eq:EPm} as $r_{tt}=F$ with $F$ from \eqref{eq:F}, the invariance condition is
\begin{equation}\label{eq:invcond}
\eta^{tt}-\bigl(F_m\,\xi+F_r\,\eta+F_{r_m}\,\eta^{m}+F_{r_{mm}}\,\eta^{mm}\bigr)\Big|_{r_{tt}=F}=0,
\end{equation}
where the prolongation coefficients are
\begin{align}
\eta^{t}&=\D_t\eta-r_t\D_t\tau-r_m\D_t\xi, & \eta^{m}&=\D_m\eta-r_t\D_m\tau-r_m\D_m\xi,\label{eq:eta1}\\
\eta^{tt}&=\D_t\eta^t-r_{tt}\D_t\tau-r_{tm}\D_t\xi, & \eta^{mm}&=\D_m\eta^m-r_{mm}\D_m\xi-r_{tm}\D_m\tau.\label{eq:eta2}
\end{align}
By $(\mathrm{H}4)$ the left side of \eqref{eq:invcond} is a function of the free jet variables $(t,m,r,r_t,r_m,r_{tm},r_{mm})$, polynomial in $r_t,r_{tm},r_{mm}$ and containing the powers $r_m^{k}$ with $k\in\{1,0\}\cup\{-\gamma+j:\,j=-2,\dots,2\}$. Splitting gives the determining equations used below.

\begin{theorem}[Classification of point symmetries]\label{thm:class}
\emph{Hypotheses:} $\gamma>1$, $(\mathrm{H}1)$ with $I$ connected, $(\mathrm{H}4)$. Every point symmetry of \eqref{eq:EPm} has the form
\begin{equation}\label{eq:X}
X=c_0\,\partial_t+c_1\,t\,\partial_t+(3c_4-2c_1)\,m\,\partial_m+c_4\,r\,\partial_r ,\qquad c_0,c_1,c_4\in\mathbb R,
\end{equation}
where the constants satisfy the \emph{classifying equation}
\begin{equation}\label{eq:classeq}
\mathcal C[S](m):=(2c_1-3c_4)\,m\,S'(m)+2\bigl[c_1(\gamma-2)+c_4\bigr]S(m)=0\qquad\text{on }I .
\end{equation}
Conversely, every such $X$ with \eqref{eq:classeq} is a point symmetry.
\end{theorem}
\begin{proof}
\emph{Step $1$ (structure).} The coefficients of the monomials $r_t^3$, $r_t^2$, $r_t^2r_m$, $r_tr_{tm}$, $r_{tm}$ ($r_m^0$), $r_tr_{mm}r_m^{-\gamma-2}$, $r_tr_{mm}r_m^{-\gamma-1}$, $r_{tm}r_m^{-\gamma}$, $r_{tm}r_m^{-\gamma-1}$, $r_{mm}r_m^{-\gamma}$, $r_{mm}r_m^{-\gamma-2}$ are, respectively,
\begin{equation}\label{eq:det1}
\begin{gathered}
-\tau_{rr},\quad \eta_{rr}-2\tau_{rt},\quad -\xi_{rr},\quad -2\xi_r,\quad -2\xi_t,\\
-\gamma(\gamma+1)r^{2-2\gamma}S\tau_m,\quad -\gamma(\gamma+3)r^{2-2\gamma}S\tau_r,\quad 2\gamma r^{2-2\gamma}S\tau_r,\quad 2\gamma r^{2-2\gamma}S\tau_m,\\
\gamma(1-\gamma)r^{2-2\gamma}S\xi_r,\quad \gamma(\gamma+1)r^{2-2\gamma}S\eta_m ,
\end{gathered}
\end{equation}
all of which must vanish. Since $S>0$ and $\gamma>1$ every numerical factor is non-zero, so
\begin{equation}\label{eq:det1b}
\tau_r=\tau_m=0,\qquad \xi_r=\xi_t=0,\qquad \eta_m=0,\qquad \eta_{rr}=0 ,
\end{equation}
i.e.\ $\tau=\tau(t)$, $\xi=\xi(m)$, $\eta=A(t)\,r+B(t)$.\\
\emph{Step $2$ (gravity).} The $r_m^{0}$ coefficient of the terms linear in $r_t$ is $3Gm\tau_r/r^2-\tau_{tt}+2\eta_{rt}$, giving
\begin{equation}\label{eq:Atau}
2A'(t)=\tau''(t).
\end{equation}
The jet-free $r_m^0$ coefficient (the gravitational balance) is
\begin{equation}\label{eq:grav}
r^{4}A''+r^{3}B''+G\,r\,\bigl(-3mA+2m\tau'+\xi\bigr)-2G\,m\,B=0 .
\end{equation}
Splitting \eqref{eq:grav} in the independent powers of $r$ gives
\begin{equation}\label{eq:gravsplit}
A''=0,\qquad B''=0,\qquad B=0\ (m\neq0),\qquad \xi=m\,(3A-2\tau').
\end{equation}
\emph{Step $3$ (time dependence).} From $A''=0$ and \eqref{eq:Atau}, $\tau=c_0+c_1t+c_2t^2$ and $A=c_4+c_2t$, hence by \eqref{eq:gravsplit}
\begin{equation}
\xi=m\bigl(3c_4-2c_1-c_2\,t\bigr).
\end{equation}
Since $\xi_t=0$ by \eqref{eq:det1b}, $c_2=0$, which gives \eqref{eq:X}.\\
\emph{Step $4$ (pressure).} With $B=0$, the coefficient of $r_{mm}r_m^{-\gamma-1}$ is
\begin{equation}\label{eq:pressure}
\gamma\,r^{2-2\gamma}\Bigl[(3\gamma-1)A\,S-(\gamma-1)\xi'\,S-2\tau'\,S-\xi\,S'\Bigr]=0 ;
\end{equation}
the remaining pressure coefficients (of $r_m^{-\gamma}$ and $r_m^{1-\gamma}$, jet-free) are, up to non-zero factors, its $m$-derivative and itself. Inserting $A=c_4$, $\tau'=c_1$, $\xi=(3c_4-2c_1)m$:
\begin{align}
&(3\gamma-1)c_4-(\gamma-1)(3c_4-2c_1)-2c_1=2c_4+2(\gamma-2)c_1,\\
&\text{so \eqref{eq:pressure}}\iff 2\bigl[c_1(\gamma-2)+c_4\bigr]S-(3c_4-2c_1)\,mS'=0,
\end{align}
which is \eqref{eq:classeq}.

\smallskip\noindent
\emph{Exponent coincidences.} The splitting treats the listed powers of $r_m$ as distinct. For $\gamma>1$ the only coincidence between $\{1,0\}$ and $\{-\gamma+j\}_{j=-2}^{2}$ is $2-\gamma=0$ at $\gamma=2$, and it affects only the jet-free coefficient $\gamma\bigl(-2\gamma r^{1-2\gamma}\xi_r+r^{2-2\gamma}\xi_{rr}\bigr)S$, which vanishes by \eqref{eq:det1b}. Hence Steps $1$-$4$ hold verbatim for $\gamma=2$; the full $\gamma=2$ system was also generated and reduced symbolically (check $\mathrm{V}19$).\\
\emph{Converse.} Each $X$ with \eqref{eq:classeq} satisfies \eqref{eq:invcond} identically (checks $\mathrm{V}6$, $\mathrm{V}7$, $\mathrm{V}14$).
\end{proof}

\begin{corollary}[Algebras]\label{cor:cases}
\emph{Hypotheses:} those of Theorem \ref{thm:class}. Table \ref{tab:class} lists the admitted algebras. In particular:
\begin{enumerate}
\item[$(\mathrm{a})$] $\partial_t$ is always admitted.
\item[$(\mathrm{b})$] If $S=Km^{\sigma}$ with $\sigma\neq2/3$, one further generator exists:
\begin{equation}\label{eq:Xsigma}
X_\sigma=t\,\partial_t+c_4\,r\,\partial_r+(3c_4-2)\,m\,\partial_m,\qquad c_4=\frac{2(\gamma+\sigma-2)}{3\sigma-2}.
\end{equation}
\item[$(\mathrm{c})$] If $S=Km^{2/3}$ and $\gamma\ne4/3$, the further generator is the mass scaling $3m\partial_m+r\partial_r$.
\item[$(\mathrm{d})$] If $\gamma=4/3$ and $S$ is \emph{arbitrary}, the further generator is $3t\partial_t+2r\partial_r$, which leaves $m$ fixed.
\item[$(\mathrm{e})$] If $\gamma=4/3$ and $S=Km^{2/3}$ the algebra is three-dimensional, spanned by
\begin{equation}
X_1=\partial_t,\qquad X_2=3t\partial_t+2r\partial_r,\qquad X_3=3m\partial_m+r\partial_r ,
\end{equation}
with $[X_1,X_2]=3X_1$ and $[X_1,X_3]=[X_2,X_3]=0$: the direct sum of the two-dimensional non-Abelian Lie algebra and $\mathbb R$.
\end{enumerate}
\end{corollary}
\begin{proof}
For $S=Km^\sigma$, \eqref{eq:classeq} reads
\begin{equation}\label{eq:powerclass}
c_1\,(2\sigma+2\gamma-4)+c_4\,(2-3\sigma)=0 .
\end{equation}
For $\sigma\ne2/3$ this fixes $c_4/c_1$. For $\sigma=2/3$ it reads $c_1(2\gamma-8/3)=0$: $c_1=0$ unless $\gamma=4/3$, and $c_4$ is free. If $S$ is not a power law (i.e.\ $mS'/S$ is not constant on $I$), \eqref{eq:classeq} forces $2c_1-3c_4=0$ and $c_1(\gamma-2)+c_4=0$, i.e.\ $c_1(\gamma-4/3)=0$. The brackets in $(\mathrm{e})$ follow by direct computation ($\mathrm{V}8$).
\end{proof}

\begin{table}[H]
\centering
\caption{Point-symmetry algebras of the mass-coordinate Euler-Poisson equation \eqref{eq:EPm} ($S>0$, $\gamma>1$).}\label{tab:class}
\small
\begin{tabularx}{\textwidth}{@{}lllX@{}}
\toprule
Case & $S(m)$ & $\gamma$ & Additional generator(s) beyond $\partial_t$ \\
\midrule
$\mathrm{A}$ & not a power law & $\ne4/3$ & none (dimension $1$) \\
$\mathrm{B}$ & $Km^{\sigma}$, $\sigma\ne2/3$ & any & $t\partial_t+c_4r\partial_r+(3c_4-2)m\partial_m$, $c_4=\frac{2(\gamma+\sigma-2)}{3\sigma-2}$ \\
$\mathrm{C}$ & $Km^{2/3}$ & $\ne4/3$ & $3m\partial_m+r\partial_r$ \\
$\mathrm{D}$ & arbitrary & $4/3$ & $3t\partial_t+2r\partial_r$ \\
$\mathrm{E}$ & $Km^{2/3}$ & $4/3$ & $3t\partial_t+2r\partial_r$ and $3m\partial_m+r\partial_r$ (dimension $3$) \\
\bottomrule
\end{tabularx}
\end{table}
\begin{figure}[H]
\centering
\includegraphics[width=0.49\textwidth]{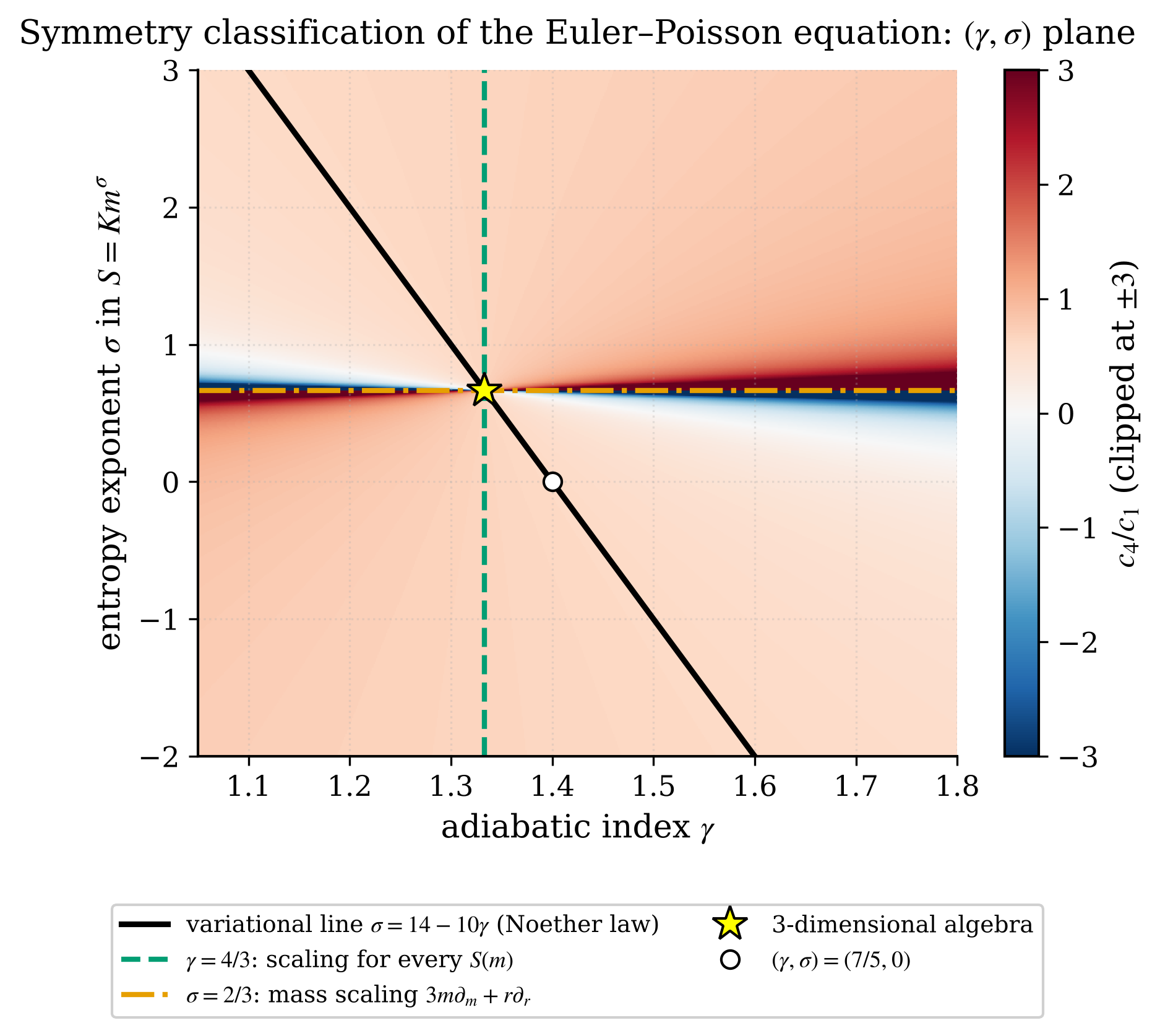}
\caption{Classification map in the $(\gamma,\sigma)$-plane for $S=Km^\sigma$. Colour: the scaling ratio $c_4/c_1$ of \eqref{eq:Xsigma}, clipped at $\pm3$. The dashed vertical line $\gamma=4/3$ carries a scaling for every entropy profile; the dash-dotted line $\sigma=2/3$ carries the mass scaling; their intersection (star) has a three-dimensional algebra. The solid line $\sigma=14-10\gamma$ is where a scaling is variational (Theorem \ref{thm:noether}); the circle marks $(\gamma,\sigma)=(7/5,0)$.}\label{fig:map}
\end{figure}

\subsection{Relation to the two-parameter scaling group of the non-isentropic system}
Let $(r,t,m,S)\mapsto(\lambda^{a}r,\lambda^{b}t,\lambda^{c}m,\lambda^{e}S)$. Requiring that $r_{tt}$, $Gm/r^2$ and $r^2\partial_m(Sr^{-2\gamma}r_m^{-\gamma})$ scale identically gives
\begin{equation}\label{eq:scalingeqs}
a-2b=c-2a=2a+e-2\gamma a-\gamma(a-c)-c ,
\end{equation}
two linear equations for four exponents; hence a two-parameter group (symbolic solution, check $\mathrm{V}20$),
\begin{equation}\label{eq:twoparam}
b=\frac{3a-c}{2},\qquad e=(3\gamma-4)\,a-(\gamma-2)\,c .
\end{equation}
This is the mass-coordinate form of the two-parameter scaling invariance of the non-isentropic Euler-Poisson system \cite[Sec. 1.1]{AHS2025}. Theorem \ref{thm:class} concerns the \emph{entropy-fixed} equation, i.e.\ the stabiliser of the profile $S(m)$ in this group.

\begin{proposition}[Entropy-fixed scalings and the parametrisation of \cite{AHS2025}]\label{prop:AHS}
\emph{Hypotheses:} $\gamma>1$, $S=Km^{\sigma}$ with $\sigma\neq2/3$ and $\gamma+\sigma\neq2$. A scaling \eqref{eq:twoparam} leaves $S=Km^\sigma$ invariant iff $e=\sigma c$, and then (with $b=1$) $r\sim t^{c_4}$ and $m\sim t^{3c_4-2}$ with $c_4$ given by \eqref{eq:Xsigma}. In the notation of \cite{AHS2025}, in which the entropy scales as $\lambda^{\alpha_E}$ under $r\mapsto\lambda r$ and, along the self-similar solutions, the entropy is a power $S\propto(\text{enclosed mass})^{n/3}$ (see \cite[(1.21), (1.37)]{AHS2025}),
\begin{equation}\label{eq:AHSmap}
\sigma=\frac{n}{3}=\frac{(2-\gamma)\,\alpha_E}{4-3\gamma+\alpha_E},\qquad
\alpha_E=\frac{\sigma\,(3\gamma-4)}{\gamma+\sigma-2},\qquad
c_4=\frac{2(2-\gamma)}{2-\alpha_E}.
\end{equation}
The scaling is mass-critical (i.e.\ the mass label does not scale) iff $\alpha_E=3\gamma-4$, and then $c_4=2/3$.
\end{proposition}
\begin{proof}
$e=\sigma c$ and \eqref{eq:twoparam} give $(3\gamma-4)a=(\sigma+\gamma-2)c$. With $b=(3a-c)/2$,
\begin{equation}
\frac{a}{b}=\frac{2a}{3a-c}=\frac{2(\sigma+\gamma-2)}{3(\sigma+\gamma-2)-(3\gamma-4)}=\frac{2(\gamma+\sigma-2)}{3\sigma-2}=c_4 ,
\end{equation}
and $c/b=(3c_4-2)$. Substituting $\sigma=(2-\gamma)\alpha_E/(4-3\gamma+\alpha_E)$ into $c_4$ gives $2(2-\gamma)/(2-\alpha_E)$ (check $\mathrm{V}20$), which is $1/b$ with $b=b(\alpha_E)=(2-\alpha_E)/(2(2-\gamma))$ the time-scaling exponent of \cite[(1.12)-(1.13)]{AHS2025}; inverting gives $\alpha_E(\sigma)$. Mass-criticality $3c_4-2=0$ means $c_4=2/3$, i.e.\ $\alpha_E=3\gamma-4$.
\end{proof}
\smallskip\noindent
Thus, in the notation of \cite{AHS2025}, the case $\gamma=4/3$ of Corollary \ref{cor:cases}$(\mathrm{d})$ is the mass-critical point $(\alpha_E,\gamma)=(0,4/3)$ where the scaling fixes the mass label and hence is compatible with an arbitrary $S(m)$; and the special exponent $\sigma=2/3$ at $\gamma=4/3$ (case $\mathrm{E}$) is the entropy exponent for which the full group \eqref{eq:twoparam} with $c\neq0$ preserves $S$: indeed at $\gamma=4/3$, \eqref{eq:twoparam} gives $e=\tfrac23c$.

\begin{proposition}[Scaling for constant entropy]\label{prop:scale}
\emph{Hypotheses:} as in Theorem \ref{thm:class} with $S\equiv K>0$. The scaling generator is $t\partial_t+(2-\gamma)r\partial_r+(4-3\gamma)m\partial_m$, i.e.\ $(t,r,m)\mapsto(\lambda t,\lambda^{2-\gamma}r,\lambda^{4-3\gamma}m)$, under which $\tilde\rho\mapsto\lambda^{-2}\tilde\rho$.
\end{proposition}
\begin{proof}
Case $\mathrm{B}$ with $\sigma=0$ gives $c_4=2-\gamma$ and $\xi=(3(2-\gamma)-2)m=(4-3\gamma)m$. Moreover $r^2r_m\mapsto\lambda^{2(2-\gamma)+(2-\gamma)-(4-3\gamma)}r^2r_m=\lambda^{2}r^2r_m$.
\end{proof}
\smallskip\noindent
The factor $\lambda^{-2}$ is the density scaling of the isothermal Larson-Penston problem \cite{GuoHadzicJang2021}. At $\gamma=4/3$, the mass exponent $4-3\gamma$ vanishes.

\begin{proposition}[Exact symmetry defect]\label{prop:defect}
\emph{Hypotheses:} $\gamma>1$, $S\in C^2(I)$ (positivity not needed), $r\in C^2(\Omega)$ with $r,r_m>0$. For $X$ as in \eqref{eq:X},
\begin{equation}\label{eq:defect}
X^{(2)}\Delta\Big|_{\Delta=0}=r^{1-2\gamma}\,r_m^{-\gamma-1}\Bigl[\gamma\,\bigl(r\,r_{mm}+2\,r_m^{2}\bigr)\,\mathcal C[S]-r\,r_m\,\bigl(\mathcal C[S]\bigr)'\Bigr].
\end{equation}
In particular, for $X_2=3t\partial_t+2r\partial_r$ and $\gamma=\tfrac43+\varepsilon$ one has $\mathcal C[S]=6\varepsilon S$ and
\begin{equation}\label{eq:defect2}
X_2^{(2)}\Delta\Big|_{\Delta=0}=6\varepsilon\,r^{1-2\gamma}r_m^{-\gamma-1}\Bigl[\gamma\,(r\,r_{mm}+2r_m^2)\,S-r\,r_m\,S'\Bigr].
\end{equation}
\end{proposition}
\begin{proof}
Symbolic evaluation of \eqref{eq:invcond} for \eqref{eq:X} with $S$ an arbitrary function (check $\mathrm{V}15$). The right side vanishes identically iff $\mathcal C[S]\equiv0$ (the coefficients of $rr_{mm}$ and of $rr_m$ are $\gamma\mathcal C$ and $-\mathcal C'$), recovering Theorem \ref{thm:class}.
\end{proof}
\begin{remark}[Near-critical index and approximate symmetry]
Adding to $X_2$ a correction $\varepsilon\,(c_1'\,t\partial_t+(3c_4'-2c_1')m\partial_m+c_4'\,r\partial_r)$ changes $\mathcal C$ at order $\varepsilon$ by $\mathcal C_{\gamma=4/3}[c_1',c_4']$, so first-order cancellation of \eqref{eq:defect2} requires $\mathcal C_{4/3}[c_1',c_4']+6S\equiv0$, i.e.\ $(2c_1'-3c_4')mS'=-2(c_4'-\tfrac23c_1')S-6S$: solvable only if $mS'/S$ is constant, i.e.\ for power laws. Hence, within the class \eqref{eq:X} no first-order approximate symmetry deforms $X_2$ for generic $S$. For the approximate-symmetry framework, see \cite{Shamaoon2023b}.
\end{remark}

\section{Conservation laws and virial identities}\label{sec:cons}
Symmetries are only as useful as the identities they produce. Because the equation is variational, Noether's theorem applies, and the question is which of the scalings of Section \ref{sec:lie} actually leave the action invariant. The answer is surprisingly restrictive. When a scaling is not variational, it still produces a balance law for integrated quantities, which we state as a virial identity.

\smallskip\noindent
Let $Q=\eta-\tau r_t-\xi r_m$ be the characteristic of $X$. $X$ is a \emph{divergence (Noether) symmetry} of $\mathcal L$ if
\begin{equation}\label{eq:Noethercond}
X^{(1)}\mathcal L+\mathcal L\,(\D_t\tau+\D_m\xi)=\D_tB^t+\D_mB^m
\end{equation}
for some $C^1$ functions $B^t,B^m$ of $(t,m,r,r_t,r_m)$. Then
\begin{equation}\label{eq:current}
T^t=\tau\mathcal L+\mathcal L_{r_t}Q-B^t,\qquad T^m=\xi\mathcal L+\mathcal L_{r_m}Q-B^m
\end{equation}
satisfy, identically in $r(t,m)\in C^2$,
\begin{equation}\label{eq:Noetheridentity}
\D_tT^t+\D_mT^m=-Q\,\frac{\delta\mathcal L}{\delta r}=Q\,\Delta[r],
\end{equation}
so $(T^t,T^m)$ is a conserved current on classical solutions \cite{Noether1918,Olver1993}. For $X=\partial_t$ ($\tau=1$, $Q=-r_t$, $B=0$) one recovers \eqref{eq:energy}.

\begin{proposition}[Weights of the scalings]\label{prop:weights}
\emph{Hypotheses:} $X$ as in \eqref{eq:X} with $c_0=0$, $S=Km^{\sigma}$, $r\in C^2$ with $r,r_m>0$. The prolongation coefficients are $\eta^{t}=(c_4-c_1)r_t$ and $\eta^{m}=2(c_1-c_4)r_m$, and $\D_t\tau+\D_m\xi=3c_4-c_1$. The three parts of $\mathcal L$ satisfy
\begin{align}
X^{(1)}\bigl(\tfrac12r_t^2\bigr)+\tfrac12r_t^2\,(\D_t\tau+\D_m\xi)&=(5c_4-3c_1)\,\tfrac12r_t^2,\label{eq:wkin}\\
X^{(1)}\Bigl(\frac{Gm}{r}\Bigr)+\frac{Gm}{r}\,(\D_t\tau+\D_m\xi)&=(5c_4-3c_1)\,\frac{Gm}{r},\label{eq:wgrav}\\
X^{(1)}(e)+e\,(\D_t\tau+\D_m\xi)&=\bigl[c_1(1-2\gamma-2\sigma)+3c_4(\sigma+1)\bigr]\,e .\label{eq:wint}
\end{align}
The weights in \eqref{eq:wkin}-\eqref{eq:wint} coincide iff \eqref{eq:powerclass} holds, and then $X^{(1)}\mathcal L+\mathcal L(\D_t\tau+\D_m\xi)=w\,\mathcal L$ with $w=5c_4-3c_1$.
\end{proposition}
\begin{proof}
From \eqref{eq:eta1}, with $\tau=c_1t$, $\xi=(3c_4-2c_1)m$, $\eta=c_4r$:
\begin{align}
\eta^t&=c_4r_t-r_t\,c_1-0=(c_4-c_1)\,r_t,\\
\eta^m&=c_4r_m-0-r_m(3c_4-2c_1)=2(c_1-c_4)\,r_m .
\end{align}
Kinetic part: $X^{(1)}(\frac12r_t^2)=r_t\eta^t=2(c_4-c_1)\cdot\frac12r_t^2$, and
\begin{equation}
2(c_4-c_1)+(3c_4-c_1)=5c_4-3c_1 .
\end{equation}
Gravitational part: $(\xi\partial_m+\eta\partial_r)(Gm/r)=[(3c_4-2c_1)-c_4]\,Gm/r=2(c_4-c_1)\,Gm/r$, and the same sum gives $5c_4-3c_1$. Internal part, $e\propto m^{\sigma}r^{2-2\gamma}r_m^{1-\gamma}$:
\begin{equation}
\frac{X^{(1)}e}{e}=\underbrace{(3c_4-2c_1)\sigma}_{\xi\partial_m}+\underbrace{c_4(2-2\gamma)}_{\eta\partial_r}+\underbrace{2(c_1-c_4)(1-\gamma)}_{\eta^m\partial_{r_m}}=(3c_4-2c_1)\sigma+2c_1(1-\gamma),
\end{equation}
and adding $3c_4-c_1$:
\begin{equation}
(3c_4-2c_1)\sigma+2c_1(1-\gamma)+3c_4-c_1=c_1(1-2\gamma-2\sigma)+3c_4(\sigma+1).
\end{equation}
Equating this with $5c_4-3c_1$ gives $c_4(2-3\sigma)=2c_1(2-\gamma-\sigma)$, which is \eqref{eq:powerclass}.
\end{proof}

\begin{theorem}[Noether point symmetries and a conservation law]\label{thm:noether}
\emph{Hypotheses:} $\gamma>1$, $S\in C^2(I)$, $S>0$, $I$ connected, $(\mathrm{H}4)$ for the symmetry generators, $\mathcal L$ as in \eqref{eq:L}. The point symmetries of \eqref{eq:EPm} that are Noether symmetries of \eqref{eq:L} are $\partial_t$ (weight $0$) and a scaling with $w=0$, which exists exactly when
\begin{equation}\label{eq:line}
S(m)=K\,m^{\,14-10\gamma},\qquad X_N=5t\,\partial_t-m\,\partial_m+3r\,\partial_r\quad(c_1=5,\ c_4=3).
\end{equation}
For $X_N$ and $Q=3r-5tr_t+mr_m$ the current
\begin{align}
T^t&=5t\,\mathcal L+r_t\,Q=3r\,r_t+m\,r_mr_t-5t\,\mathcal E,\label{eq:Tt}\\
T^m&=-m\,\mathcal L+S\,r^{2-2\gamma}r_m^{-\gamma}\,Q,\label{eq:Tm}
\end{align}
satisfies, identically in $r\in C^2$ with $r,r_m>0$, $\D_tT^t+\D_mT^m=Q\,\Delta[r]$; hence $\D_tT^t+\D_mT^m=0$ on classical solutions.
\end{theorem}
\begin{proof}
By Theorem \ref{thm:class} every point symmetry has the form \eqref{eq:X}. If $X^{(1)}\mathcal L+\mathcal L(\D_t\tau+\D_m\xi)=w\mathcal L$ with $w\ne0$ and \eqref{eq:Noethercond} holds, then $w\mathcal L=\D_tB^t+\D_mB^m$, and applying the Euler operator (which annihilates divergences),
\begin{equation}
w\,\frac{\delta\mathcal L}{\delta r}=0,
\end{equation}
contradicting $\delta\mathcal L/\delta r=-\Delta[r]\not\equiv0$. So $w=0$, i.e.\ $c_4=\tfrac35c_1$. Inserting this in \eqref{eq:Xsigma} (for $\sigma\neq2/3$):
\begin{align}
\frac{2(\gamma+\sigma-2)}{3\sigma-2}&=\frac35\\
\iff 10(\gamma+\sigma-2)&=3(3\sigma-2)\iff \sigma=14-10\gamma .
\end{align}
At $(\gamma,\sigma)=(4/3,2/3)$ the line meets the degenerate case and $5c_4=3c_1$ is realised inside the three-dimensional algebra of Corollary \ref{cor:cases}$(\mathrm{e})$. At $\gamma=4/3$ with $S$ arbitrary, $c_4=2c_1/3$ gives $w=c_1/3\ne0$: not variational. Conversely, on the line $w=0$, so \eqref{eq:Noethercond} holds with $B=0$ and \eqref{eq:Noetheridentity} gives the stated identity. With $Q=\eta-\tau r_t-\xi r_m$, $\tau=5t$, $\xi=-m$, $\eta=3r$, and $\mathcal L_{r_t}=r_t$, $\mathcal L_{r_m}=Sr^{2-2\gamma}r_m^{-\gamma}$, the currents \eqref{eq:current} are \eqref{eq:Tt}-\eqref{eq:Tm}; the second form of $T^t$ follows from $5t\mathcal L+r_t(3r-5tr_t+mr_m)=3rr_t+mr_mr_t-5t(\tfrac12r_t^2-\tfrac{Gm}{r}+e)$. The identity was verified symbolically for $(\gamma,\sigma)=(7/5,0),(6/5,2),(3/2,-1)$ and fails off the line (check $\mathrm{V}13$).
\end{proof}
\begin{remark}[Position of the variational line]
In the parametrisation of Proposition \ref{prop:AHS} the variational scaling corresponds to $\alpha_E=(10\gamma-14)/3$, and $\alpha_E-(3\gamma-4)=(\gamma-2)/3<0$ for $\gamma<2$: it lies in the mass-\emph{sub}critical region of \cite{AHS2025} (the supercritical collapse solutions of \cite{GHJS2022,AHS2025} have $\alpha_E>3\gamma-4$). The conservation law is therefore a property of the stable/expanding regime, e.g.\ the constant-entropy diatomic gas $(\gamma,\sigma)=(7/5,0)$.
\end{remark}

\begin{proposition}[Virial identities]\label{prop:virial}
\emph{Hypotheses:} $\gamma>1$, $(\mathrm{H}1)$-$(\mathrm{H}3)$, $r$ a classical solution with $r\in C^2(J\times(0,M))$. Define
\begin{equation}\label{eq:virialdefs}
\begin{aligned}
K_{\rm kin}&=\tfrac12\!\int_0^M\! r_t^2\dd m, & W&=-G\!\int_0^M\!\frac{m}{r}\dd m, & U&=\int_0^M e\dd m,\\
I_{\rm in}&=\int_0^M r^2\dd m, & \mathcal Q&=\int_0^M r\,r_t\dd m, & E&=K_{\rm kin}+W+U .
\end{aligned}
\end{equation}
Then
\begin{align}
\tfrac12\ddot I_{\rm in}&=2K_{\rm kin}+W+3(\gamma-1)U,\label{eq:virial1}\\
&=3(\gamma-1)E+(5-3\gamma)K_{\rm kin}+(4-3\gamma)W .\label{eq:virial2}
\end{align}
For $\gamma=4/3$ the scaling $3t\partial_t+2r\partial_r$ gives
\begin{equation}\label{eq:scaleid}
\frac{\mathrm d}{\mathrm dt}\bigl(2\mathcal Q-3tE\bigr)=\int_0^M\mathcal L\dd m=K_{\rm kin}-W-U .
\end{equation}
\end{proposition}
\begin{proof}
Pointwise, on solutions,
\begin{equation}\label{eq:pointvirial}
r\,r_{tt}=-\frac{Gm}{r}-\bigl(r^{3}P\bigr)_m+3(\gamma-1)\,e,
\end{equation}
because $r\cdot r^2P_m=(r^3P)_m-3r^2r_mP$ and $r^2r_mP=S\,r^{2-2\gamma}r_m^{1-\gamma}=(\gamma-1)e$ (check $\mathrm{V}3$). By $(\mathrm{H}3)$ we may differentiate under the integral and integrate \eqref{eq:pointvirial} with $[r^3P]_0^M=0$:
\begin{equation}
\tfrac12\ddot I_{\rm in}=\int_0^M(r_t^2+rr_{tt})\dd m=2K_{\rm kin}+W+3(\gamma-1)U,
\end{equation}
which is \eqref{eq:virial1}; substituting $U=E-K_{\rm kin}-W$ gives \eqref{eq:virial2}. For $\gamma=4/3$ the same computation gives $\dot{\mathcal Q}=\int(r_t^2+rr_{tt})\dd m=2K_{\rm kin}+W+U$, and $\dot E=0$ (Proposition \ref{prop:var}), so
\begin{equation}
\frac{\mathrm d}{\mathrm dt}\bigl(2\mathcal Q-3tE\bigr)=2(2K_{\rm kin}+W+U)-3(K_{\rm kin}+W+U)=K_{\rm kin}-W-U .
\end{equation}
\end{proof}
\smallskip\noindent
Identity \eqref{eq:scaleid} is the Noether-type identity of the non-variational critical scaling: its right-hand side is the integrated Lagrangian, i.e.\ the weight $w=5c_4-3c_1=1$ (for $c_1=3$, $c_4=2$) times $\int\mathcal L\dd m$.

\section{Exact solutions}\label{sec:exact}
We now collect explicit solutions. They serve two purposes: they show concretely what the symmetries of Section \ref{sec:lie} select, and they provide exact backgrounds against which the numerical scheme of Section \ref{sec:num} and the perturbation theory of Section \ref{sec:pert} are tested.

\begin{proposition}[Einstein-de Sitter-type collapse, any $\gamma$]\label{prop:eds}
\emph{Hypotheses:} $\gamma>1$, $S\equiv K>0$ constant, domain $\{t>0,\ m>0\}$. Then
\begin{equation}\label{eq:EdS}
r(t,m)=A\,m^{1/3}\,t^{2/3},\qquad A^3=\tfrac92\,G,
\end{equation}
is a classical solution of \eqref{eq:EPm}; so is $r=Am^{1/3}(t_*-t)^{2/3}$ on $t<t_*$, collapsing to $r=0$ at $t=t_*$. The physical density is $\rho=1/(6\pi G\,t^{2})$, uniform in $m$, and the pressure force vanishes identically.
\end{proposition}
\begin{proof}
$r^2r_m=A^3t^2/3$ is independent of $m$, so $P$ is independent of $m$ and $P_m=0$. Then
\begin{equation}
r_{tt}=-\tfrac29Am^{1/3}t^{-4/3},\qquad -\frac{Gm}{r^2}=-\frac{G}{A^{2}}\,m^{1/3}t^{-4/3},
\end{equation}
which agree iff $A^3=9G/2$; and $\rho=1/(4\pi r^2r_m)=3/(4\pi A^3t^2)=1/(6\pi Gt^2)$ (check $\mathrm{V}4$). The time-reversed solution follows because \eqref{eq:EPm} contains no first-order time derivative.
\end{proof}

\begin{figure}[H]
\begin{minipage}[t]{0.49\textwidth}
\centering
\includegraphics[width=\linewidth]{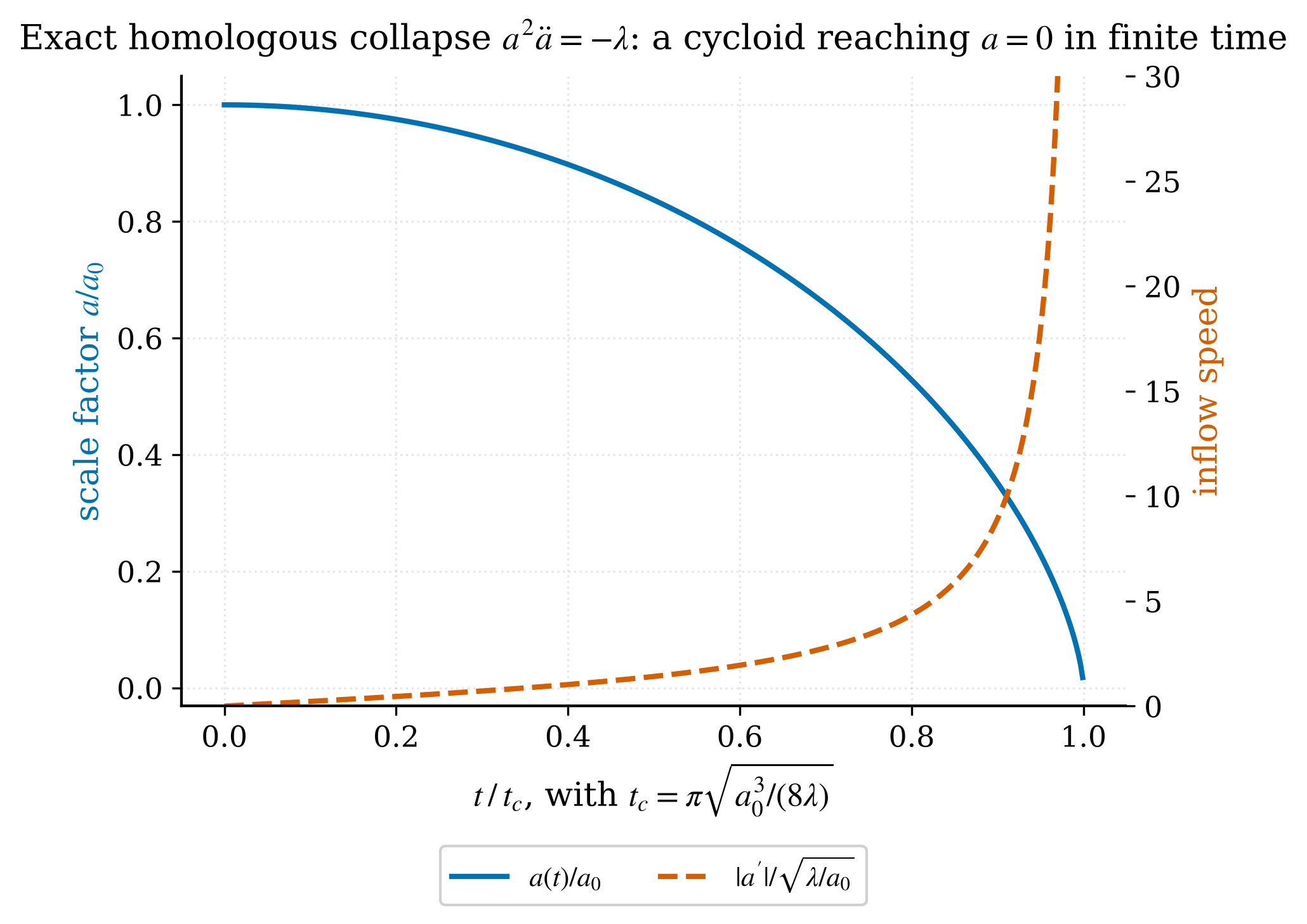}
\captionof{figure}{Exact homologous collapse of Proposition \ref{prop:uniform} ($\gamma=4/3$, $\lambda=1$): scale factor $a(t)/a_0$ (left axis, solid) and inflow speed $|\dot a|/\sqrt{\lambda/a_0}$ (right axis, dashed, from $\dot a^2=2\lambda(1/a-1/a_0)$) versus $t/t_c$. The cycloid reaches $a=0$ in finite time with diverging speed.}\label{fig:cyc}
\end{minipage}\hfill
\begin{minipage}[t]{0.49\textwidth}
\centering
\includegraphics[width=\linewidth]{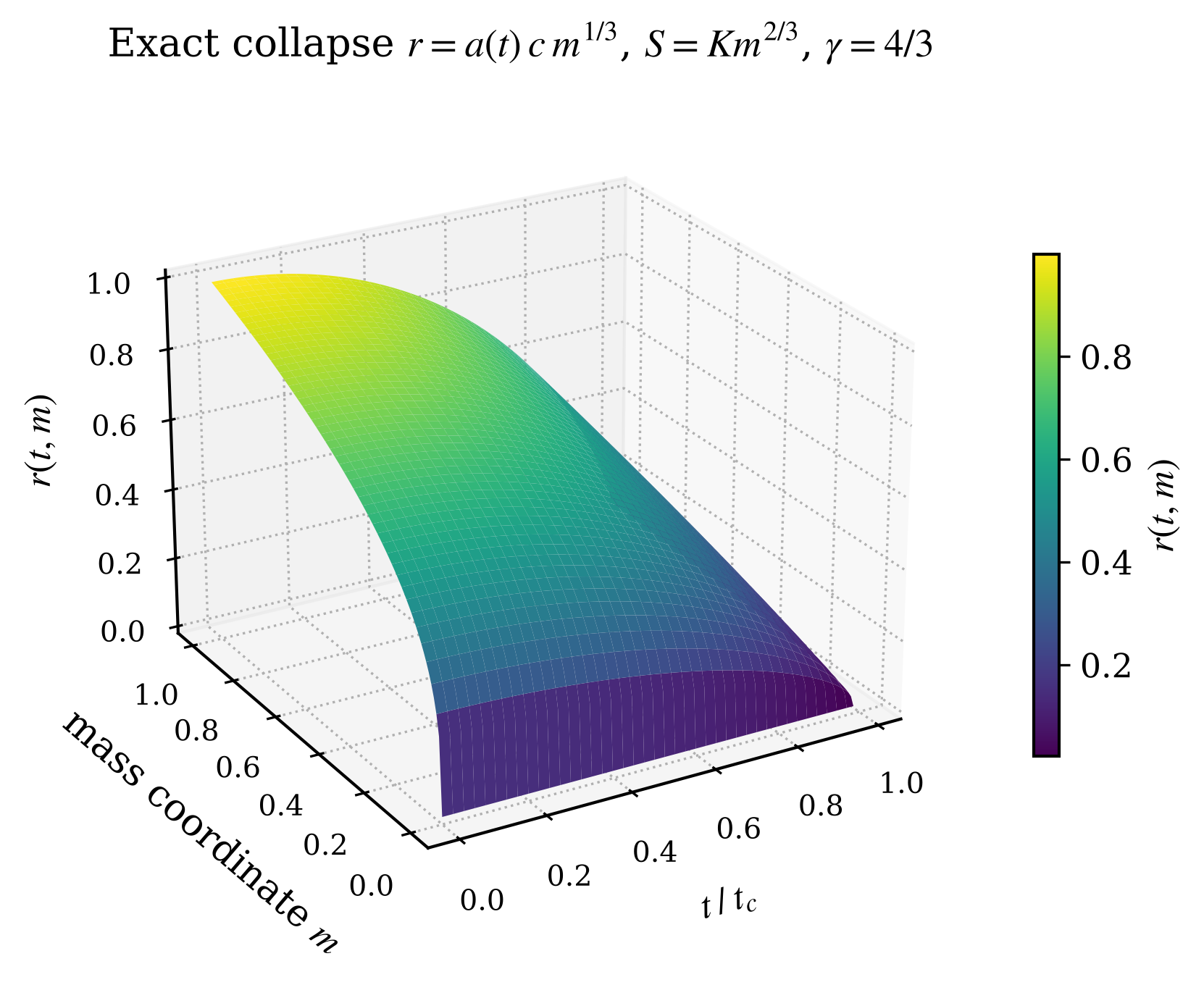}
\captionof{figure}{Surface $r(t,m)=a(t)\,c\,m^{1/3}$ ($S=Km^{2/3}$, $\gamma=4/3$, $c=a_0=G=K=1$) over $(t/t_c,m)$: all shells reach $r=0$ simultaneously at $t=t_c$.}\label{fig:uni3d}
\end{minipage}
\end{figure}

\begin{proposition}[Uniform-density reduction, any $\gamma$]\label{prop:uniform}
\emph{Hypotheses:} $\gamma>1$, $S=Km^{2/3}$ with $K>0$, $c>0$, $a\in C^2(J)$, $a>0$. Then $r=a(t)\,c\,m^{1/3}$ is a classical solution on $J\times(0,\infty)$ iff
\begin{equation}\label{eq:aode}
\ddot a=-\frac{G}{c^3a^{2}}-\tfrac23\,3^{\gamma}K\,c^{1-3\gamma}\,a^{2-3\gamma}.
\end{equation}
For $\gamma=4/3$, $a^2\ddot a=-\lambda$ with $\lambda=(G+2\cdot3^{1/3}K)/c^{3}$, and the solution with $a(0)=a_0$, $\dot a(0)=0$ is the cycloid
\begin{equation}\label{eq:cyc}
a=\frac{a_0}{2}(1+\cos\vartheta),\qquad t=\sqrt{\frac{a_0^{3}}{8\lambda}}\,(\vartheta+\sin\vartheta),\qquad t_c=\pi\sqrt{\frac{a_0^{3}}{8\lambda}},\qquad \vartheta\in[0,\pi).
\end{equation}
\end{proposition}
\begin{proof}
With $r=acm^{1/3}$: $r_m=\frac{ac}{3}m^{-2/3}$ and $S\,r^{-2\gamma}r_m^{-\gamma}=K\,3^\gamma(ac)^{-3\gamma}m^{2/3}$, so
\begin{equation}
r^2\partial_m\bigl(Sr^{-2\gamma}r_m^{-\gamma}\bigr)=\tfrac23\,3^{\gamma}K\,(ac)^{2-3\gamma}\,m^{1/3},
\end{equation}
and dividing \eqref{eq:EPm} by $c\,m^{1/3}$ gives \eqref{eq:aode} (check $\mathrm{V}5$). For $\gamma=4/3$, $a^{2-3\gamma}=a^{-2}$ and $\tfrac23 3^{4/3}=2\cdot3^{1/3}$. For the cycloid, $\dot a=\mathrm da/\mathrm d\vartheta\,/\,(\mathrm dt/\mathrm d\vartheta)$ and one verifies $a^2\ddot a=-\lambda$ (check $\mathrm{V}10$); the energy integral is $\tfrac12\dot a^2-\lambda/a=-\lambda/a_0$.
\end{proof}
\begin{remark}
For $S=Km^{2/3}$ the pressure \emph{increases} outward at uniform density, so this is an exact but non-stellar solution (a Lagrangian ball under prescribed external pressure). See Figures \ref{fig:cyc} and \ref{fig:uni3d}. Near $t_c$, $a\simeq(9\lambda/2)^{1/3}(t_c-t)^{2/3}$, the zero-energy behaviour consistent with Proposition \ref{prop:eds}, so $\tilde\rho\propto a^{-3}\propto(t_c-t)^{-2}$.
\end{remark}

\begin{proposition}[Self-similar reduction]\label{prop:ss}
\emph{Hypotheses:} $\gamma>1$, $S\equiv K>0$, $R\in C^2(\mathcal S)$ on an open interval $\mathcal S\subset(0,\infty)$ with $R,R_s>0$, $t>0$ and $m=s\,t^{4-3\gamma}$ ranging in $I$. Let $r=t^{2-\gamma}R(s)$, $s=m\,t^{3\gamma-4}$ (the invariants of Proposition \ref{prop:scale}), and $\Pi:=K(R^2R_s)^{-\gamma}$. Then $P=t^{-2\gamma}\Pi(s)$ and $r$ is a classical solution of \eqref{eq:EPm} iff
\begin{equation}\label{eq:ssode}
(3\gamma-4)^2s^2R_{ss}+(3\gamma-4)(\gamma-1)\,sR_s+(\gamma-1)(\gamma-2)R+\frac{Gs}{R^{2}}+R^{2}\,\Pi_s=0 ,
\end{equation}
equivalently
\begin{equation}\label{eq:ssode2}
(3\gamma-4)^2s^2R_{ss}+(3\gamma-4)(\gamma-1)\,sR_s+(\gamma-1)(\gamma-2)R+\frac{Gs}{R^{2}}=K\gamma\,(R^2R_s)^{-\gamma}\Bigl(\frac{R^2R_{ss}}{R_s}+2RR_s\Bigr).
\end{equation}
The Einstein-de Sitter solution \eqref{eq:EdS} is $R=As^{1/3}$.
\end{proposition}
\begin{proof}
With $r=t^{2-\gamma}R(s)$ and $s=mt^{3\gamma-4}$:
\begin{equation}
r_m=t^{2\gamma-2}R_s,\qquad r^{-2\gamma}r_m^{-\gamma}=t^{-2\gamma(2-\gamma)-\gamma(2\gamma-2)}(R^2R_s)^{-\gamma}=t^{-2\gamma}(R^2R_s)^{-\gamma},
\end{equation}
\smallskip\noindent
so $P=t^{-2\gamma}\Pi(s)$ and $r^2P_m=t^{4-2\gamma}R^2\,t^{-2\gamma}t^{3\gamma-4}\Pi_s=t^{-\gamma}R^2\Pi_s$. Likewise $r_{tt}$ and $Gm/r^2$ equal $t^{-\gamma}$ times the first four terms of \eqref{eq:ssode}; dividing by $t^{-\gamma}$ removes all $t$-dependence (symbolic derivation). Differentiating $\Pi$ gives $R^2\Pi_s=-K\gamma(R^2R_s)^{-\gamma}(R^2R_{ss}/R_s+2RR_s)$, hence \eqref{eq:ssode2}. $R=As^{1/3}$ satisfies \eqref{eq:ssode2} identically ($\mathrm{V}11$).
\end{proof}
\smallskip\noindent
At $\gamma=4/3$ the derivative terms $s^2R_{ss}$, $sR_s$ drop out and $s=m$: critical collapse has no genuine similarity variable, and \eqref{eq:ssode} degenerates into the profile problem \eqref{eq:prof}. For $\gamma\in(1,4/3)$, \eqref{eq:ssode} is the Lagrangian counterpart of the self-similar problem whose smooth solutions were constructed in \cite{GHJS2022} (Yahil-type solutions \cite{Yahil1983}); the existence theory is not re-derived here.

\section{Series analysis: analytic centre expansion and surface expansion}\label{sec:series}
Global profiles of collapsing stars are usually found by shooting from the centre. For that one needs to know what a regular centre looks like, and how the profile ends at the surface. In mass coordinates both questions have clean answers, which we establish here.

\subsection{Homologous collapse of \texorpdfstring{$\gamma=4/3$}{gamma=4/3} polytropes}\label{sec:homol}
Let $\gamma=4/3$, $S\equiv K>0$ and seek $r=a(t)x(m)$ with $a\in C^2$, $a>0$, $x\in C^2(I)$, $x,x_m>0$. Then \eqref{eq:EPm} separates \cite{GoldreichWeber1980}:
\begin{align}
a^2\ddot a&=-\lambda,\label{eq:sepa}\\
K\,x^{2}\,\frac{\mathrm d}{\mathrm dm}\!\bigl(x^{-8/3}x_m^{-4/3}\bigr)+\frac{Gm}{x^{2}}&=\lambda x\quad\text{on }I ,\label{eq:prof}
\end{align}
for a separation constant $\lambda\in\mathbb R$ (substitute $r=ax$, use $r^{-2\gamma}r_m^{-\gamma}=a^{-4}x^{-8/3}x_m^{-4/3}$ and divide by $a^{-2}x$). With the comoving pressure $\Pi:=K\,x^{-8/3}x_m^{-4/3}$ the profile equation is the first-order system
\begin{equation}\label{eq:sys}
x_m=\Bigl(\frac{K}{\Pi\,x^{8/3}}\Bigr)^{3/4},\qquad \Pi_m=\frac{\lambda}{x}-\frac{Gm}{x^{4}} ,
\end{equation}
(the second-order form is numerically ill-conditioned at the centre because of cancellation; \eqref{eq:sys} removes it). The density is exactly $\tilde\rho=1/(x^2x_m)=(\Pi/K)^{3/4}$. Units $G=K=1$ are used for all numerical values, with $x\simeq c_0m^{1/3}$ and, by the scaling $(x,a,\lambda)\mapsto(\varsigma x,a/\varsigma,\lambda/\varsigma^3)$ which leaves $r=ax$ invariant, the normalisation $c_0=1$. The value $\lambda=1$ is exactly $x=m^{1/3}$, the Einstein-de Sitter solution \eqref{eq:EdS}.

\subsection{Analytic expansion at the regular centre (Briot-Bouquet)}
Put $u=m^{1/3}$ (homologous case) or $u=s^{1/3}$ (self-similar case), $y:=R/u$ (resp.\ $x/u$) and let $\Pi$ be the corresponding comoving pressure. The definition of $\Pi$ gives $y^2(y+uy_u)/3=(K/\Pi)^{1/\gamma}$ in both cases (since $R^2R_s=y^2(y+uy_u)/3$ for $R=uy$, $s=u^3$, and likewise for $x^2x_m$ with $m=u^3$), that is,
\begin{equation}\label{eq:BB1}
u\,y_u=F_1(y,\Pi):=3\Bigl(\frac{K}{\Pi}\Bigr)^{1/\gamma}y^{-2}-y ,
\end{equation}
and \eqref{eq:ssode} (self-similar) or \eqref{eq:sys} (homologous) gives
\begin{equation}\label{eq:BB2}
u\,\Pi_u=u^{2}H(u^{2},y,\Pi),
\end{equation}
where, writing $\hat N(y,\Pi)=2(3\gamma-4)^2(K/\Pi)^{2/\gamma}y^{-5}-(3\gamma-4)(\gamma-1)(K/\Pi)^{1/\gamma}y^{-2}-(\gamma-1)(\gamma-2)y-Gy^{-2}$,
\begin{align}
H_{\rm ss}(w,y,\Pi)&=\frac{3\,\hat N(y,\Pi)}{y^{2}-(3\gamma-4)^2\,w\,(K/\Pi)^{1/\gamma}/(\gamma\,\Pi\,y^{2})}\qquad(\text{self-similar}),\label{eq:Hss}\\
H_{\rm hom}(w,y,\Pi)&=3\Bigl(\frac{\lambda}{y}-\frac{G}{y^{4}}\Bigr)\qquad(\gamma=4/3,\ \text{homologous}).\label{eq:Hhom}
\end{align}
(Derivation of \eqref{eq:Hss}: $R=uy$, $s=u^3$, $R_s=W=(K/\Pi)^{1/\gamma}u^{-2}y^{-2}$; the numerator of the right side of \eqref{eq:sssys} below equals $u\hat N$ and its bracket equals $u^2[y^2-(3\gamma-4)^2u^2(K/\Pi)^{1/\gamma}/(\gamma\Pi y^2)]$; and $u\Pi_u=3s\Pi_s$. For \eqref{eq:Hhom}, $u\Pi_u=3u^3\Pi_m$ with \eqref{eq:sys}.)

\begin{theorem}[Analytic centre expansion]\label{thm:BB}
\emph{Hypotheses:} $\gamma>1$, $K,G>0$, $\lambda\in\mathbb R$, $A_0>0$, and $\Pi_0:=K\,(A_0^3/3)^{-\gamma}$. In the self-similar case $\gamma>1$ is arbitrary; in the homologous case $\gamma=4/3$. Then there exist $\nu>0$ and a \emph{unique} pair $(y,\Pi)$ of functions analytic on $(-\nu,\nu)$ with $y(0)=A_0$, $\Pi(0)=\Pi_0$, solving \eqref{eq:BB1}-\eqref{eq:BB2}; $y$ and $\Pi$ are even in $u$. Consequently
\begin{equation}\label{eq:centreseries}
R(s)=s^{1/3}\sum_{k\ge0}a_k\,s^{2k/3},\qquad\Bigl(\text{resp.}\ x(m)=m^{1/3}\sum_{k\ge0}c_k\,m^{2k/3}\Bigr),
\end{equation}
converges absolutely for $0<s<\nu^3$ (resp.\ $0<m<\nu^3$), with $a_0=A_0$ and $R_s>0$, and is a classical solution of \eqref{eq:ssode2} (resp.\ \eqref{eq:prof}).
\end{theorem}
\begin{proof}
Write $Y=(y,\Pi)$. Equations \eqref{eq:BB1}-\eqref{eq:BB2} have the Briot-Bouquet form $u\,Y_u=F(u,Y)$ with $F=(F_1,u^2H)$ analytic in a neighbourhood of $(0,Y_0)$, $Y_0=(A_0,\Pi_0)$, $y>0$, $\Pi>0$ (for $H_{\rm ss}$ the denominator equals $y^2\neq0$ at $u=0$). Since $(K/\Pi_0)^{1/\gamma}=A_0^3/3$,
\begin{equation}
F_1(A_0,\Pi_0)=3\cdot\frac{A_0^3}{3}A_0^{-2}-A_0=0,\qquad F(0,Y_0)=0 .
\end{equation}
The Jacobian of $F(0,\cdot)$ at $Y_0$ is
\begin{equation}
\partial_YF(0,Y_0)=\begin{pmatrix}\partial_yF_1&\partial_\Pi F_1\\0&0\end{pmatrix},\qquad
\partial_yF_1=-6\Bigl(\frac{K}{\Pi_0}\Bigr)^{1/\gamma}A_0^{-3}-1=-3,\qquad \partial_\Pi F_1=-\frac{A_0}{\gamma\Pi_0},
\end{equation}
with eigenvalues $\{-3,0\}$ (check $\mathrm{V}21$), none of which is a positive integer. By the Briot-Bouquet theorem \cite{Hille1976} (the nonresonance condition is exactly that no eigenvalue of $\partial_YF(0,Y_0)$ equals a positive integer, so that the coefficient recursion $(k-\partial_YF(0,Y_0))Y_k=\ldots$ is uniquely solvable for $k\ge1$ and the majorant series converges), there is a unique analytic solution with $Y(0)=Y_0$. $F$ depends on $u$ only through $u^2$ and $u\,\mathrm d/\mathrm du$ is invariant under $u\mapsto-u$, so $Y(-u)$ is also a solution with the same value at $0$; uniqueness gives evenness. Finally $R=uy$ satisfies $R_s=W>0$ by \eqref{eq:BB1} and solves \eqref{eq:ssode2} because \eqref{eq:BB2} is equivalent to \eqref{eq:ssode} for $u\ne0$.
\end{proof}
\begin{remark}
The eigenvalue $0$ reflects the free central amplitude $A_0$, and the eigenvalue $-3$ the singular solution $y\sim u^{-3}$, i.e.\ $R\sim u^{-2}\propto s^{-2/3}$, which is not regular at the centre. The theorem therefore identifies the regular centre of \emph{both} the self-similar and the homologous flows as a one-parameter analytic family, in agreement with the smoothness selection principle used in \cite{GHJS2022,AHS2025}. For comparison, in the non-isentropic case with $\gamma>4/3$ and $n>2$ (that is $S\propto m^{\sigma}$, $\sigma=n/3>2/3$), smoothness at the centre forces the central density to take the Einstein-de Sitter value, $\rho(0)=1/(6\pi)$ for $G=1$ \cite[Lemma 1.10]{AHS2025}; this is exactly the value $A_0^3=9G/2$, $\rho=1/(6\pi G t^2)$, of Proposition \ref{prop:eds} at which $a_1$ in \eqref{eq:a1} vanishes, i.e.\ where the pressure drops out of the leading-order balance.
\end{remark}

\begin{theorem}[Closed-form coefficients]\label{thm:ssseries}
\emph{Hypotheses:} those of Theorem \ref{thm:BB}, self-similar case. The coefficients of \eqref{eq:centreseries} satisfy
\begin{align}
a_1&=\frac{A_0^{\,3\gamma-3}\,\bigl(9G-2A_0^{3}\bigr)}{30\cdot3^{\gamma}\,K\,\gamma},\label{eq:a1}\\
a_2&=\frac{A_0^{\,6\gamma-7}}{12600\cdot9^{\gamma}K^{2}\gamma^{2}}\Bigl[A_0^{6}\,(-360\gamma^2+1000\gamma-536)\nonumber\\
&\qquad+A_0^{3}G\,(1620\gamma^2-4950\gamma+2574)+G^{2}(2025\gamma-729)\Bigr],\label{eq:a2}
\end{align}
and higher coefficients follow from the recursion of Theorem \ref{thm:BB} (the symbolic expression of $a_3$ is in the supplementary scripts).
\end{theorem}
\begin{proof}
Expand $y=A_0+a_1u^2+\cdots$, $\Pi=\Pi_0+\Pi_1u^2+\cdots$. At order $u^2$, \eqref{eq:BB1} and \eqref{eq:BB2} give
\begin{align}
2a_1&=-3a_1+\partial_\Pi F_1\,\Pi_1\ \Longrightarrow\ 5a_1=-\frac{A_0}{\gamma\Pi_0}\Pi_1,\\
2\Pi_1&=H_{\rm ss}(0,A_0,\Pi_0)=\frac{3\hat N(A_0,\Pi_0)}{A_0^2}.
\end{align}
At the centre, $(K/\Pi_0)^{1/\gamma}=A_0^3/3$ and
\begin{equation}
\hat N(A_0,\Pi_0)=A_0\Bigl[\tfrac{2}{9}(3\gamma-4)^2-\tfrac13(3\gamma-4)(\gamma-1)-(\gamma-1)(\gamma-2)\Bigr]-\frac{G}{A_0^2}=\frac{2}{9}A_0-\frac{G}{A_0^{2}},
\end{equation}
because $2(3\gamma-4)^2/9-(3\gamma-4)(\gamma-1)/3-(\gamma-1)(\gamma-2)=2/9$. Eliminating $\Pi_1$,
\begin{equation}
a_1=-\frac{A_0}{5\gamma\Pi_0}\cdot\frac{3}{2A_0^2}\Bigl(\frac29A_0-\frac{G}{A_0^2}\Bigr)=\frac{9G-2A_0^3}{30\,\gamma\,A_0^{3}\,\Pi_0},
\end{equation}
and $\Pi_0=K3^{\gamma}A_0^{-3\gamma}$ gives \eqref{eq:a1}. Formula \eqref{eq:a2} is the order-$u^4$ equation solved symbolically; it was checked against the symbolic recursion at a random parameter point to $13$ digits ($\mathrm{V}18$).
\end{proof}
\smallskip\noindent
In particular $a_1=0$ iff $A_0^3=9G/2$, the Einstein-de Sitter core of Proposition \ref{prop:eds}.

\begin{figure}[H]
\centering
\includegraphics[width=0.49\textwidth]{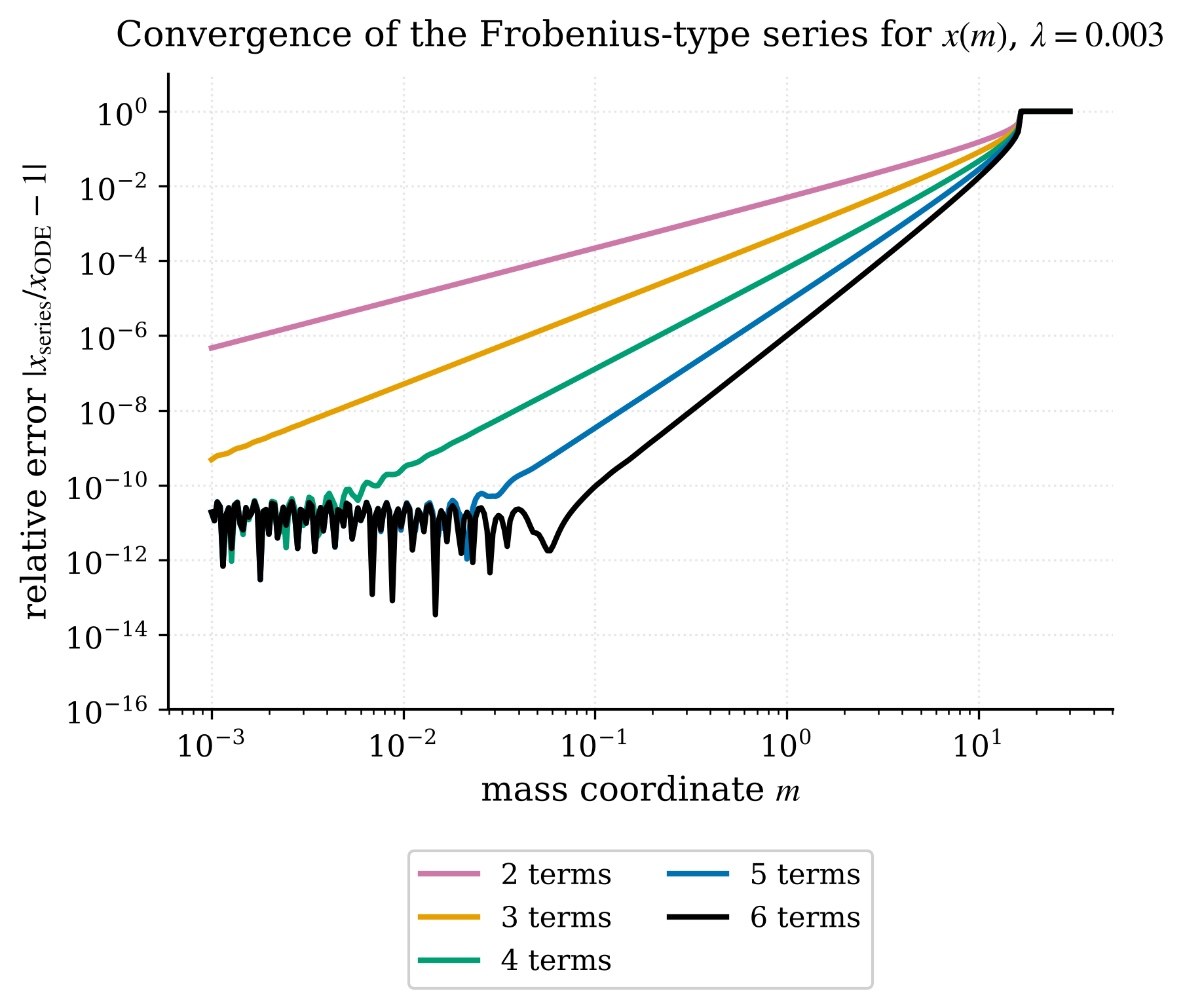}
\caption{Relative error $|x_{\rm series}/x_{\rm ODE}-1|$ of the truncated central series of Theorem \ref{thm:series} ($2$ to $6$ terms) at $\lambda=0.003$, against the numerical solution of \eqref{eq:sys}. The plateau near $10^{-11}$ is the numerical error floor; at large $m$ the series leaves its disc of convergence, and the numerical solution should be used.}\label{fig:series}
\end{figure}

\begin{theorem}[Homologous profile coefficients]\label{thm:series}
\emph{Hypotheses:} those of Theorem \ref{thm:BB}, homologous case, $G=K=1$, $A_0=c_0=1$. The expansion $x=m^{1/3}\sum_kc_km^{2k/3}$ has $c_0=1$,
\begin{align}
c_1&=\frac{3^{2/3}(1-\lambda)}{40},\qquad
c_2=\frac{3^{1/3}\,(118\lambda^2-191\lambda+73)}{22400},\label{eq:c12}\\
c_3&=-\frac{437\lambda^3}{302400}+\frac{2671\lambda^2}{806400}-\frac{3827\lambda}{1612800}+\frac{2447}{4838400},\label{eq:c3}
\end{align}
and $c_4,c_5$ in Table \ref{tab:series}. All $c_k$ ($k\ge1$) vanish at $\lambda=1$.
\end{theorem}
\begin{proof}
By Theorem \ref{thm:BB} the expansion exists and is analytic. With $w=m^{2/3}$, $y(w)=\sum c_kw^k$, $x=m^{1/3}y$, in units $G=K=1$:
\begin{equation}
x_m=\frac{y+2wy'}{3m^{2/3}},\qquad \Pi=3^{4/3}\,y^{-8/3}(y+2wy')^{-4/3},\qquad 2\,\frac{\mathrm d\Pi}{\mathrm dw}=3\Bigl(\frac{\lambda}{y}-\frac{1}{y^{4}}\Bigr).
\end{equation}
Matching powers of $w$ determines $c_k$ recursively (symbolic recursion, five terms). For $\lambda=1$, $y\equiv1$ solves the last equation, whence $c_k(1)=0$ by uniqueness.
\end{proof}
\begin{table}[H]
\centering
\caption{Series coefficients of $x(m)=m^{1/3}\sum c_km^{2k/3}$ ($G=K=c_0=1$). $c_k(1)=0$ for $k\ge1$ is verified exactly.}\label{tab:series}
\small
\begin{tabular}{@{}lcccccc@{}}
\toprule
$\lambda$ & $c_0$ & $c_1$ & $c_2$ & $c_3$ & $c_4$ & $c_5$ \\
\midrule
$0$ & $1$ & $5.2002\times10^{-2}$ & $4.7002\times10^{-3}$ & $5.0575\times10^{-4}$ & $5.9490\times10^{-5}$ & $7.3904\times10^{-6}$ \\
$0.003$ & $1$ & $5.1846\times10^{-2}$ & $4.6634\times10^{-3}$ & $4.9866\times10^{-4}$ & $5.8230\times10^{-5}$ & $7.1758\times10^{-6}$ \\
$1$ & $1$ & $0$ & $0$ & $0$ & $0$ & $0$ \\
\bottomrule
\end{tabular}
\end{table}
\smallskip\noindent
For $\lambda=0.3$ the six-term series agrees with the numerical solution of \eqref{eq:sys} (tolerance $10^{-12}$) to $2\times10^{-12}$ (relative) at $m=0.05$ and to $6\times10^{-8}$ at $m=1$ (truncation error); the numerical solution does not depend on the integration start point ($m_0=10^{-6},\dots,10^{-2}$; agreement to $8$ digits at $m=1$). Figure \ref{fig:series} shows the error for increasing truncation order.

\begin{proposition}[Exact cross-validation at $\gamma=4/3$]\label{prop:cross}
\emph{Hypotheses:} $\gamma=4/3$, $G=K=1$, $A_0^3=\tfrac92\lambda$, $\lambda>0$. The coefficients of Theorems \ref{thm:ssseries} and \ref{thm:series} satisfy
\begin{equation}\label{eq:cross}
a_k=A_0\,c_k(\lambda),\qquad k=1,2,3,
\end{equation}
\smallskip\noindent
as exact identities in $\lambda$.
\end{proposition}
\begin{proof}
At $\gamma=4/3$ the similarity variable is $s=m$ and $r=t^{2/3}R(m)$, i.e.\ $a(t)=t^{2/3}$ with $a^2\ddot a=-2/9$. The gauge $(x,a,\lambda)\mapsto(\varsigma x,a/\varsigma,\lambda/\varsigma^{3})$ leaves $r=ax$ invariant; $\varsigma^3=9\lambda/2$ maps the profile of Theorem \ref{thm:series} (normalised by $c_0=1$) to $R=A_0x$, $A_0=\varsigma$. Substituting $\gamma=4/3$, $A_0^3=9\lambda/2$ in \eqref{eq:a1}, \eqref{eq:a2} and the symbolic $a_3$ gives $a_k/A_0=c_k(\lambda)$ identically for $k=1,2,3$ ($\mathrm{V}17$).
\end{proof}
For example, \eqref{eq:a1} at $\gamma=4/3$ gives $a_1/A_0=(9-2A_0^3)/(40\cdot3^{4/3})=3^{2/3}(1-\lambda)/40=c_1$. The two series come from different reductions of the same PDE, so the identity is a nontrivial consistency test of both.

\begin{figure}[H]
\begin{minipage}[t]{0.49\textwidth}
\centering
\includegraphics[width=\linewidth]{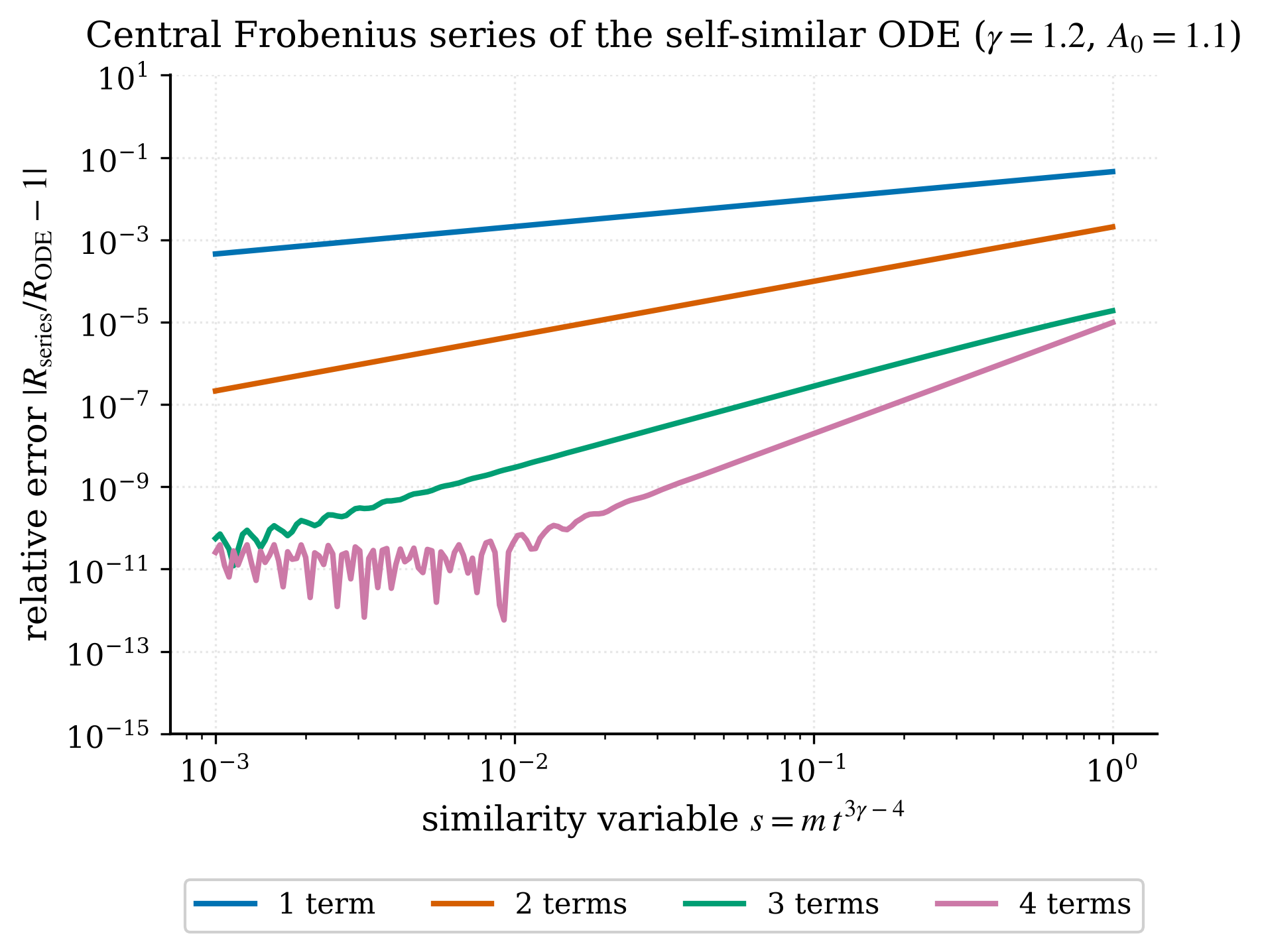}
\captionof{figure}{Relative error of the central series \eqref{eq:centreseries} of the self-similar flow ($\gamma=1.2$, $A_0=1.1$, $G=K=1$) for one to four terms against the numerical solution of \eqref{eq:sssys}. The plateau near $10^{-11}$ is the numerical error floor.}\label{fig:sserr}
\end{minipage}\hfill
\begin{minipage}[t]{0.49\textwidth}
\centering
\includegraphics[width=\linewidth]{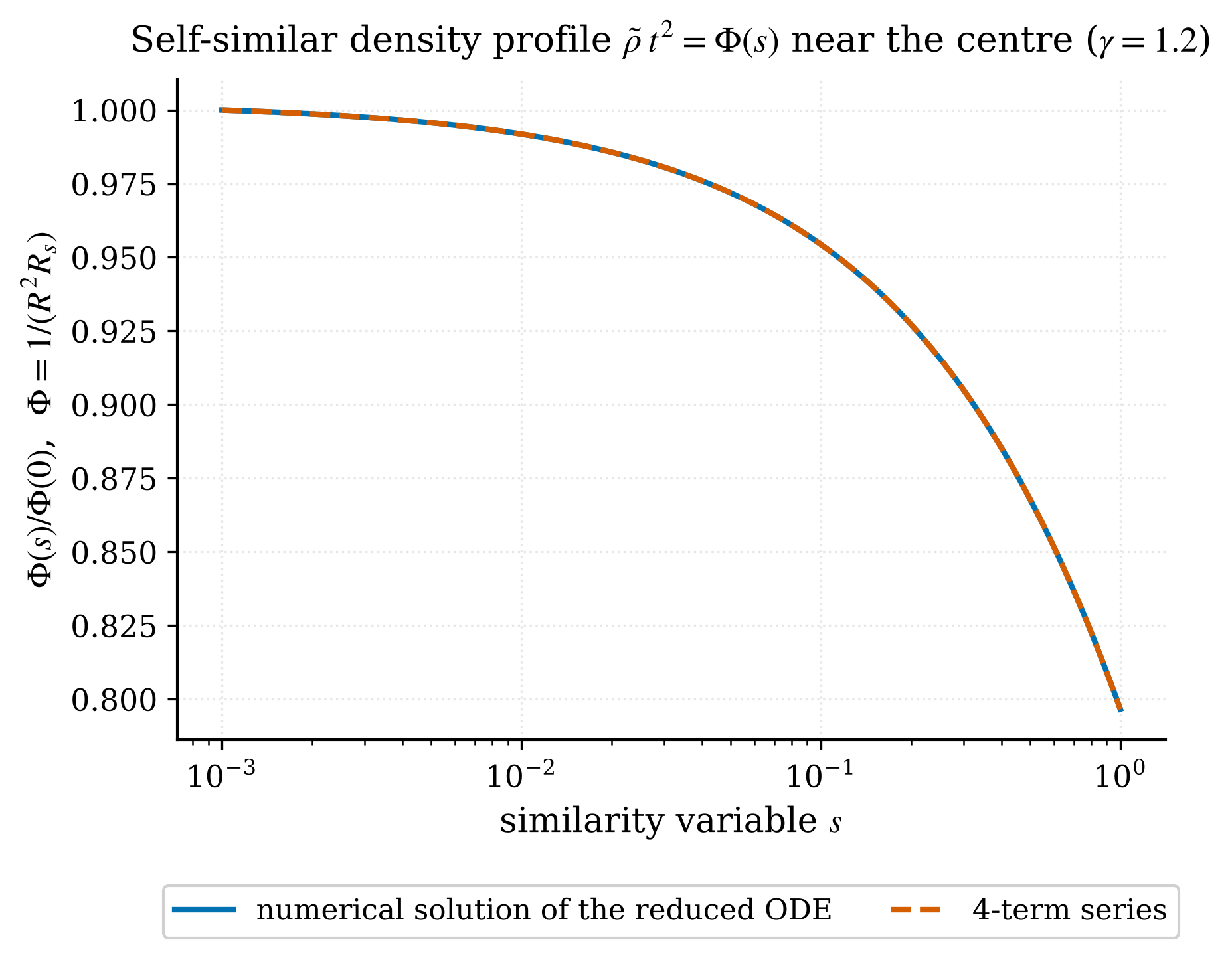}
\captionof{figure}{Self-similar density $\Phi(s)/\Phi(0)$, $\Phi=1/(R^2R_s)$ ($\tilde\rho\,t^{2}=\Phi(s)$), for $\gamma=1.2$, $A_0=1.1$: numerical solution (solid) and four-term series (dashed). $\Phi$ decreases monotonically away from the uniform core.}\label{fig:ssdens}
\end{minipage}
\end{figure}

\paragraph{Numerical solution of the self-similar flow.} Writing $W=R_s=(K/\Pi)^{1/\gamma}R^{-2}$, equation \eqref{eq:ssode} becomes a first-order system in $(R,\Pi)$ without cancellation at the centre,
\begin{equation}\label{eq:sssys}
\begin{aligned}[b]
R_s&=W,\\
\Pi_s\Bigl[R^{2}-\frac{(3\gamma-4)^2s^2W}{\gamma\,\Pi}\Bigr]&=\frac{2(3\gamma-4)^2s^2W^{2}}{R}-(3\gamma-4)(\gamma-1)sW\\ &\quad-(\gamma-1)(\gamma-2)R-\frac{Gs}{R^{2}},
\end{aligned}
\end{equation}
where the bracket vanishes at the sonic points (not reached in the range computed here). Table \ref{tab:ssseries} gives the coefficients for $\gamma=1.2$, $A_0=1.1$, $G=K=1$.
\begin{table}[H]
\centering
\caption{Coefficients $a_k$ of Theorem \ref{thm:ssseries} for $\gamma=1.2$, $A_0=1.1$, $G=K=1$, and relative error of the truncated series against the numerical solution of \eqref{eq:sssys} (started at $s_0=10^{-5}$).}\label{tab:ssseries}
\small
\begin{tabular}{@{}lcccc@{}}
\toprule
 & $a_0$ & $a_1$ & $a_2$ & $a_3$ \\
\midrule
value & $1.1$ & $4.9882\times10^{-2}$ & $2.3479\times10^{-3}$ & $3.3082\times10^{-5}$ \\
\midrule
terms kept & $1$ & $2$ & $3$ & $4$ \\
error at $s=0.1$ & $9.8\times10^{-3}$ & $9.8\times10^{-5}$ & $2.8\times10^{-7}$ & $2.0\times10^{-8}$ \\
error at $s=1$ & $4.5\times10^{-2}$ & $2.1\times10^{-3}$ & $1.9\times10^{-5}$ & $9.8\times10^{-6}$ \\
\bottomrule
\end{tabular}
\end{table}
\smallskip\noindent
The computed $R(1)=1.152251216105$ is independent of the start point $s_0\in\{10^{-4},10^{-5},10^{-6}\}$ to $10^{-13}$. Figures \ref{fig:sserr} and \ref{fig:ssdens} show the series error and the density profile $\Phi(s)=1/(R^2R_s)$.

\subsection{Puiseux expansion at the stellar surface}\label{sec:surface}
\begin{figure}[H]
\centering
\includegraphics[width=0.49\textwidth]{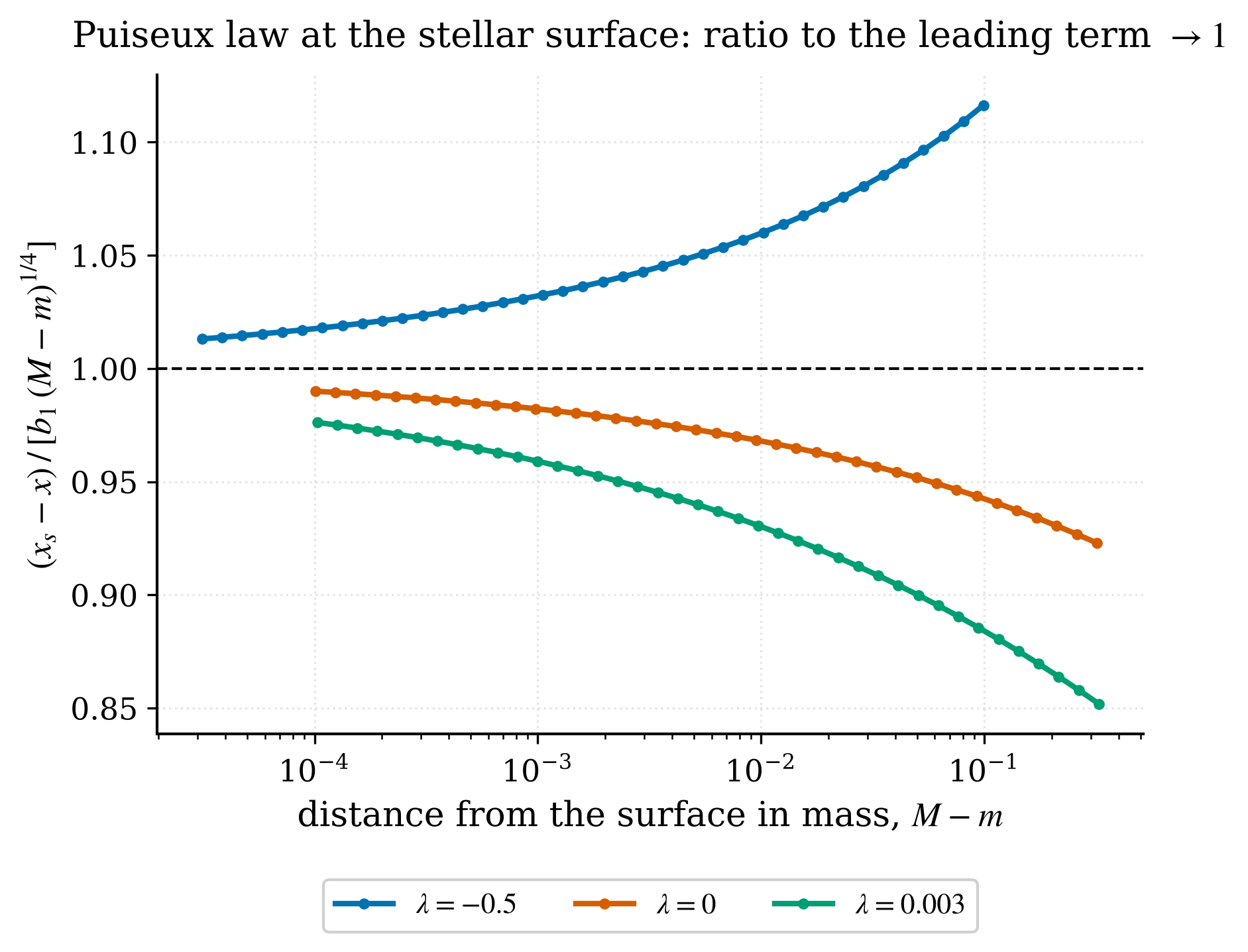}
\caption{Test of the Puiseux law \eqref{eq:puiseux}: ratio of $x_s-x$ to $b_1(M-m)^{1/4}$ versus $M-m$ for three values of $\lambda$; the dashed line is the limit $1$.}\label{fig:surface}
\end{figure}

\begin{proposition}[Surface expansion]\label{prop:surface}
\emph{Hypotheses:} $\gamma=4/3$, $x\in C^1((0,M))$ and $\Pi\in C^1((0,M))$ solve \eqref{eq:sys} with $\Pi>0$ on $(0,M)$, $\Pi(m)\to0$ and $x(m)\to x_s\in(0,\infty)$ as $m\to M^-$, and $q:=(GM-\lambda x_s^{3})/x_s^{4}>0$. Put $\varepsilon:=M-m$. Then
\begin{align}
\Pi&=q\,\varepsilon+O(\varepsilon^{5/4}),\label{eq:Pisurf}\\
x&=x_s-b_1\,\varepsilon^{1/4}+O(\varepsilon^{1/2}),\qquad b_1=4\Bigl(\frac{K}{q\,x_s^{8/3}}\Bigr)^{3/4},\label{eq:puiseux}\\
\tilde\rho&=\Bigl(\frac{\Pi}{K}\Bigr)^{3/4}=\Bigl(\frac{q}{K}\Bigr)^{3/4}\varepsilon^{3/4}\bigl(1+O(\varepsilon^{1/4})\bigr)\ \propto\ (x_s-x)^{3}=(x_s-x)^{1/(\gamma-1)}.\label{eq:surfrho}
\end{align}
\end{proposition}
\begin{proof}
\smallskip\noindent
By \eqref{eq:sys}, $\Pi_m=\lambda/x-Gm/x^4$ is continuous up to $m=M$ with value $-q$, so $\Pi=q\varepsilon+o(\varepsilon)$. Then
\begin{equation}
x_m=\Bigl(\frac{K}{\Pi x^{8/3}}\Bigr)^{3/4}=\Bigl(\frac{K}{qx_s^{8/3}}\Bigr)^{3/4}\varepsilon^{-3/4}\bigl(1+o(1)\bigr),
\end{equation}
which is integrable at $M$ (this also shows that $x$ has a finite limit, consistently with the hypothesis). Integrating, with $\int_m^M(M-\mu)^{-3/4}\dd\mu=4\varepsilon^{1/4}$, gives $x_s-x=b_1\varepsilon^{1/4}(1+o(1))$. Inserting this into $\Pi_m=\lambda/x-Gm/x^4$ gives $\Pi_m=-q+O(\varepsilon^{1/4})$, hence \eqref{eq:Pisurf}; repeating the integration with $x^{-8/3}=x_s^{-8/3}(1+O(\varepsilon^{1/4}))$ gives the remainder in \eqref{eq:puiseux}. Finally $\tilde\rho=1/(x^2x_m)=(\Pi/K)^{3/4}$ exactly, and $\varepsilon\propto(x_s-x)^4$.
\end{proof}
The leading law is confirmed numerically (Table \ref{tab:surface}, Figure \ref{fig:surface}): the ratio of $x_s-x$ to the leading term tends to $1$ with deviations decreasing like $\varepsilon^{1/4}$ (for $\lambda=-0.5$ the deviations $0.083,\,0.044,\,0.024$ at $\varepsilon=3.1\times10^{-2},3.1\times10^{-3},3.1\times10^{-4}$ shrink by $0.53$-$0.54$ per decade, against $10^{-1/4}=0.56$).
\begin{table}[H]
\centering
\caption{Surface data and test of \eqref{eq:puiseux} ($G=K=1$). The ratio is $(x_s-x)/[b_1\varepsilon^{1/4}]$ at the distance $\varepsilon=M-m$ shown.}\label{tab:surface}
\small
\begin{tabular}{@{}rccccrc@{}}
\toprule
$\lambda$ & $M$ & $x_s$ & $q$ & $b_1$ & $\varepsilon$ & ratio \\
\midrule
$-0.5$ & $5.022986$ & $3.693731$ & $0.162348$ & $1.14629$ & $3.1\times10^{-4}$ & $1.0237$ \\
$0$ & $16.145888$ & $9.564015$ & $0.001930$ & $4.74956$ & $1.0\times10^{-3}$ & $0.9821$ \\
$0.003$ & $16.457292$ & $10.295977$ & $0.001173$ & $5.95276$ & $1.0\times10^{-3}$ & $0.9590$ \\
\bottomrule
\end{tabular}
\end{table}

\subsection{Static limit, mass relation and profile identity}
\begin{figure}[H]
\centering
\includegraphics[width=0.49\textwidth]{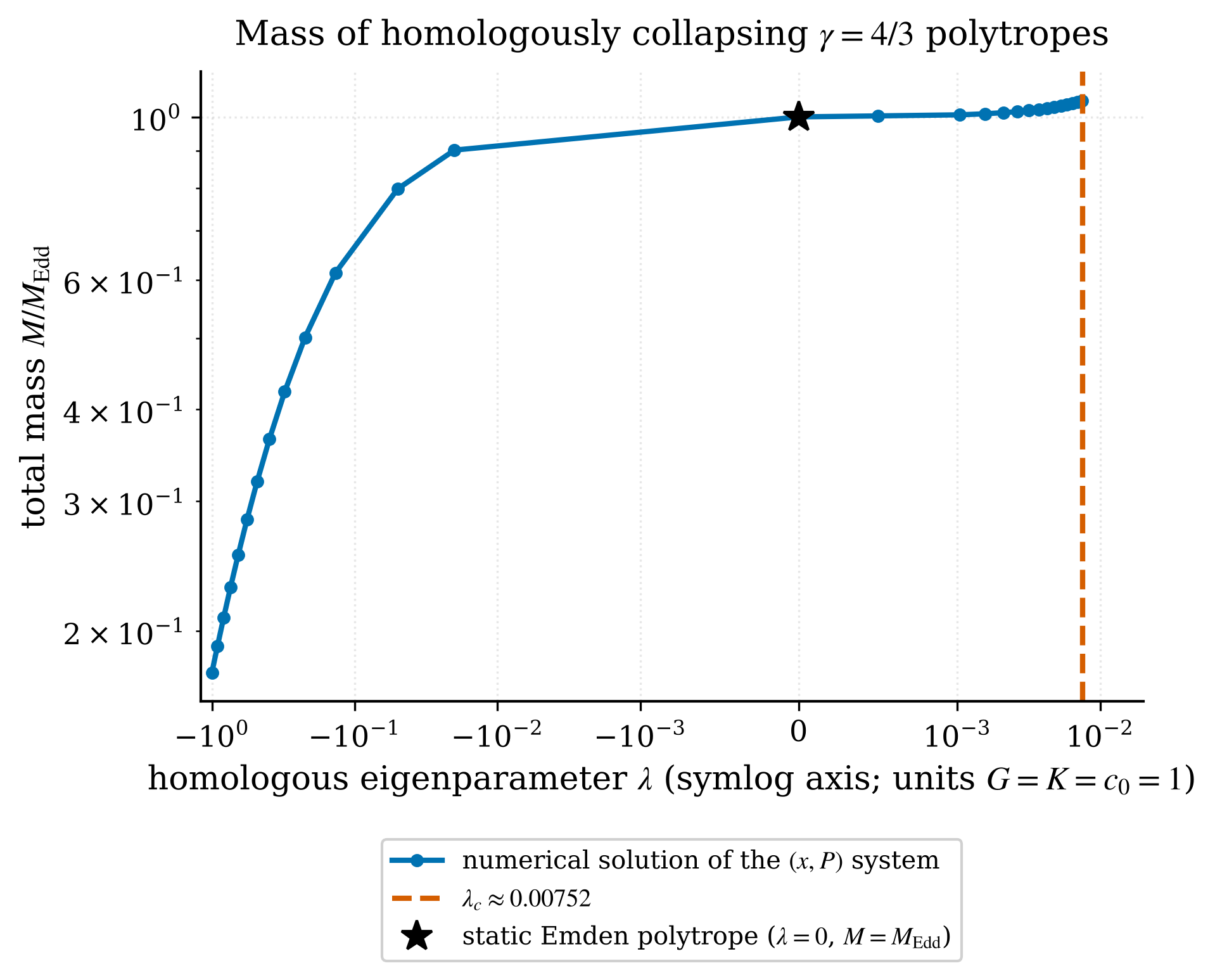}
\caption{Total mass $M/\Edd$ of the homologous profile versus $\lambda$ (symmetric-logarithmic $\lambda$-axis; $G=K=c_0=1$). The star marks the static Emden polytrope. Beyond the numerical threshold $\lambda_c\approx0.00752$ (dashed) no finite surface was found for $m\le2\times10^4$.}\label{fig:mass}
\end{figure}

\begin{figure}[H]
\begin{minipage}[t]{0.49\textwidth}
\centering
\includegraphics[width=\linewidth]{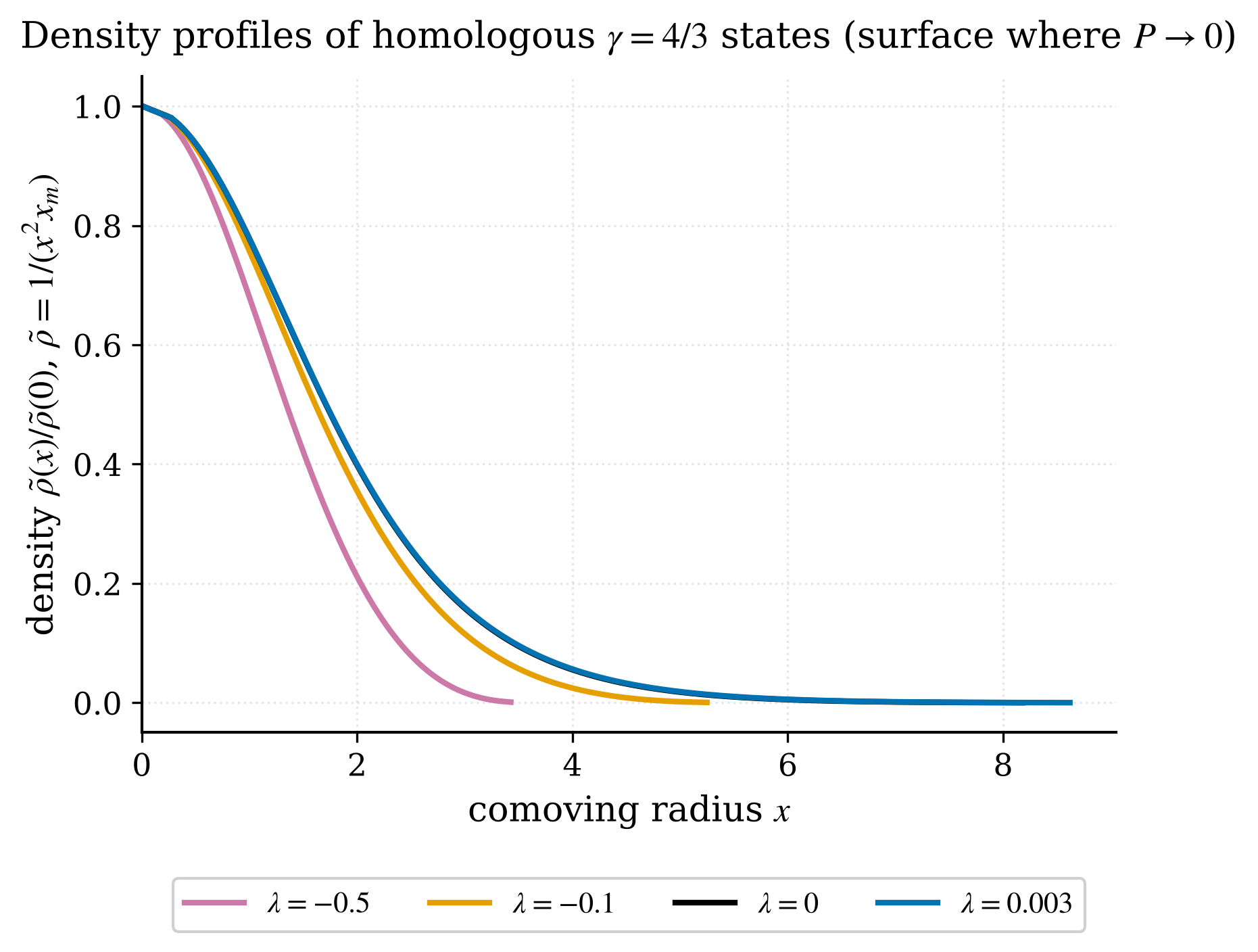}
\captionof{figure}{Normalised density $\tilde\rho(x)/\tilde\rho(0)$, $\tilde\rho=1/(x^2x_m)=(\Pi/K)^{3/4}$, of homologous states for four values of $\lambda$; the surface is where $\Pi\to0$ and the density vanishes like $(x_s-x)^3$ (Proposition \ref{prop:surface}).}\label{fig:rho}
\end{minipage}\hfill
\begin{minipage}[t]{0.49\textwidth}
\centering
\includegraphics[width=\linewidth]{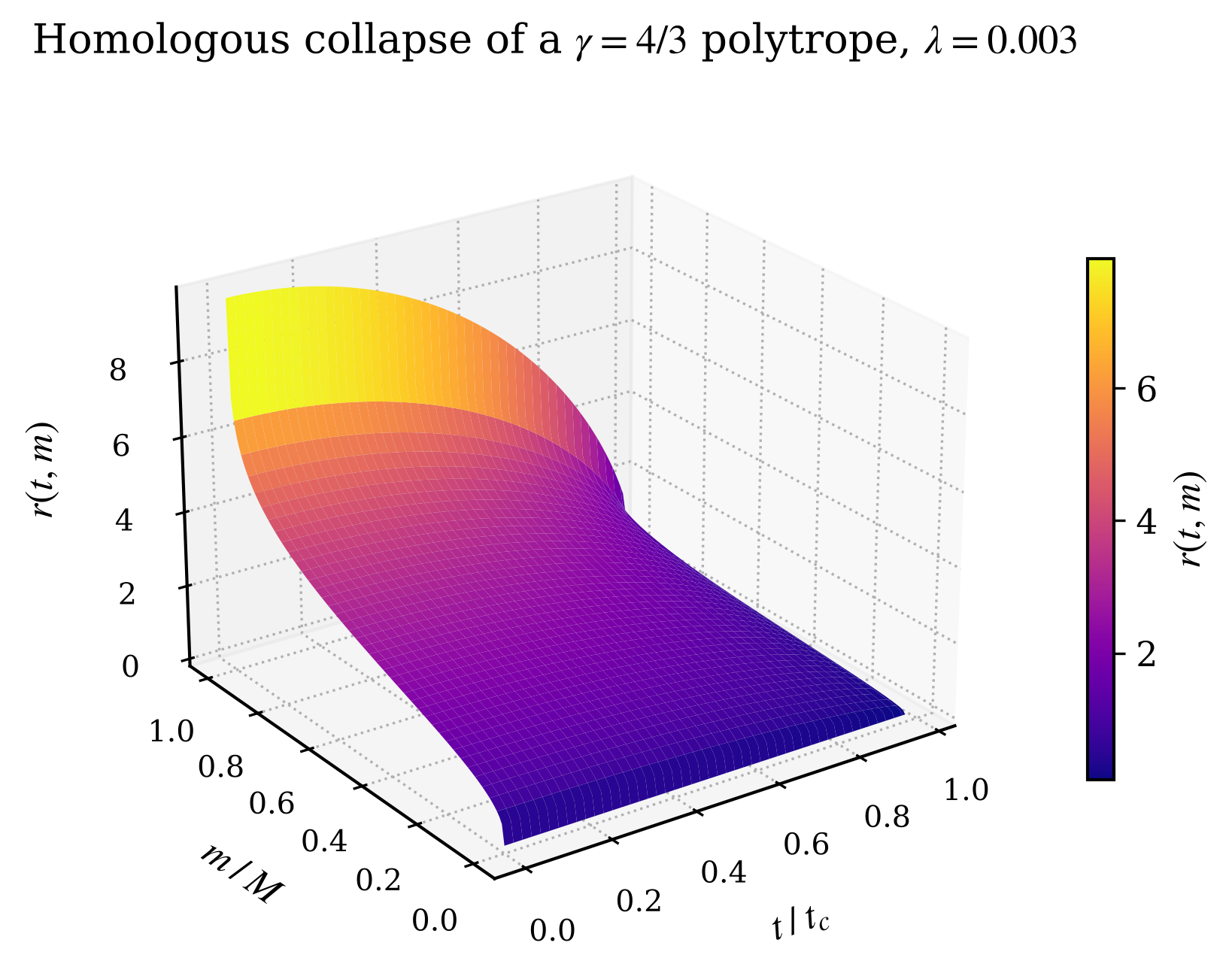}
\captionof{figure}{Surface $r(t,m)=a(t)x(m)$ of the homologous collapse of a $\gamma=4/3$ polytrope ($\lambda=0.003$), with $a(t)$ the cycloid \eqref{eq:cyc}.}\label{fig:poly3d}
\end{minipage}
\end{figure}

\begin{figure}[H]
\centering
\includegraphics[width=0.49\textwidth]{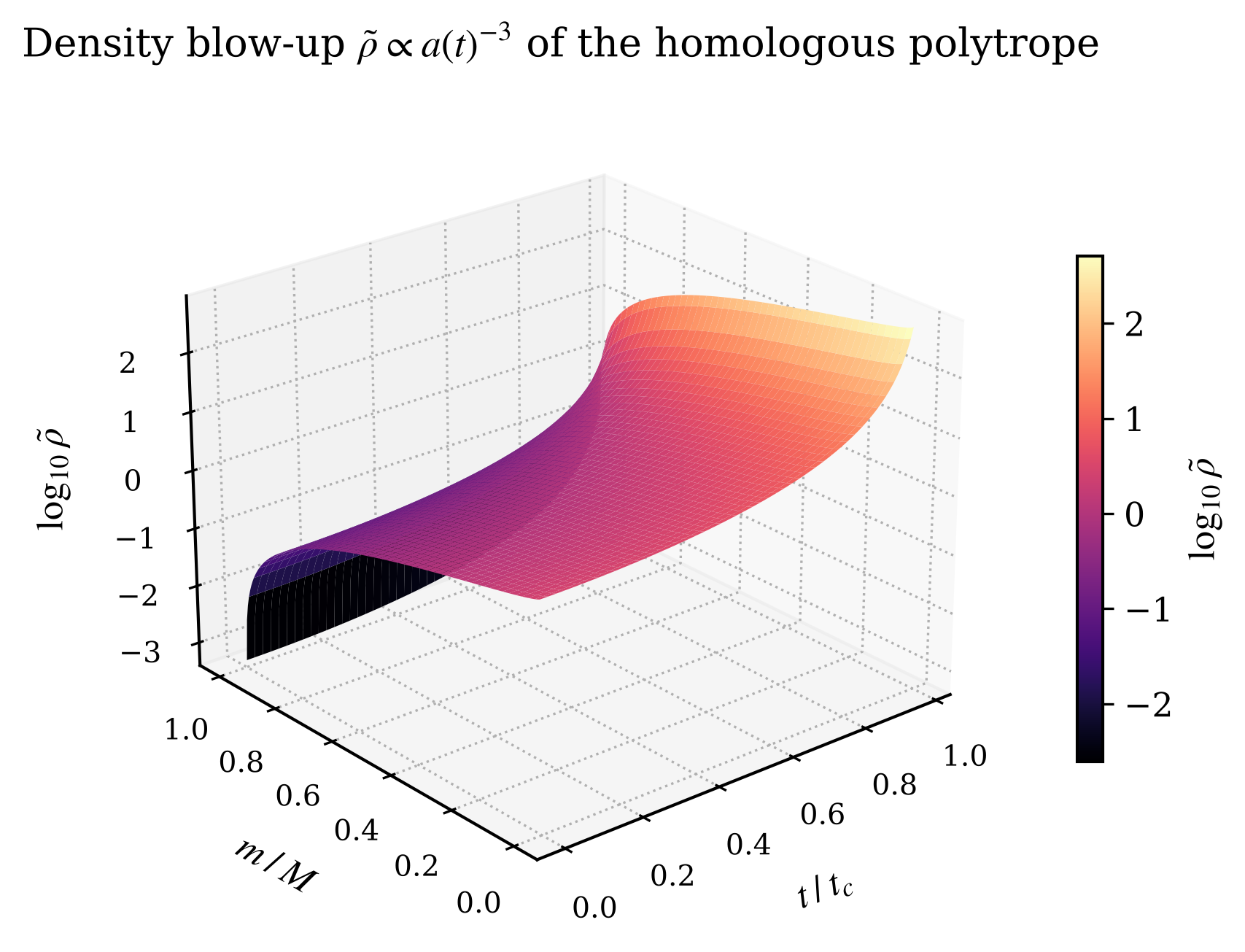}
\caption{Density $\tilde\rho(t,m)=\tilde\rho_0(m)/a(t)^3$ of the same solution on a logarithmic scale: it grows by the factor $a^{-3}$ uniformly in $m$ as $t\to t_c$.}\label{fig:rho3d}
\end{figure}

For $\lambda=0$ the profile is the static $n=3$ polytrope. Integrating the Lane-Emden equation $\theta''+2\theta'/\xi+\theta^3=0$ independently gives $\xi_1=6.896849$ and $-\xi_1^2\theta'(\xi_1)=2.018236$ (the classical values \cite{Chandrasekhar1939}), hence the Eddington mass
\begin{equation}
\Edd=4\pi\,(2.018236)\Bigl[\frac{(4\pi)^{1/3}K}{\pi G}\Bigr]^{3/2}=16.14589\quad(G=K=1).
\end{equation}
The mass of the profile from \eqref{eq:sys}, refined with Proposition \ref{prop:surface}, is $M=16.145888$ at $\lambda=0$, equal to $\Edd$ to the six digits printed. For $\lambda\ne0$ the surface mass is shown in Figure \ref{fig:mass} and Table \ref{tab:mass}.
\begin{table}[H]
\centering
\caption{Surface mass of the homologous $\gamma=4/3$ profiles ($G=K=c_0=1$, $\Edd=16.1459$).}\label{tab:mass}
\small
\begin{tabular}{@{}rcccccccc@{}}
\toprule
$\lambda$ & $-1$ & $-0.5$ & $-0.2$ & $-0.05$ & $0$ & $0.003$ & $0.005$ & $0.0075$ \\
\midrule
$M/\Edd$ & $0.1755$ & $0.3111$ & $0.5267$ & $0.7980$ & $1.0000$ & $1.0193$ & $1.033$ & $1.053$ \\
\bottomrule
\end{tabular}
\end{table}
\begin{remark}
For $\lambda\in(0,\lambda_c)$ the mass slightly exceeds $\Edd$ and increases with $\lambda$. As $\lambda\uparrow\lambda_c$ the surface pressure slope $q$ of Proposition \ref{prop:surface} tends to zero (estimated at the point where $\Pi=10^{-7}$: $q\approx1.2\times10^{-3},\,3.2\times10^{-4},\,4.6\times10^{-5}$ at $\lambda=0.004,\,0.007,\,0.0075$) while the radius at which $\Pi$ falls to $10^{-7}$ grows ($10.0,\,11.6,\,12.8$; these underestimate $x_s$) and $M/\Edd$ approaches $\approx1.053$; for $\lambda\ge\lambda_c$ the pressure $\Pi$ stayed positive on the whole range computed ($m<2\times10^{4}$), i.e.\ no finite surface was found. The threshold $\lambda_c\approx0.0075174$ (bisection) is therefore a numerical observation on the integration range, associated with the degeneration $q\to0$ of the surface, and not a proved bound; it is consistent with the known requirement that homologous collapse needs a mass of order $\Edd$ or above \cite{GoldreichWeber1980}.
\end{remark}
\paragraph{Profile identity.} Multiply the second equation of \eqref{eq:sys} written as $x^2\Pi_m=\lambda x-Gm/x^2$ by $x$ and integrate over $(0,M)$; with $x^3\Pi\to0$ at both ends (since $\Pi\to0$ at $M$ and $x\to0$ at $0$) and $x^2x_m\Pi=Kx^{-2/3}x_m^{-1/3}$,
\begin{equation}
\int_0^M x^3\Pi_m\dd m=-3\int_0^Mx^2x_m\Pi\dd m=-3K\int_0^Mx^{-2/3}x_m^{-1/3}\dd m ,
\end{equation}
so that
\begin{equation}\label{eq:profid}
\lambda\int_0^M x^2\dd m-G\int_0^M\frac{m}{x}\dd m+3K\int_0^Mx^{-2/3}x_m^{-1/3}\dd m=0 ,
\end{equation}
the $\gamma=4/3$ virial balance of the profile. Numerically the relative residual is $1.5\times10^{-7}$ ($\lambda=-0.5$), $9.5\times10^{-7}$ ($\lambda=0$) and $1.1\times10^{-6}$ ($\lambda=0.003$), consistent with the trapezoid quadrature error. Figures \ref{fig:rho}-\ref{fig:rho3d} show the density profiles, the collapse surface $r(t,m)=a(t)x(m)$ and the density growth $\tilde\rho\propto a^{-3}$.

\section{Gradient catastrophe and singularity formation in mass coordinates}\label{sec:catastrophe}
A reader familiar with the Hopf equation will recognise the mechanism below: neighbouring characteristics meet and the gradient blows up in finite time. In Lagrangian variables the same event is the vanishing of $r_m$, and it is independent of the self-similar blow-up of the previous sections. We treat it in the two cases that can be solved exactly.

\smallskip\noindent
In mass coordinates the density is $\tilde\rho=1/(r^2r_m)$, so a density singularity is either $r\to0$ (collapse to the centre) or $r_m\to0$ (neighbouring shells meet: shell crossing, the Lagrangian form of a gradient catastrophe). This links the model to the gradient catastrophe of the Hopf equation and of hydrodynamic-type systems \cite{Whitham1974,Sideris1985}: for $u_t+uu_r=0$ the velocity is constant along trajectories, which in Lagrangian variables is $r_{tt}=0$.

\begin{proposition}[Hopf-type fold]\label{prop:hopf}
\emph{Hypotheses:} $r(t,m)=r_0(m)+v_0(m)\,t$ (free streaming, $r_{tt}=0$) with $r_0,v_0\in C^3(I)$, $r_0'>0$; $\varphi(m):=-r_0'(m)/v_0'(m)$ on $\{v_0'<0\}$ attains an interior minimum $t_*=\varphi(m_*)>0$ with $\varphi''(m_*)>0$, and $r_*:=r_0(m_*)+v_0(m_*)t_*>0$. Then $t_*$ is the first time at which $r_m$ vanishes (on the set where $v_0'<0$), and
\begin{equation}\label{eq:fold}
\begin{aligned}[b]r_m(t,m)&=|v_0'(m)|\,\bigl(\varphi(m)-t\bigr)\\&=|v_0'(m_*)|\Bigl[(t_*-t)+\tfrac12\varphi''(m_*)(m-m_*)^2\Bigr]\bigl(1+o(1)\bigr)\quad\text{as }(t,m)\to(t_*,m_*).\end{aligned}
\end{equation}
Consequently, with $c_*:=r_*^2|v_0'(m_*)|$, the density $\tilde\rho=1/(r^2r_m)$ satisfies
\begin{align}
\tilde\rho&\sim\frac{1}{c_*\,(t_*-t)}&&\text{at }m=m_*,\\
\tilde\rho&\sim\frac{2}{c_*\,\varphi''(m_*)}\,|m-m_*|^{-2}&&\text{at }t=t_*.
\end{align}
\end{proposition}
\begin{proof}
$r_m=r_0'+v_0't=v_0'(t-\varphi)=|v_0'|(\varphi-t)$ where $v_0'<0$, which first vanishes at $t=\min\varphi=t_*$. Taylor expansion of $\varphi\in C^2$ at the nondegenerate minimum, $\varphi(m)=t_*+\tfrac12\varphi''(m_*)(m-m_*)^2+o((m-m_*)^2)$, and continuity of $|v_0'|$ give \eqref{eq:fold}; the density statements follow from $r\to r_*>0$.
\end{proof}

\begin{figure}[H]
\begin{minipage}[t]{0.49\textwidth}
\centering
\includegraphics[width=\linewidth]{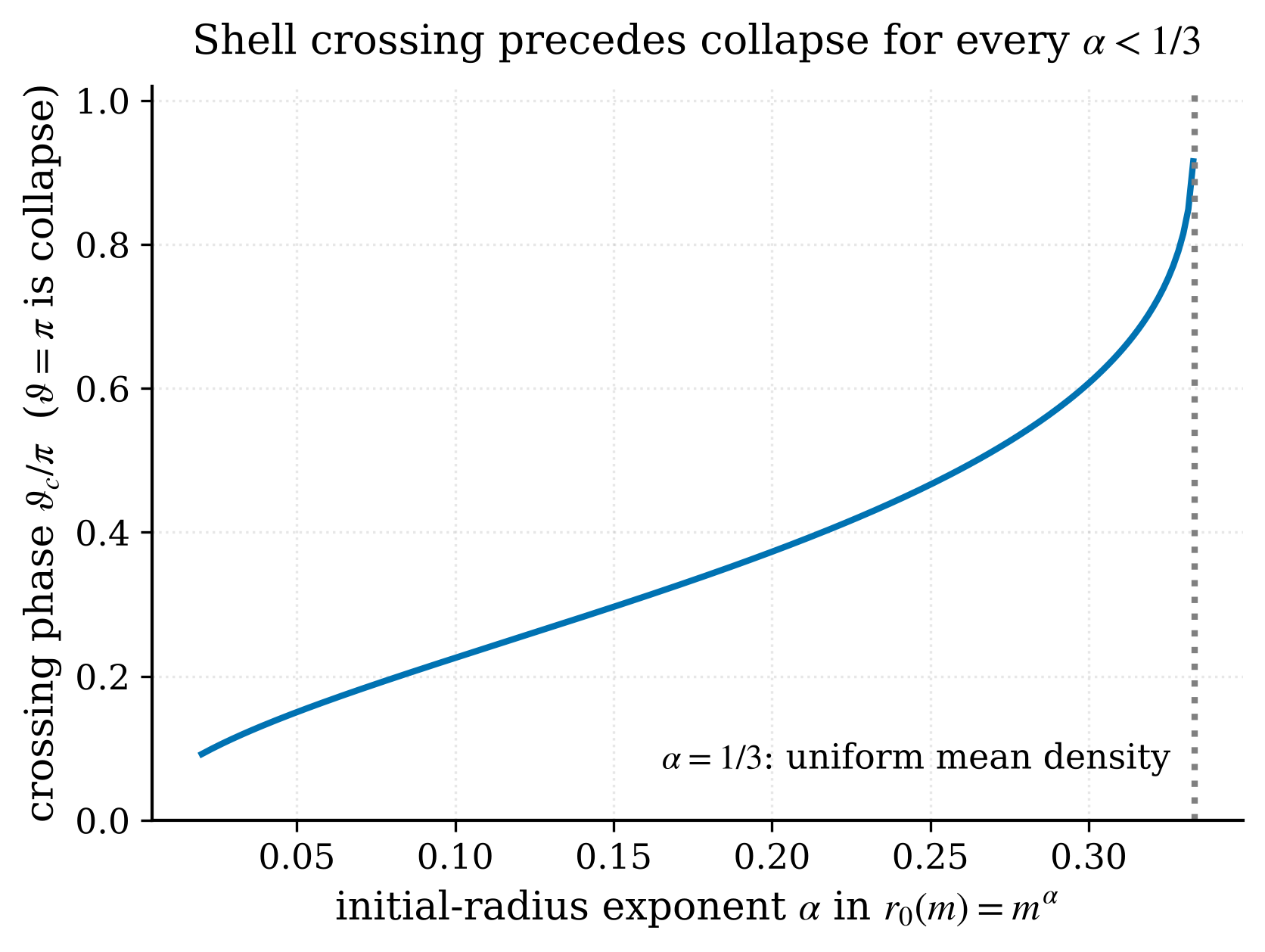}
\captionof{figure}{Crossing phase $\vartheta_c/\pi$ from \eqref{eq:thetac} versus the exponent $\alpha$ of $r_0=m^{\alpha}$ ($\vartheta=\pi$ is collapse): shell crossing precedes collapse for every $\alpha<1/3$.}\label{fig:cross}
\end{minipage}\hfill
\begin{minipage}[t]{0.49\textwidth}
\centering
\includegraphics[width=\linewidth]{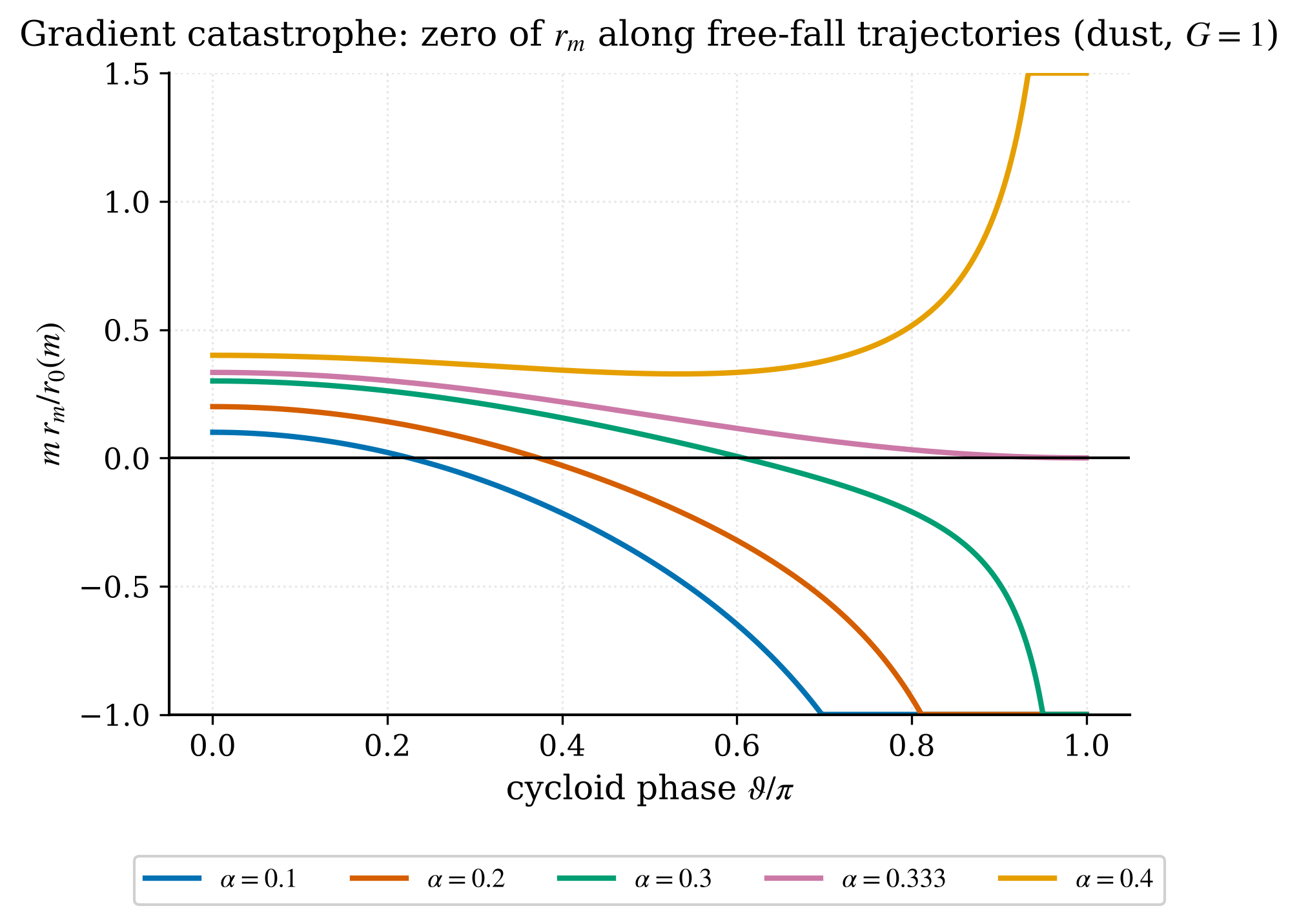}
\captionof{figure}{The normalised $m\,r_m/r_0(m)$ along the free-fall cycloid for several $\alpha$ ($G=1$). Curves with $\alpha<1/3$ cross zero transversally before $\vartheta=\pi$ (gradient catastrophe); $\alpha=1/3$ reaches zero only at $\vartheta=\pi$; $\alpha>1/3$ stays positive.}\label{fig:rmdust}
\end{minipage}
\end{figure}

\begin{proposition}[Free fall of dust: exact shell-crossing criterion]\label{prop:dust}
\emph{Hypotheses:} $S\equiv0$ (pressureless), $G>0$, $r_0\in C^1(I)$ with $r_0>0$, $r_0'>0$; shells are released from rest at $t=0$. Set $T(m):=\sqrt{r_0^3/(8Gm)}$. Then for $0\le t<\pi T(m)$, $r(t,m)$ is given parametrically by $\vartheta\in[0,\pi)$:
\begin{equation}\label{eq:cycdust}
r=\tfrac12r_0\,(1+\cos\vartheta),\qquad t=T(m)\,(\vartheta+\sin\vartheta),\qquad t_c(m)=\pi\,T(m),
\end{equation}
and at fixed $t$,
\begin{equation}\label{eq:rmdust}
r_m=\frac{r_0'}{2}\,(1+\cos\vartheta)+\frac{r_0\,\eta\,\sin\vartheta\,(\vartheta+\sin\vartheta)}{2\,(1+\cos\vartheta)},\qquad \eta:=\frac{T'}{T}=\frac{3r_0'}{2r_0}-\frac{1}{2m}.
\end{equation}
Consequently, at a given $m$, $r_m>0$ for all $\vartheta\in(0,\pi)$ iff $\eta(m)\ge0$, i.e.\ iff the mean enclosed density $\bar\rho(m)\propto m/r_0(m)^3$ is non-increasing at $m$; if $\bar\rho$ increases outward at $m$ ($\eta<0$), $r_m$ vanishes at some $\vartheta_c(m)<\pi$: shell crossing \emph{precedes} collapse.
For $r_0=m^{\alpha}$ the phase $\vartheta_c$ is independent of $m$ and solves
\begin{equation}\label{eq:thetac}
\alpha\,(1+\cos\vartheta)^2+\tfrac12(3\alpha-1)\sin\vartheta\,(\vartheta+\sin\vartheta)=0,
\end{equation}
the crossing time is $t_\times(m)=T(m)(\vartheta_c+\sin\vartheta_c)$ with $T=m^{(3\alpha-1)/2}/\sqrt{8G}$, and for $\alpha=1/3$ (uniform mean density) all shells collapse simultaneously.
\end{proposition}
\begin{proof}
Radial Kepler free fall from rest gives \eqref{eq:cycdust}. At fixed $t$, differentiation of $t=T(\vartheta+\sin\vartheta)$ in $m$ gives
\begin{equation}
\vartheta_m=-\frac{T'(\vartheta+\sin\vartheta)}{T\,(1+\cos\vartheta)},
\end{equation}
and then, with $r_\vartheta=-\tfrac12r_0\sin\vartheta$,
\begin{equation}
r_m=\tfrac12r_0'(1+\cos\vartheta)+r_\vartheta\vartheta_m=\tfrac12r_0'(1+\cos\vartheta)+\frac{r_0\,\eta\,\sin\vartheta\,(\vartheta+\sin\vartheta)}{2(1+\cos\vartheta)},
\end{equation}
which is \eqref{eq:rmdust}; $\eta=T'/T=\tfrac32r_0'/r_0-\tfrac1{2m}$ since $T\propto r_0^{3/2}m^{-1/2}$ (check $\mathrm{V}16$). If $\eta\ge0$ both terms of \eqref{eq:rmdust} are positive on $(0,\pi)$. If $\eta<0$ the first term is positive for small $\vartheta$, while the second tends to $-\infty$ as $\vartheta\to\pi$ because $\sin\vartheta/(1+\cos\vartheta)=\tan(\vartheta/2)\to\infty$ and $\vartheta+\sin\vartheta\to\pi$; hence $r_m$ has a zero. For $r_0=m^\alpha$, $r_0'=\alpha r_0/m$ and $\eta=(3\alpha-1)/(2m)$; $r_m=0$ multiplied by $4m(1+\cos\vartheta)/r_0$ gives \eqref{eq:thetac}. Finally
\begin{equation}
\eta\ge0\iff3mr_0'\ge r_0\iff\Bigl(\frac{r_0^3}{m}\Bigr)'=\frac{3r_0^2r_0'm-r_0^3}{m^2}\ge0\iff\bar\rho'\le0 .
\end{equation}
\end{proof}
\smallskip\noindent
For self-gravitating dust in Lagrangian form, see also \cite[Sec. 2.4.1]{Hadzic2023}. Proposition \ref{prop:dust} is the Newtonian, free-fall counterpart of the necessary and sufficient no-shell-crossing conditions of the Tolman (Lema\^{\i}tre-Tolman) model \cite{HellabyLake1985}, whose collapse-time monotonicity condition $t_c'\ge0$ it reproduces ($\eta=T'/T=t_c'/t_c$); the explicit crossing phase \eqref{eq:thetac} and the confirmation below are specific to the free-fall setting.
\begin{example}[$\alpha=0.3$, $G=1$]
Equation \eqref{eq:thetac} gives $\vartheta_c=1.907379$ ($\vartheta_c+\sin\vartheta_c=2.851268$); for $\alpha=0.1$ and $0.2$ one gets $\vartheta_c=0.708479$ and $1.171603$ (Figure \ref{fig:cross}). Since $T=m^{-0.05}/\sqrt8$ decreases with $m$, the first crossing occurs at the outer edge of the shells considered, $t_\times(1)=2.851268/\sqrt8=1.008075$. Direct integration of $2001$ shells $m\in[0.5,1]$ ($\mathrm{DOP}853$, tolerance $10^{-12}$) gives the first overlap at $t=1.008353$, in agreement within the time sampling ($3\times10^{-4}$).
\end{example}
\begin{figure}[H]
\begin{minipage}[t]{0.49\textwidth}
\centering
\includegraphics[width=\linewidth]{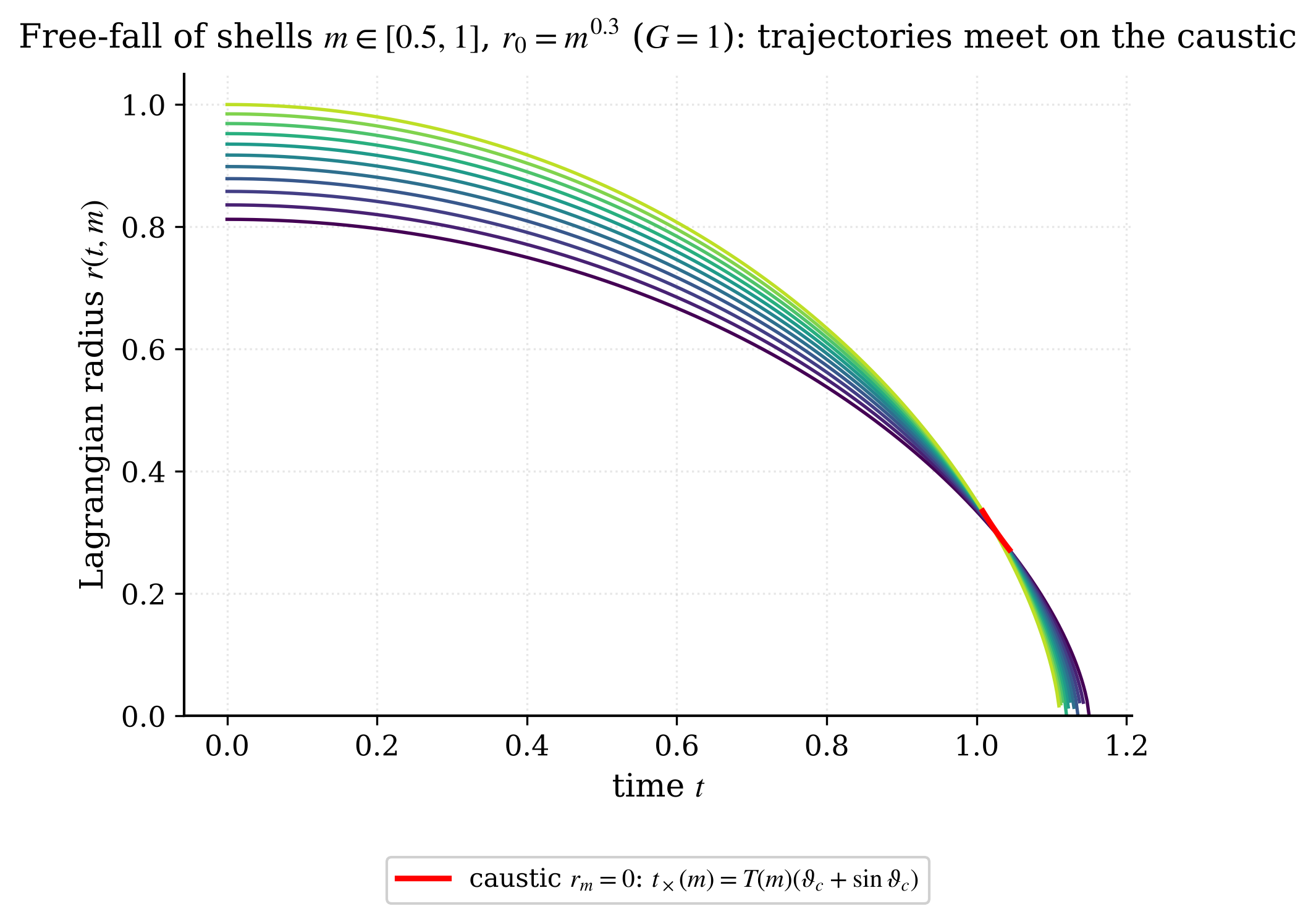}
\captionof{figure}{Free-fall trajectories $r(t,m)$ of eleven shells $m\in[0.5,1]$ with $r_0=m^{0.3}$, $G=1$; the red curve is the caustic $r_m=0$, $t_\times(m)=T(m)(\vartheta_c+\sin\vartheta_c)$, where neighbouring shells meet.}\label{fig:traj}
\end{minipage}\hfill
\begin{minipage}[t]{0.49\textwidth}
\centering
\includegraphics[width=\linewidth]{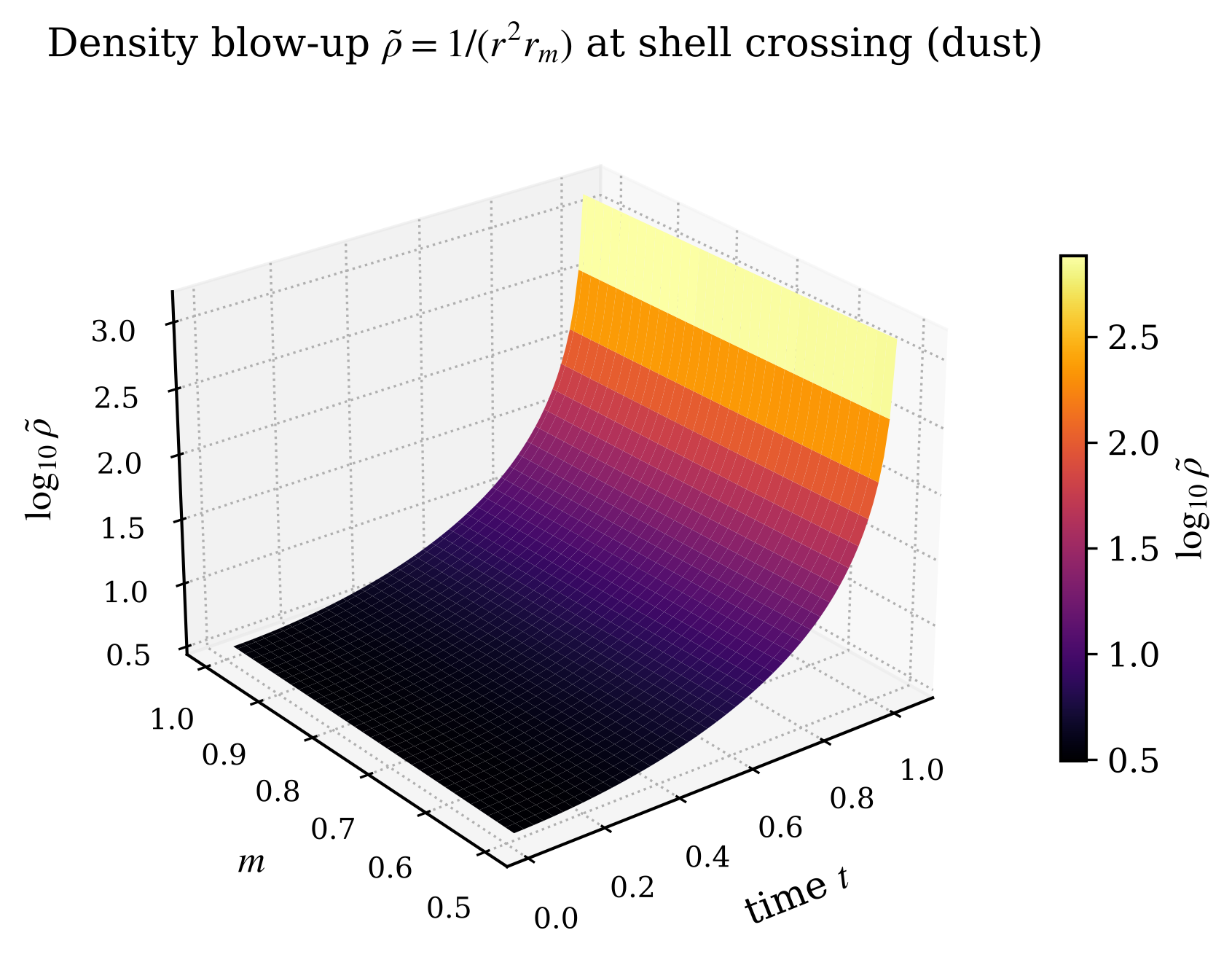}
\captionof{figure}{Density $\log_{10}\tilde\rho$, $\tilde\rho=1/(r^2r_m)$, of the dust solution of Figure \ref{fig:traj} over $(t,m)$ up to $99.5\%$ of the crossing time of each shell: the density diverges as $r_m\to0$.}\label{fig:dust3d}
\end{minipage}
\end{figure}

\begin{table}[H]
\centering
\caption{Blow-up scenarios in mass coordinates ($\tilde\rho=1/(r^2r_m)$).}\label{tab:scen}
\small
\begin{tabularx}{\textwidth}{@{}lXXl@{}}
\toprule
Scenario & Mechanism & Density law & Reference \\
\midrule
Hopf-type fold ($r_{tt}=0$) & $r_m\to0$ at interior $m_*$ & $\tilde\rho\sim(t_*-t)^{-1}$ at $m_*$ & Prop. \ref{prop:hopf} \\
Dust, $\bar\rho$ increasing outward & shell crossing, $r_m\to0$ before $r\to0$ & $\tilde\rho\to\infty$ at $t_\times(m)$ & Prop. \ref{prop:dust} \\
Homologous collapse, $\gamma=4/3$ & simultaneous $a\to0$ & $\tilde\rho\propto a^{-3}\sim(t_c-t)^{-2}$ & Sec. \ref{sec:homol} \\
Self-similar collapse, $\gamma<4/3$ & $\tilde\rho\,t^2=\Phi(s)$ & $\tilde\rho\sim t^{-2}$ & Prop. \ref{prop:scale}, \cite{GHJS2022} \\
\bottomrule
\end{tabularx}
\end{table}

\section{Exact linear perturbation theory at the critical index}\label{sec:pert}
At $\gamma=4/3$ the Einstein-de Sitter background has no intrinsic time scale left once the sound speed is taken into account, and this is why the linearised problem can be solved in closed form. The result is of Jeans type and its structure, in particular the log-periodic regime, is a useful benchmark for numerical schemes.

\begin{figure}[H]
\centering
\includegraphics[width=0.49\textwidth]{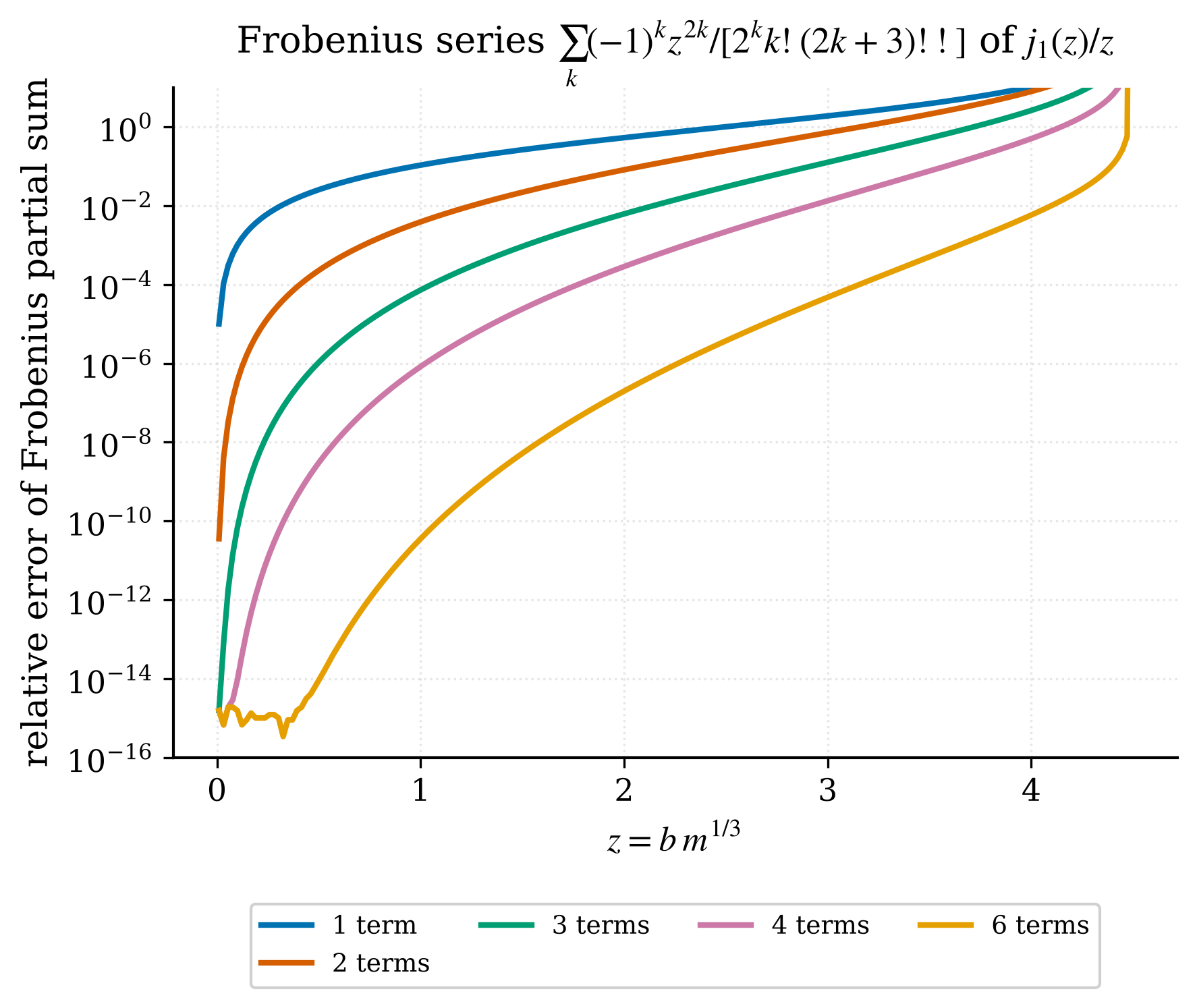}
\caption{Relative error of the Frobenius partial sums \eqref{eq:modeseries} of $j_1(z)/z$ for $1$, $2$, $3$, $4$ and $6$ terms on $0<z\le z_1=4.4934$.}\label{fig:modeseries}
\end{figure}

\smallskip\noindent
We linearise \eqref{eq:EPm} for $\gamma=4/3$, $S\equiv K>0$ about the collapsing Einstein-de Sitter solution $r_0=Am^{1/3}T^{2/3}$, $A^3=9G/2$, where $T=t_*-t>0$ is the time to blow-up (\eqref{eq:EPm} is invariant under $t\mapsto t_*-t$ since it contains no first-order time derivative). Set
\begin{equation}\label{eq:ell}
r^\varepsilon=r_0\bigl(1+\varepsilon\,T^{\kappa}f(m)\bigr),\qquad \ell:=\frac{24K}{3^{2/3}G}.
\end{equation}
\begin{theorem}[Separable linearisation and exact modes]\label{thm:modes}
\emph{Hypotheses:} $\gamma=4/3$, $S\equiv K>0$, $\kappa\in\mathbb R$ (for complex $\kappa$ take real parts: the linearised operator has real coefficients), $f\in C^2((0,M])$, domain $\{0<T<T_0,\ 0<m\le M\}$, and $|\varepsilon|\,T_0^{\kappa}\sup|f|$ small enough that $r^\varepsilon$ and $r^\varepsilon_m$ stay positive. Then, uniformly on compact subsets,
\begin{equation}
\Delta[r^\varepsilon]=\varepsilon\,\frac{m^{1/3}T^{\kappa-4/3}}{6\,G^{2/3}}\Bigl[6^{2/3}G\,(3\kappa^2+\kappa-2)f-24\cdot2^{2/3}K\,\bigl(m^{4/3}f''+2m^{1/3}f'\bigr)\Bigr]+O(\varepsilon^2),
\end{equation}
so the $O(\varepsilon)$ equation separates exactly in $T$ and $m$:
\begin{equation}\label{eq:modeode}
\ell\bigl(m^{4/3}f''+2m^{1/3}f'\bigr)=(3\kappa-2)(\kappa+1)\,f,\qquad\text{equivalently}\qquad \ell\,m^{-2/3}\,(m^2f')'=(3\kappa-2)(\kappa+1)f .
\end{equation}
For $-1<\kappa<2/3$ the solution regular at $m=0$ is
\begin{equation}\label{eq:j1}
f(m)=\frac{j_1(z)}{z}=\frac{\sin z-z\cos z}{z^{3}},\qquad z=b\,m^{1/3},\quad b^2=\frac{9(2-3\kappa)(\kappa+1)}{\ell},
\end{equation}
with $f(0)=1/3$; the second solution $y_1(z)/z\sim m^{-1}$ is singular. For $\kappa>2/3$ or $\kappa<-1$ the regular solution is $i_1(z)/z$ (modified spherical Bessel function \cite{DLMF2010}).
\end{theorem}
\begin{proof}
$\Delta$ is a smooth function of $(r,r_m,r_{mm},r_{tt},m)$ for $r,r_m>0$; Taylor expansion in $\varepsilon$ with remainder bounded on compact sets gives the $O(\varepsilon^2)$ term, and the $O(\varepsilon^0)$ term vanishes by Proposition \ref{prop:eds}. The $O(\varepsilon)$ coefficient is the displayed expression (symbolic computation). Setting the bracket to zero and dividing by $6^{2/3}G$ gives \eqref{eq:modeode}, since $24\cdot2^{2/3}/6^{2/3}=24/3^{2/3}$. With $f=j_1(z)/z$ the left side of \eqref{eq:modeode} equals the right side identically ($\mathrm{V}9$); this is the half-integer Bessel case $\nu=3/2$ of $y''+2y'/m=-c\,m^{-4/3}y$ ($c=-(3\kappa-2)(\kappa+1)/\ell$).
\end{proof}
\paragraph{Frobenius structure at the centre.} Near $m=0$ put $f\sim m^{p}$ in \eqref{eq:modeode}: the leading term is $\ell\,p(p+1)\,m^{p-2/3}$, so the indicial exponents are $p=0,-1$, in agreement with the regular solution $f(0)=1/3$ and the singular solution $\sim m^{-1}$. The regular solution has the series in $z^2=b^2m^{2/3}$
\begin{equation}\label{eq:modeseries}
\frac{j_1(z)}{z}=\sum_{k\ge0}\frac{(-1)^{k}\,z^{2k}}{2^{k}\,k!\,(2k+3)!!},\qquad
\frac{f_{k+1}}{f_k}=-\frac{z^2}{2(k+1)(2k+5)},
\end{equation}
\cite{DLMF2010}; the series converges for all $z$ (ratio $\to0$) and Figure \ref{fig:modeseries} shows its truncation error up to $z_1=4.4934$.

\begin{figure}[H]
\begin{minipage}[t]{0.49\textwidth}
\centering
\includegraphics[width=\linewidth]{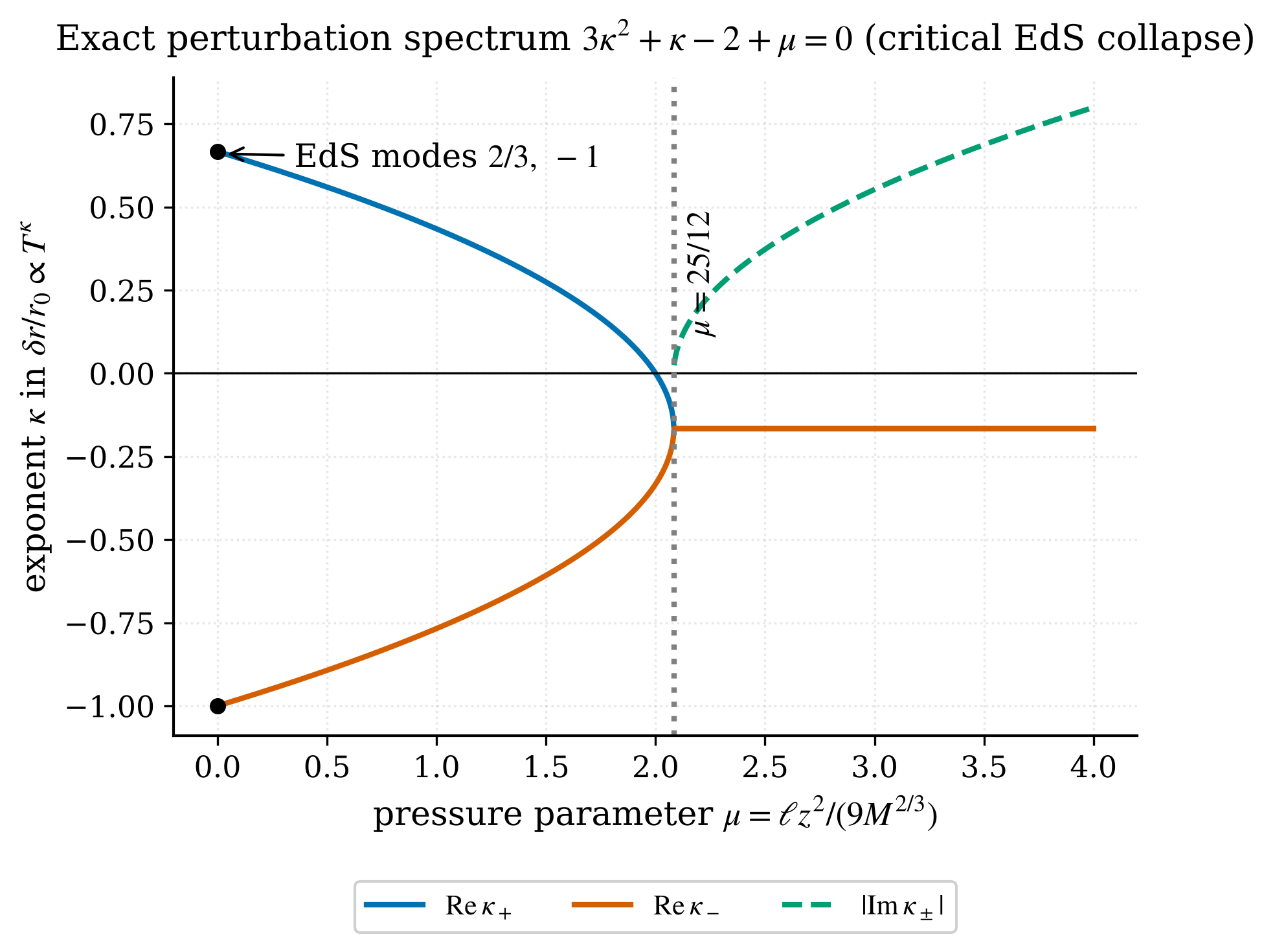}
\captionof{figure}{Exact exponent spectrum \eqref{eq:kap} versus $\mu$. Solid: $\mathrm{Re}\,\kappa_\pm$; dashed: $|\mathrm{Im}\,\kappa|$ for $\mu>25/12$. Dots: the pressureless modes $2/3$ and $-1$ at $\mu=0$.}\label{fig:spec}
\end{minipage}\hfill
\begin{minipage}[t]{0.49\textwidth}
\centering
\includegraphics[width=\linewidth]{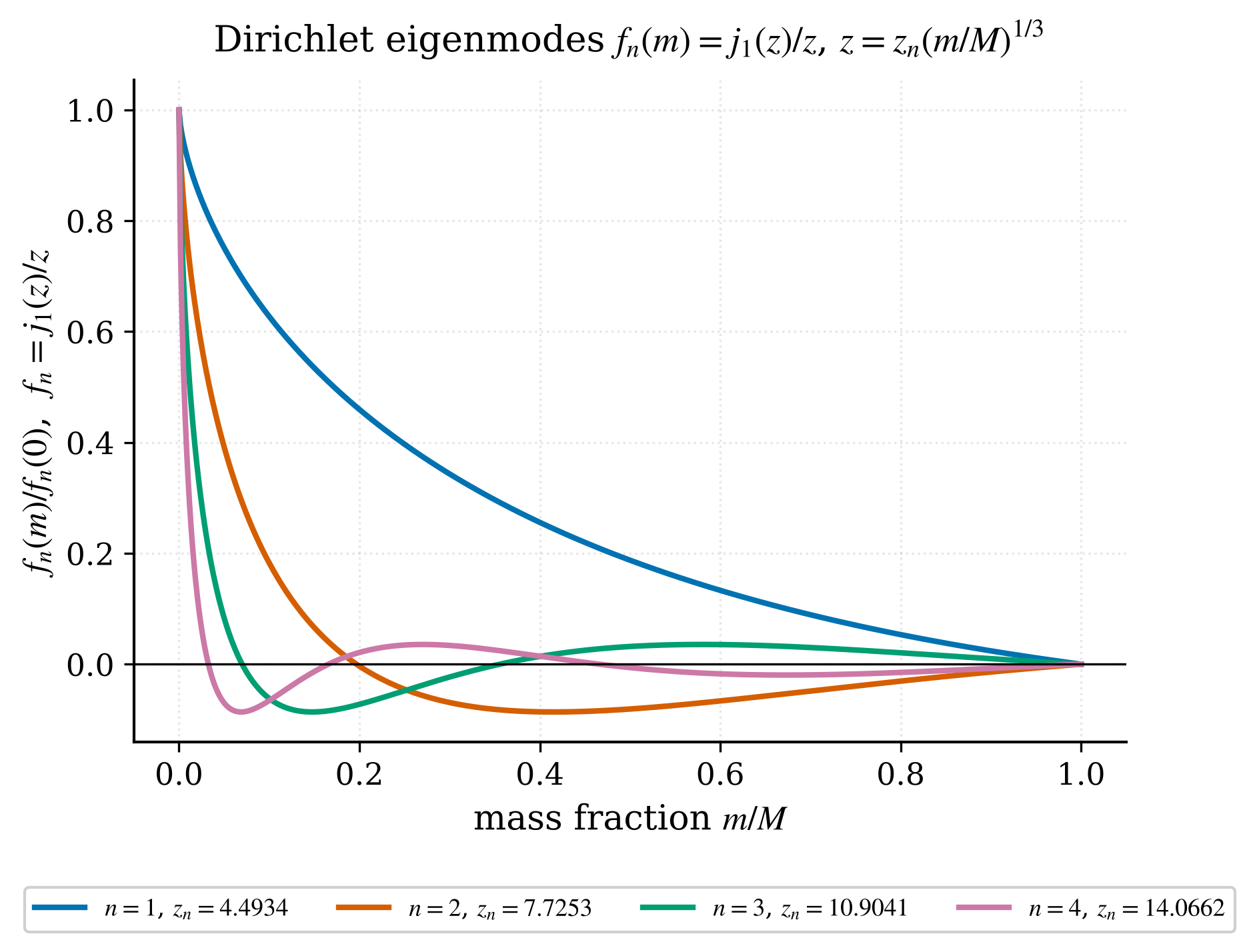}
\captionof{figure}{Dirichlet eigenfunctions $f_n(m)=j_1(z)/z$, $z=z_n(m/M)^{1/3}$, normalised by $f_n(0)$, for $n=1,\dots,4$; $f_n$ has $n-1$ interior zeros.}\label{fig:modes}
\end{minipage}
\end{figure}

\begin{corollary}[Spectrum for a piston boundary]\label{cor:spec}
\emph{Hypotheses:} those of Theorem \ref{thm:modes}, and the boundary motion $r^\varepsilon(t,M)=r_0(t,M)$ is prescribed (so $f(M)=0$). Then $z_n=bM^{1/3}$ is a positive root of $\tan z=z$, $z_n=4.4934,\,7.7253,\,10.9041,\,14.0662,\dots$, and the exponents satisfy
\begin{equation}\label{eq:disp}
3\kappa^2+\kappa-2+\mu_n=0,\qquad \mu_n=\Theta z_n^2,\quad \Theta=\frac{\ell}{9M^{2/3}}=\frac{8K}{3^{5/3}G\,M^{2/3}},
\end{equation}
that is,
\begin{equation}\label{eq:kap}
\kappa_\pm(\mu)=\frac{-1\pm\sqrt{25-12\mu}}{6}.
\end{equation}
At $\mu=0$ one recovers the pressureless Einstein-de Sitter modes $\kappa=2/3,\,-1$. For $\mu<2$ one branch is positive (relative perturbation decreasing as $T\to0$) and one is negative (relative growth); for $2<\mu<25/12$ both are negative; for $\mu>25/12$
\begin{equation}
\kappa=-\tfrac16\pm i\omega,\qquad\omega=\frac{\sqrt{12\mu-25}}{6},
\end{equation}
so the relative perturbation grows like $T^{-1/6}$ with log-periodic oscillation $\cos(\omega\ln T+\varphi)$. In particular, every mode with $\mu>0$ has a branch that grows relative to the background as $T\to0$.
\end{corollary}
\begin{proof}
The $z_n$ are the zeros of $j_1(z)=(\sin z-z\cos z)/z^2$, i.e.\ $\tan z=z$; $b^2M^{2/3}=z_n^2$ and \eqref{eq:j1} give $(2-3\kappa)(\kappa+1)=\mu_n$, i.e.\ $3\kappa^2+\kappa-2+\mu_n=0$. The roots are \eqref{eq:kap}; the discriminant is $25-12\mu$, and $\kappa_+>0\iff\sqrt{25-12\mu}>1\iff\mu<2$; for $\mu>25/12$ the roots are complex with real part $-1/6$.
\end{proof}
\smallskip\noindent
For $\Theta=0.05$ the first four values are $\mu_n=1.0095,\,2.9840,\,5.9450,\,9.8929$ (Table \ref{tab:spec}, Figures \ref{fig:spec} and \ref{fig:modes}).
\begin{table}[H]
\centering
\caption{Exponents for $\Theta=0.05$ ($z_n$ roots of $\tan z=z$).}\label{tab:spec}
\small
\begin{tabular}{@{}cccc@{}}
\toprule
$n$ & $z_n$ & $\mu_n$ & $\kappa_\pm$ \\
\midrule
$1$ & $4.4934$ & $1.0095$ & $0.4316,\;-0.7649$ \\
$2$ & $7.7253$ & $2.9840$ & $-0.1667\pm0.5479\,i$ \\
$3$ & $10.9041$ & $5.9450$ & $-0.1667\pm1.1346\,i$ \\
$4$ & $14.0662$ & $9.8929$ & $-0.1667\pm1.6134\,i$ \\
\bottomrule
\end{tabular}
\end{table}
\begin{remark}[Relation to Jeans theory and scope]
For the Einstein-de Sitter background $\rho\propto T^{-2}$ and the scale factor is $a\propto T^{2/3}$, so the squared sound speed $c_s^2\propto\rho^{\gamma-1}$ satisfies $c_s^2/(G\rho a^2)\propto T^{-2(\gamma-1)+2-4/3}=T^{(4-3\gamma)\cdot 2/3}$, which is independent of $T$ exactly when $\gamma=4/3$; this is what makes the Jeans problem scale-free and why power-law time dependence exists; in the Eulerian plane-wave form this is classical \cite{Peebles1980}. Theorem \ref{thm:modes} gives the spherical, Lagrangian form with explicit Bessel eigenfunctions. It is a statement about the linearised problem with a prescribed (piston) boundary; nonlinear stability around self-similar collapse is the subject of \cite{GHJS2025}, where the corresponding spectral problem is non-self-adjoint and treated by monotonicity arguments.
\end{remark}

\section{Numerical method and validation}\label{sec:num}
All numerical results in this paper come from one scheme, the discretisation of the Lagrangian itself. We state it, test it against exact linear modes, and then against an exact nonlinear collapse with a free surface, which is the harder test because the density vanishes at the boundary.

\begin{figure}[H]
\centering
\includegraphics[width=0.49\textwidth]{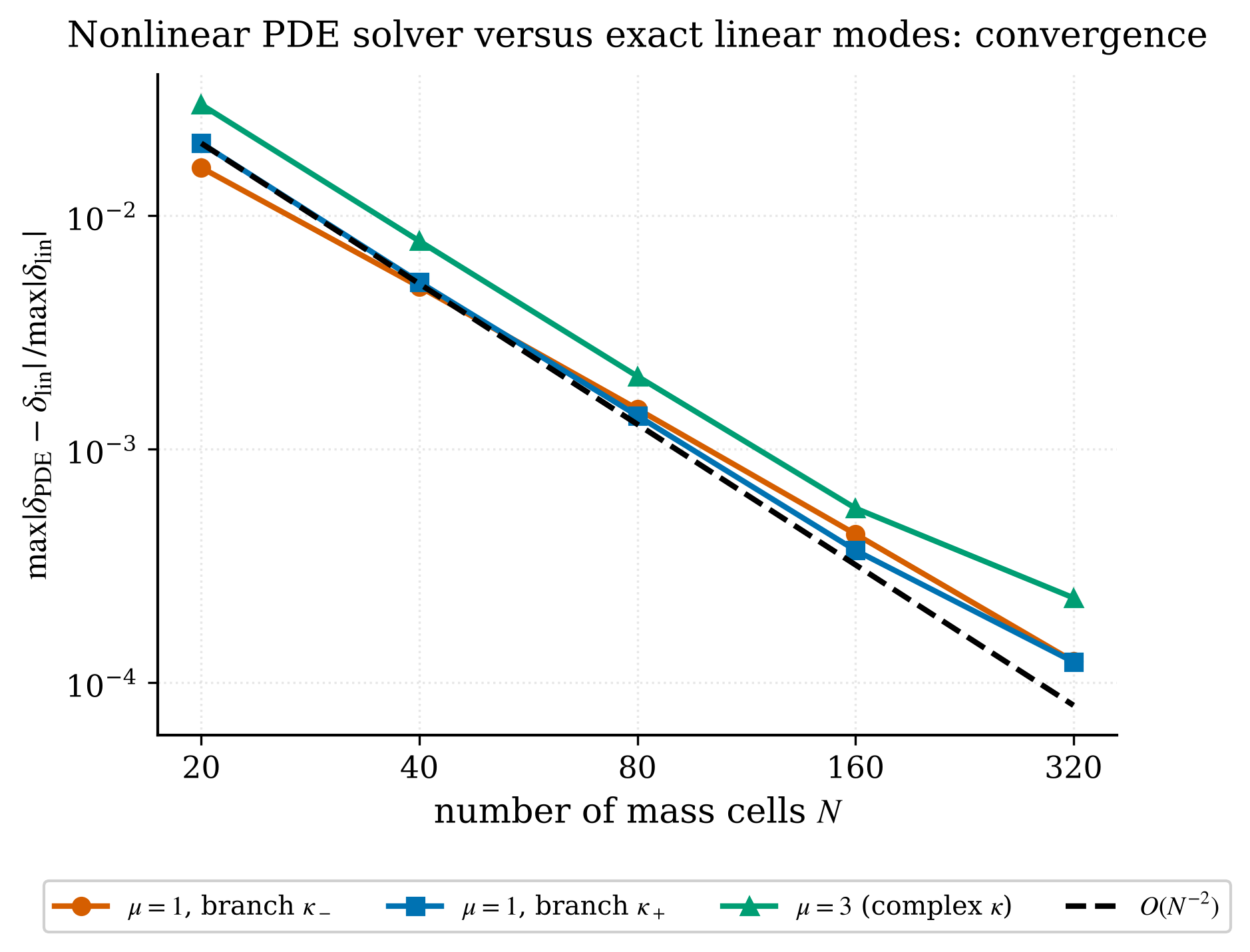}
\caption{Convergence of the nonlinear Lagrangian solver to the exact modes of Theorem \ref{thm:modes}; the dashed line has slope $-2$.}\label{fig:conv}
\end{figure}

\begin{figure}[H]
\begin{minipage}[t]{0.49\textwidth}
\centering
\includegraphics[width=\linewidth]{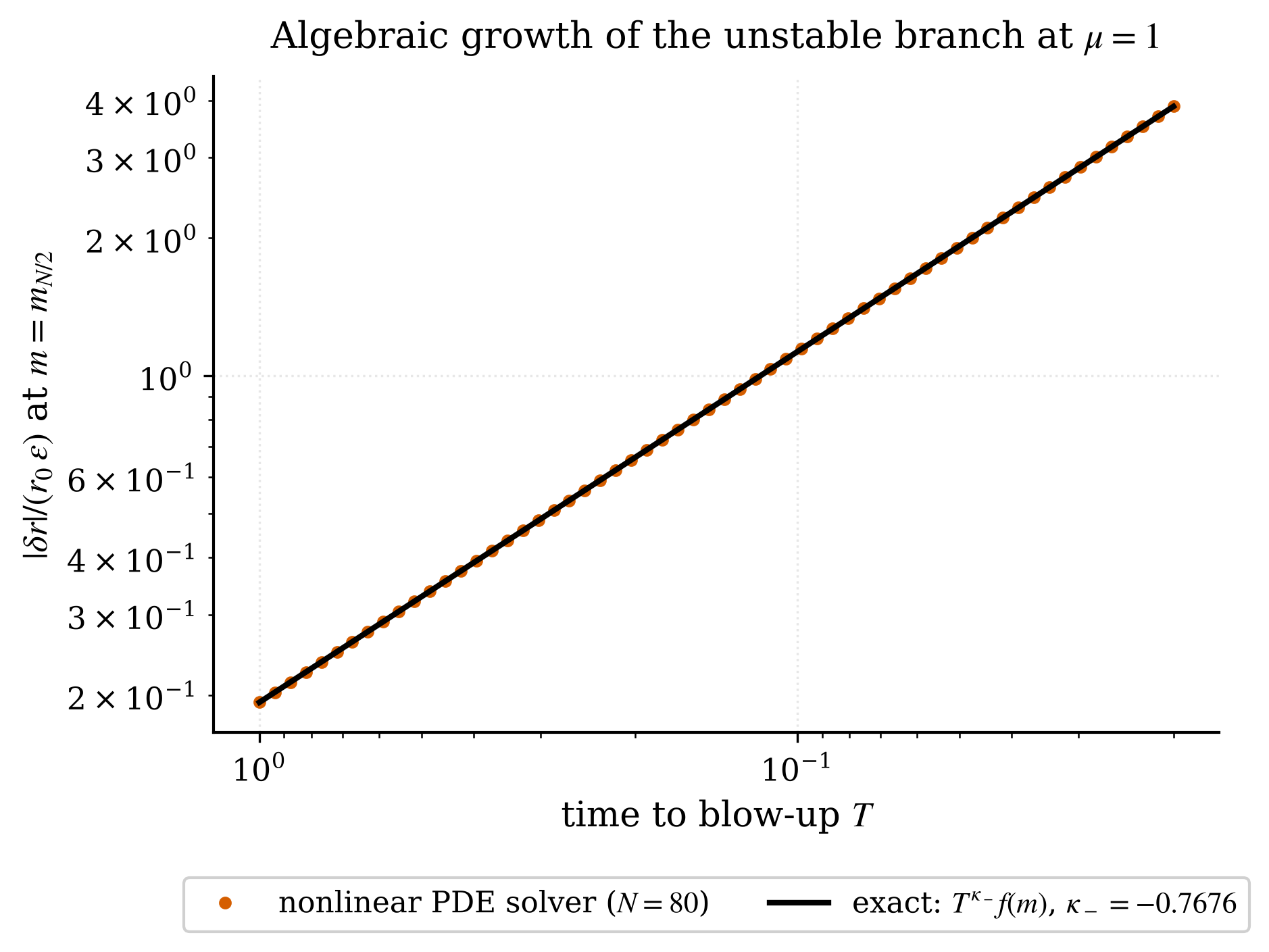}
\captionof{figure}{Relative perturbation at the middle node for $\mu=1$ on the branch $\kappa_-=-0.7676$: nonlinear solver ($N=80$, dots) versus the exact mode $T^{\kappa_-}f(m)$ (line), log-log axes, $T$ decreasing towards collapse.}\label{fig:unst}
\end{minipage}\hfill
\begin{minipage}[t]{0.49\textwidth}
\centering
\includegraphics[width=\linewidth]{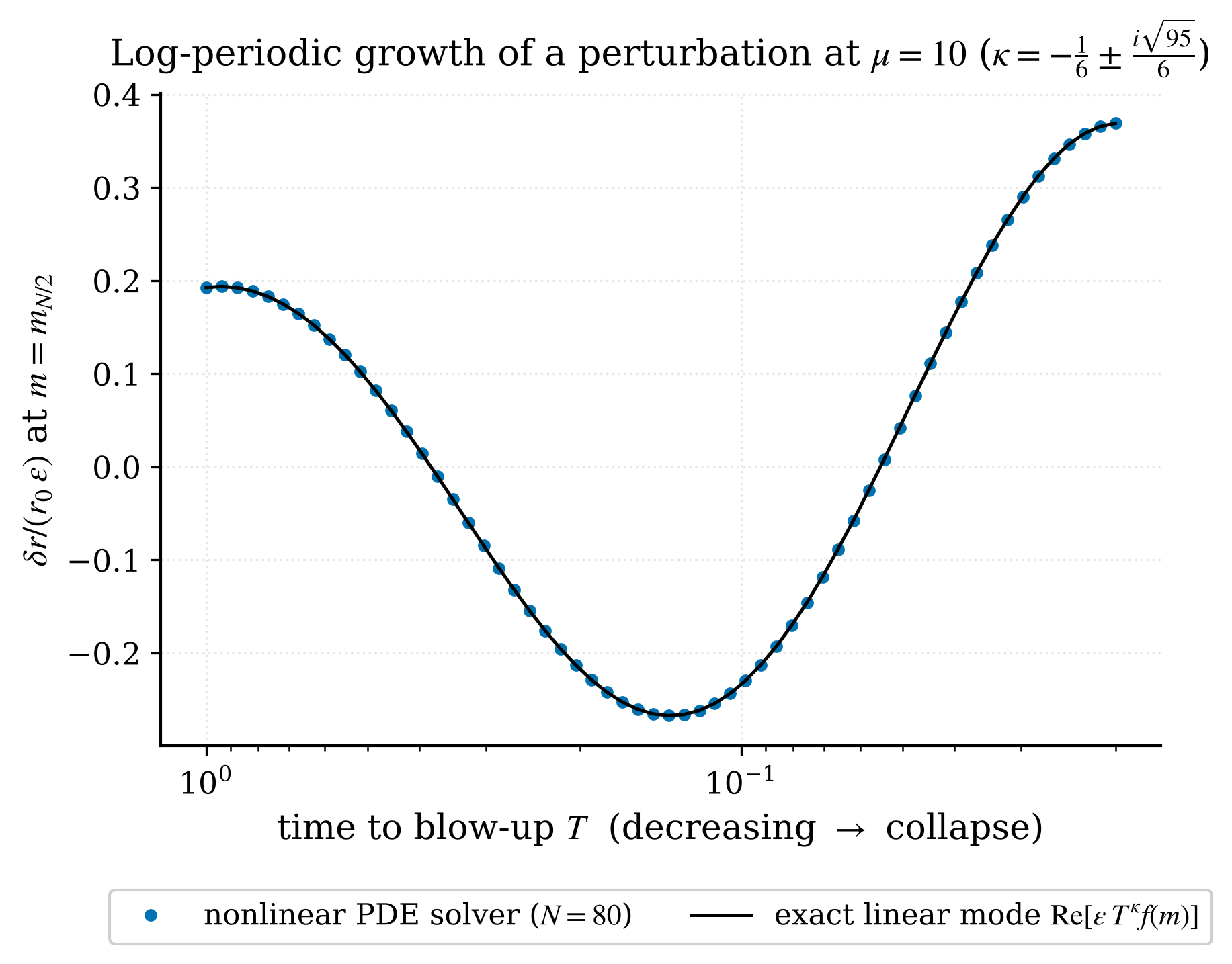}
\captionof{figure}{Log-periodic perturbation for $\mu=10$ (complex exponent) at the middle node: nonlinear solver ($N=80$) versus $\mathrm{Re}[\varepsilon T^\kappa f(m)]$.}\label{fig:osc}
\end{minipage}
\end{figure}

\begin{figure}[H]
\centering
\includegraphics[width=0.49\textwidth]{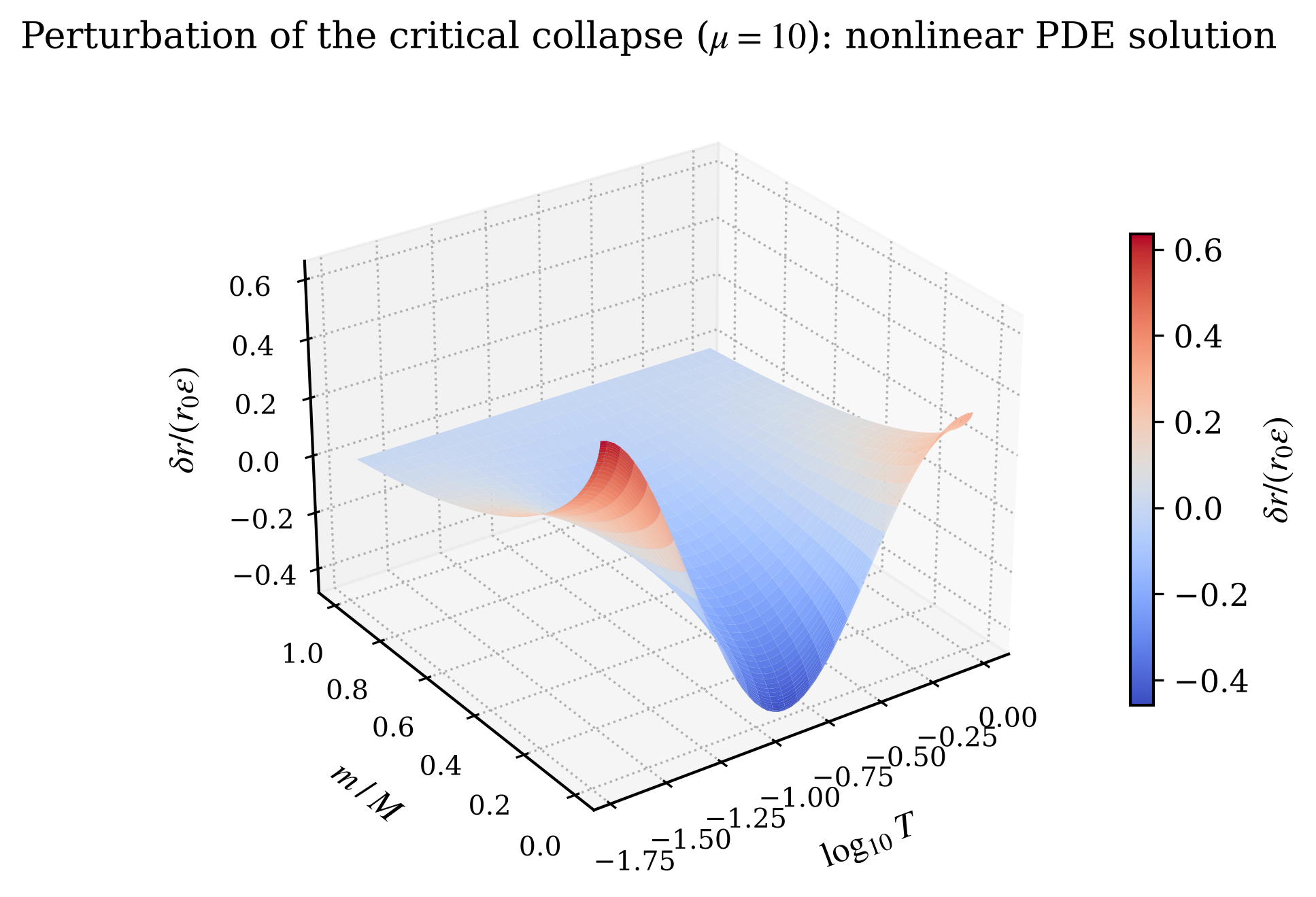}
\caption{Nonlinear solution $\delta r/(r_0\varepsilon)$ over $(\log_{10}T,m/M)$ for $\mu=10$: the spatial profile $j_1(z)/z$ modulated by the log-periodic time dependence.}\label{fig:pert3d}
\end{figure}

\begin{figure}[H]
\begin{minipage}[t]{0.49\textwidth}
\centering
\includegraphics[width=\linewidth]{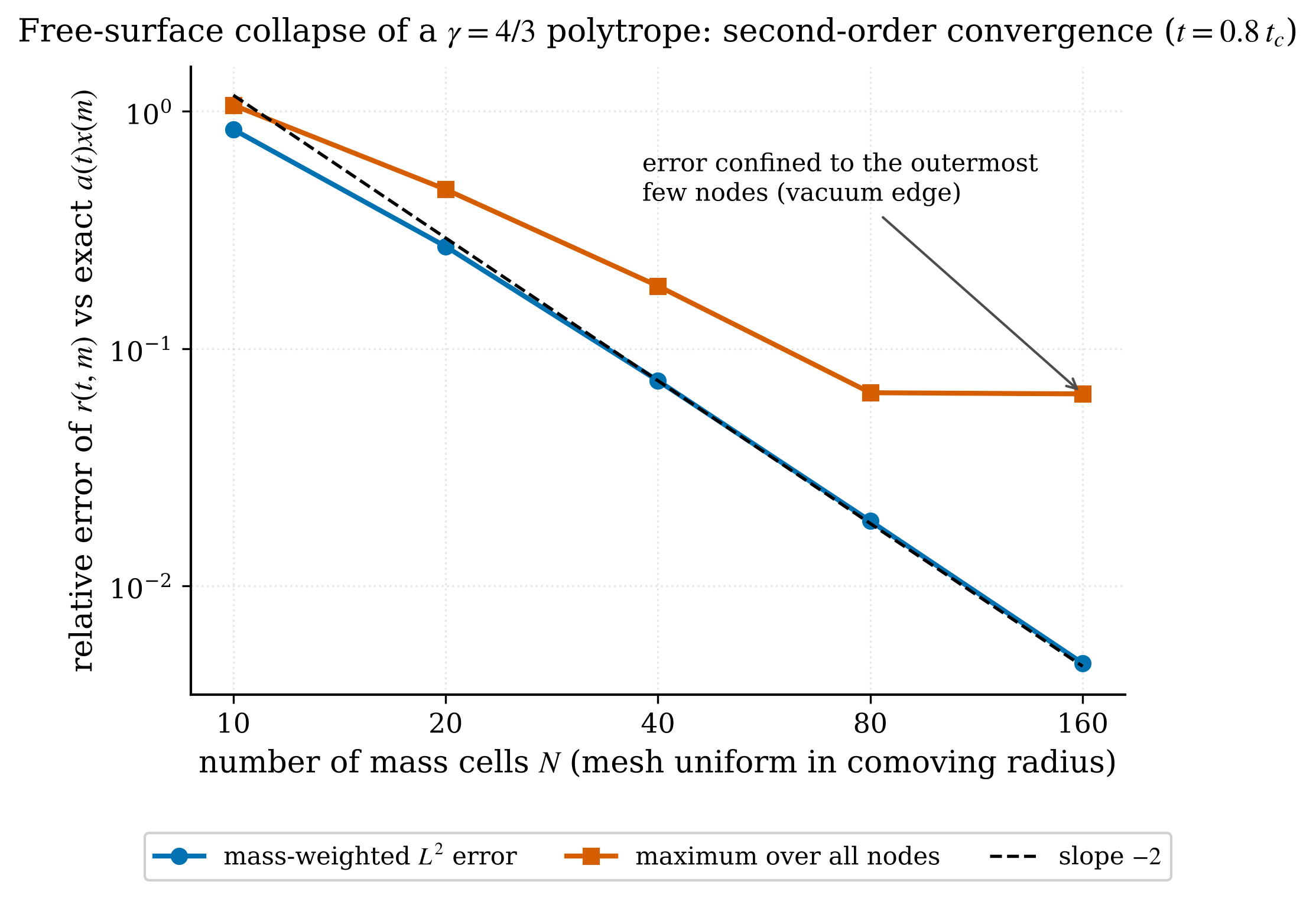}
\captionof{figure}{Convergence of the free-surface Lagrangian scheme to the exact homologous collapse: mass-weighted $L^2$ error (circles, order $\approx2$) and maximum over nodes (squares, saturating at the outermost nodes); dashed line: slope $-2$.}\label{fig:surfconv}
\end{minipage}\hfill
\begin{minipage}[t]{0.49\textwidth}
\centering
\includegraphics[width=\linewidth]{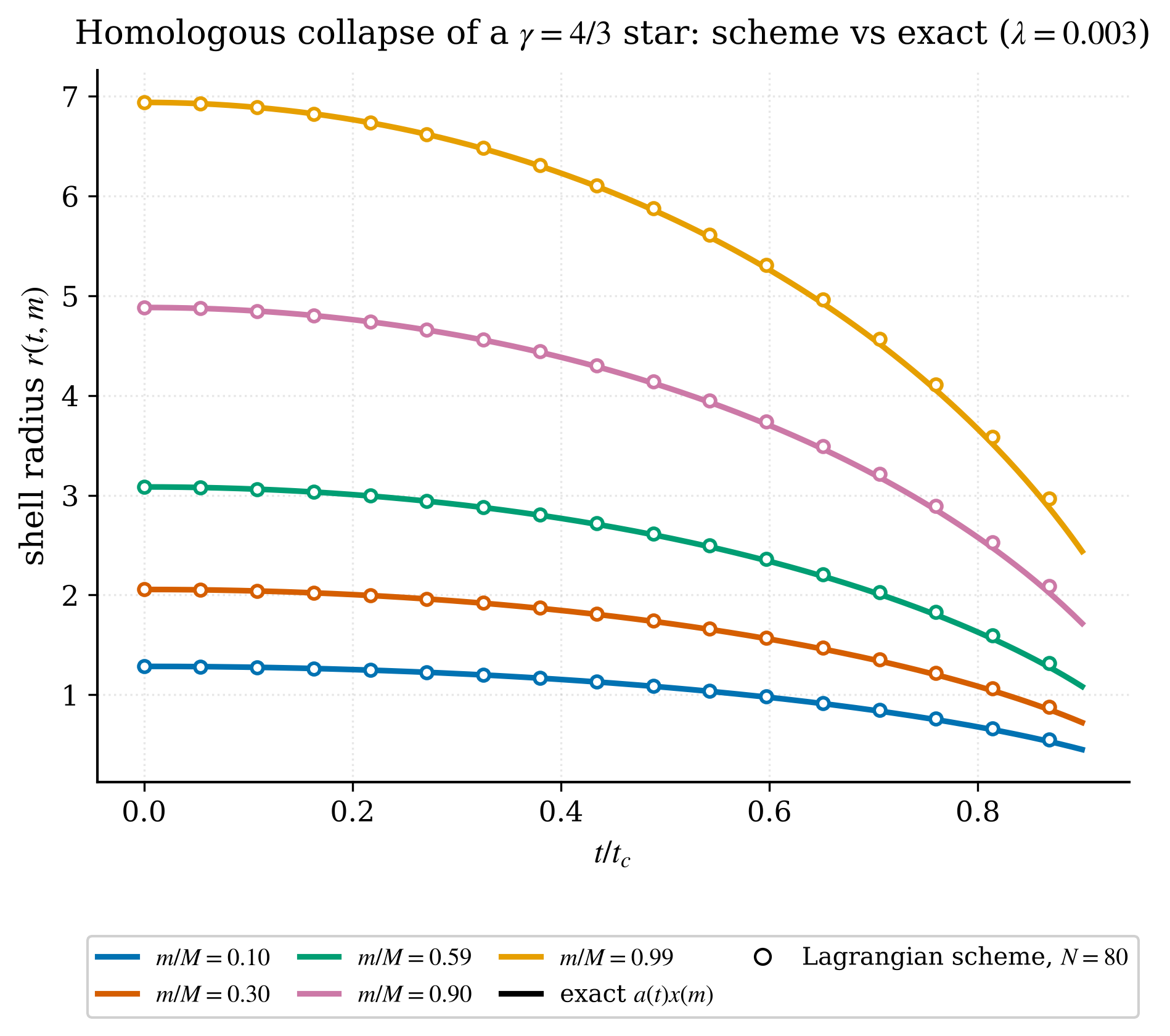}
\captionof{figure}{Shell radii $r(t,m)$ of the free-surface collapse: exact $a(t)x(m)$ (lines) and Lagrangian scheme with $N=80$ (open circles) for the shells with mass fractions $m/M=0.10,\,0.30,\,0.59,\,0.90,\,0.99$ up to $t=0.9\,t_c$. The legend is placed below the axes so that no curve is hidden.}\label{fig:surftraj}
\end{minipage}
\end{figure}

\paragraph{Scheme.} We discretise \eqref{eq:L} on nodes $0=m_0<\dots<m_N=M$ with $V_j=(r_{j+1}^3-r_j^3)/3$, cell pressure $P_j=K(V_j/\Delta m_j)^{-\gamma}$ and node mass $\mu_i=(\Delta m_{i-1}+\Delta m_i)/2$:
\begin{equation}\label{eq:scheme}
\mu_i\ddot r_i=-\frac{Gm_i\mu_i}{r_i^{2}}+r_i^{2}\,(P_{i-1}-P_i),\qquad i=1,\dots,N,
\end{equation}
with $r_0=0$ and $P_N:=0$ at a free surface (or $r_N$ prescribed for a piston). It is the Euler-Lagrange system of the discrete Lagrangian $\sum_i\bigl(\tfrac12\mu_i\dot r_i^2+Gm_i\mu_i/r_i\bigr)-\sum_j\tfrac{K}{\gamma-1}\Delta m_j(V_j/\Delta m_j)^{1-\gamma}$, hence conserves a discrete energy up to the time integrator ($\mathrm{DOP}853$, tolerances $10^{-12}/10^{-14}$).
\paragraph{Validation against the exact linear modes.} Initial data are $r=r_0(1+\varepsilon T^{\kappa}f)$ with $\varepsilon=10^{-6}$, $f=j_1(z)/z$, $z_1=4.4934$, mesh uniform in $m^{1/3}$, evolved from $T=1$ to $T=0.1$. Table \ref{tab:conv} and Figure \ref{fig:conv} give the error $\max|\delta_{\rm PDE}-\delta_{\rm lin}|/\max|\delta_{\rm lin}|$ over interior nodes and time samples: it decreases with observed order $1.7$-$2.0$ for $N\le160$ (successive doublings from $N=20$), then flattens at $1$-$2\times10^{-4}$ at $N=320$, where the residual nonlinear $O(\varepsilon)$ and time-integration contributions become comparable.
\begin{table}[H]
\centering
\caption{Relative error of the nonlinear solver against the exact linear modes versus the number of cells $N$.}\label{tab:conv}
\small
\begin{tabular}{@{}lccccc@{}}
\toprule
Case & $N=20$ & $40$ & $80$ & $160$ & $320$ \\
\midrule
$\mu=1$, branch $\kappa_-=-0.7676$ & $1.6\times10^{-2}$ & $5.0\times10^{-3}$ & $1.5\times10^{-3}$ & $4.3\times10^{-4}$ & $1.2\times10^{-4}$ \\
$\mu=1$, branch $\kappa_+=0.4343$ & $2.0\times10^{-2}$ & $5.2\times10^{-3}$ & $1.4\times10^{-3}$ & $3.7\times10^{-4}$ & $1.2\times10^{-4}$ \\
$\mu=3$ (complex $\kappa$) & $3.0\times10^{-2}$ & $7.8\times10^{-3}$ & $2.1\times10^{-3}$ & $5.6\times10^{-4}$ & $2.3\times10^{-4}$ \\
\bottomrule
\end{tabular}
\end{table}
\smallskip\noindent
Figures \ref{fig:unst}-\ref{fig:pert3d} compare the numerical perturbation with the exact mode: algebraic growth $T^{\kappa_-}$ at $\mu=1$ and log-periodic oscillation at $\mu=10$ ($\kappa=-1/6\pm i\sqrt{95}/6$; relative error $1.4\times10^{-3}$ at $N=80$).
\paragraph{Free-surface homologous collapse.} The exact solution $r=a(t)x(m)$ of Section \ref{sec:homol} ($\lambda=0.003$, $M=16.457292$, true surface radius $x_s\approx10.296$; the mesh ends at $x=10.2779$, where $\Pi<10^{-13}$ and the remaining mass is below $10^{-7}$) provides a nonlinear test with a vacuum boundary. We use the smooth system in the comoving radius, $\dd m/\dd x=(\Pi x^{8/3}/K)^{3/4}$, $\dd\Pi/\dd x=\Pi_m\,\dd m/\dd x$, to place nodes uniformly in $x$, start from rest ($a(0)=1$, $\dot a(0)=0$), and compare with $a(t)x_i$ at $t=0.8\,t_c$ (where $a=0.53$). Table \ref{tab:surfconv} and Figure \ref{fig:surfconv} show that the mass-weighted $L^2$ error, $\bigl(\sum_i\mu_i e_i^2/M\bigr)^{1/2}$ with $e_i=|r_i-ax_i|/(ax_i)$, converges with the observed order $1.9$-$2.0$ (orders $1.64,1.88,1.97,1.99$ for successive doublings), as does the error at the middle node ($7.4\times10^{-2},1.9\times10^{-2},4.8\times10^{-3}$ for $N=40,80,160$; orders $1.97,1.99$); the discrete energy is conserved to $2.5\times10^{-13}$ for $N\ge20$. The maximum-norm error is attained in the outermost nodes, where the density vanishes like $(x_s-x)^3$ (Proposition \ref{prop:surface}), and is $6.4\times10^{-2}$ for $N=80,160$; the error at the node $N-8$ is $1.8\times10^{-2},4.9\times10^{-3}$ for $N=80,160$ (order $1.9$). The mass-weighted and interior errors are the natural measures for energy-type quantities, since the outermost cell carries a vanishing fraction of the mass. Figure \ref{fig:surftraj} shows shell trajectories against the exact solution.
\begin{table}[H]
\centering
\caption{Free-surface homologous collapse ($\gamma=4/3$, $\lambda=0.003$, $t=0.8t_c$): errors of the Lagrangian scheme on a mesh uniform in comoving radius.}\label{tab:surfconv}
\small
\begin{tabular}{@{}rccc@{}}
\toprule
$N$ & mass-weighted $L^2$ error & maximum over nodes & energy drift \\
\midrule
$10$ & $8.39\times10^{-1}$ & $1.06$ & $4.9\times10^{-11}$ \\
$20$ & $2.70\times10^{-1}$ & $4.69\times10^{-1}$ & $2.5\times10^{-13}$ \\
$40$ & $7.32\times10^{-2}$ & $1.84\times10^{-1}$ & $1.9\times10^{-13}$ \\
$80$ & $1.87\times10^{-2}$ & $6.51\times10^{-2}$ & $1.4\times10^{-13}$ \\
$160$ & $4.71\times10^{-3}$ & $6.44\times10^{-2}$ & $2.4\times10^{-13}$ \\
\bottomrule
\end{tabular}
\end{table}

\section{Discussion}\label{sec:disc}
\paragraph{Criticality in the symmetry algebra.} For $\gamma\neq4/3$ and a generic entropy profile, the only point symmetry is time translation (case $\mathrm{A}$ of Table \ref{tab:class}), and Proposition \ref{prop:defect} shows that the critical scaling is violated at order $\gamma-4/3$. At $\gamma=4/3$ the scaling $3t\partial_t+2r\partial_r$ is available for every $S(m)$: in the language of \cite{AHS2025} this is the mass-critical point, where the scaling does not move the mass label and is therefore compatible with any entropy profile. The same fact shows up in the similarity variable $s=mt^{3\gamma-4}$, which degenerates to $m$, and in the self-similar equation \eqref{eq:ssode}, which loses its derivative terms. That the same borderline shows up in three independent places is, to us, the clearest sign that $\gamma=4/3$ is special already at the level of the equation and not only for particular solutions.

\paragraph{Completeness.} Theorem \ref{thm:class} says that the scalings behind the isentropic and non-isentropic implosions of \cite{GHJS2022,AHS2025} are all the point symmetries the entropy-fixed equation has. A further symmetry reduction would therefore need non-point symmetries. We have not explored contact or nonlocal symmetries.

\paragraph{The variational line.} The conservation law of Theorem \ref{thm:noether} lives on $\sigma=14-10\gamma$, which in the parametrisation of \cite{AHS2025} lies in the mass-subcritical region. We read this as saying that the extra conserved quantity belongs to the stable, expanding side of the problem and not to the collapse solutions studied in \cite{GHJS2022,AHS2025}. Whether it can be used, for instance, to control perturbations of Goldreich-Weber type stars is an open question.

\paragraph{Series and shooting.} The analytic centre family of Theorem \ref{thm:BB} and the surface law of Proposition \ref{prop:surface} are the two regular endpoints of a profile. Existence proofs construct global profiles from the centre outwards by shooting or bootstrap (continuation) arguments \cite{GuoHadzicJang2021,GHJS2022,AHS2025}. The explicit coefficients, in particular $a_1\propto 9G-2A_0^3$, which vanishes exactly for the Einstein-de Sitter core, are the kind of input such arguments need, and they can be used directly to start a numerical shooting code.

\paragraph{Gradient catastrophe.} Centrally condensed dust ($\alpha>1/3$) collapses without shell crossing, whereas data whose mean density increases outward cross first. This is the free-fall analogue of the Hopf catastrophe, and it is what one expects physically; the Lagrangian formula \eqref{eq:thetac} makes the threshold and the crossing time explicit.

\paragraph{Open problems.} We see five natural continuations: the inclusion of pressure in the shell-crossing criterion; extending the exact linear analysis to the Larson-Penston and Yahil backgrounds, where mode stability is the central difficulty \cite{GHJS2025}; relating the classification to weak solutions with concentration at the origin \cite{Schrecker2020}; non-spherical and Einstein-Euler extensions \cite{GuoHadzicJang2023}; and approximate symmetries near $\gamma=4/3$ beyond the class considered in Proposition \ref{prop:defect} \cite{Shamaoon2023b}.

\section{Conclusions}
In Lagrangian mass coordinates the spherically symmetric Euler-Poisson equations for a polytropic gas reduce to one variational equation for the shell radius, and its point symmetries can be classified completely. Besides time translation there is at most one scaling, except for $\gamma=4/3$ with $S\propto m^{2/3}$, where the algebra is three-dimensional; at $\gamma=4/3$ a scaling exists for every entropy profile. These scalings are the entropy-fixed stabilisers of the two-parameter scaling group of the non-isentropic system, and the symmetry defect is exactly of order $\gamma-4/3$. A scaling is variational only on the line $S=Km^{14-10\gamma}$, where it gives a conservation law beyond energy; otherwise it gives virial identities. Among the exact solutions are the Einstein-de Sitter collapse, the homologous cycloid at $\gamma=4/3$ and the self-similar reduction to a single ODE. The Briot-Bouquet theorem shows that the regular centre of self-similar and homologous collapse is a one-parameter analytic family with explicit leading coefficients, which agree exactly at $\gamma=4/3$, and the surface of the homologous star has a Puiseux expansion with density vanishing like the power $1/(\gamma-1)$ of the distance to the surface. In mass coordinates density singularities are zeros of $r$ or $r_m$: we gave a Hopf-type fold law and an exact criterion for shell crossing of free-falling dust, namely that it precedes collapse iff the mean density increases outward. Linear perturbations of the critical collapse separate exactly into spherical Bessel modes whose exponents solve $3\kappa^2+\kappa-2+\mu=0$. The identities were verified by computer algebra, and a second-order Lagrangian scheme confirmed the exact modes and a free-surface collapse. Open problems are contact and non-local symmetries, pressure in the shell-crossing criterion, the linear analysis around the Larson-Penston and Yahil solutions, and non-spherical and Einstein-Euler extensions.

\section*{Declarations}
\noindent{\bf Conflict of interest.} The author affirms that there is no conflict of interest regarding the publication of this paper.
\\
\noindent{\bf Funding.} This research did not receive any specific grant from funding agencies in the public, commercial, or not-for-profit sectors.
\\
\noindent{\bf Use of computational and AI tools.} Symbolic computations were performed with SymPy $1.14$, numerical computations with NumPy and SciPy, and all figures were produced directly by Matplotlib scripts that are listed in Appendix \ref{app:ver}. A large language model was used to check my own numerical code validations to remove some errors from numerical codes for drafting and editing the text. Also used to check grammar mistakes and paraphrasing. All mathematical statements, derivations, numerical results, and references were studied, derived, written, and reviewed by the author, who takes full responsibility for the content. 
\\
\noindent{\bf Data availability.} No external datasets were used. The SymPy and Python scripts used for every symbolic verification, series, profile computation, figure, and convergence study are available from the corresponding author on request.

\appendix
\section{Computer-algebra and numerical verification}\label{app:ver}
All items were verified with SymPy $1.14$ (scripts \texttt{verify\_all.py}, \texttt{noether.py}, \texttt{noether2.py}, \texttt{lin2.py}, \texttt{detsys.py}, \texttt{ss\_check.py}, \texttt{dust.py}, \texttt{g2full.py}, \texttt{ahs.py}, \texttt{bb.py}, \texttt{chk\_new.py}, \texttt{polycheck.py}); numerical work (\texttt{prof2.py}, \texttt{recompute.py}, \texttt{surf2.py}, \texttt{crit.py}, \texttt{sim3.py}, \texttt{surfdiag.py}, \texttt{figs\_*.py}) used NumPy, SciPy ($\mathrm{DOP}853$) and Matplotlib. Maple or Mathematica can reproduce the identities from the formulas in the text.
\begingroup\normalsize\renewcommand{\arraystretch}{1.05}
\begin{longtable}{@{}>{\raggedright\arraybackslash}p{1.05cm}>{\raggedright\arraybackslash}p{6.9cm}>{\raggedright\arraybackslash}p{7.1cm}@{}}
\caption{Symbolic verification (computer algebra, all PASS).}\label{tab:ver}\\
\toprule
Id & Statement & Result \\
\midrule
\endfirsthead
\multicolumn{3}{@{}l}{\small\emph{Table \thetable\ (continued)}}\\
\toprule
Id & Statement & Result \\
\midrule
\endhead
\bottomrule
\endfoot
$\mathrm{V}1$ & $\delta\mathcal L/\delta r=-\Delta[r]$ (Prop. \ref{prop:var}) & identically zero difference  \\
$\mathrm{V}2$ & energy identity \eqref{eq:energy}, $\D_t\mathcal E+\D_m(r^2Pr_t)=+r_t\Delta[r]$ (the opposite sign fails) & identically zero difference  \\
$\mathrm{V}3$ & pointwise virial identity \eqref{eq:pointvirial} & identically zero difference  \\
$\mathrm{V}4$ & EdS solution \eqref{eq:EdS}, any $\gamma$ & residual $0$  \\
$\mathrm{V}5$ & uniform-density ODE \eqref{eq:aode}, any $\gamma$ & difference $0$  \\
$\mathrm{V}6$ & mass scaling at $S=Km^{2/3}$, $\gamma=3/2$ & prolonged invariance $0$  \\
$\mathrm{V}7$ & scaling for $S=K$, $\gamma=7/5$ & prolonged invariance $0$  \\
$\mathrm{V}8$ & brackets of Cor. \ref{cor:cases}$(\mathrm{e})$ & $[X_1,X_2]=3X_1$, others $0$  \\
$\mathrm{V}9$ & $j_1(z)/z$ solves \eqref{eq:modeode} & residual $0$  \\
$\mathrm{V}10$ & cycloid \eqref{eq:cyc} & $a^2\ddot a+\lambda=0$  \\
$\mathrm{V}11$ & EdS in reduced ODE \eqref{eq:ssode2} & residual $0$  \\
$\mathrm{V}12$ & series $c_1,\dots,c_5$ at $\lambda=1$ & all vanish  \\
$\mathrm{V}13$ & identity \eqref{eq:Noetheridentity} for $X_N$ & $0$ at $(7/5,0),(6/5,2),(3/2,-1)$; non-zero for $(\gamma,\sigma)=(7/5,1)$  \\
$\mathrm{V}14$ & generators of Table \ref{tab:class} & prolonged invariance $0$; a wrong projective symmetry fails  \\
$\mathrm{V}15$ & symmetry defect \eqref{eq:defect} & difference $0$  \\
$\mathrm{V}16$ & dust $r_m$ formula \eqref{eq:rmdust} and $\eta=T'/T$ & difference $0$  \\
$\mathrm{V}17$ & cross-validation \eqref{eq:cross}, $k=1,2,3$ & exact identities in $\lambda$  \\
$\mathrm{V}18$ & closed form \eqref{eq:a2} of $a_2$ & matches symbolic $a_2$ at a random parameter point to $13$ digits  \\
$\mathrm{V}19$ & full $\gamma=2$ determining system & same first-level equations as generic $\gamma$; same classifying equation  \\
$\mathrm{V}20$ & two-parameter group \eqref{eq:twoparam}; map \eqref{eq:AHSmap} & $b=(3a-c)/2$, $e=(3\gamma-4)a-(\gamma-2)c$; $c_4-2(2-\gamma)/(2-\alpha_E)=0$; $\alpha_E=\sigma(3\gamma-4)/(\gamma+\sigma-2)$  \\
$\mathrm{V}21$ & Briot-Bouquet data, Thm. \ref{thm:BB} & $F_1(A_0,\Pi_0)=0$; eigenvalues $\{0,-3\}$; $F_2=O(u^2)$, analytic  \\
$\mathrm{V}22$ & $H_{\rm ss}$ of \eqref{eq:Hss} & equals the symbolic $3s\Pi_s/u^2$ at random points  \\
$\mathrm{V}23$ & Thm. \ref{thm:class}, independent check & numerical nullspace dimension of polynomial symmetries (degree $\le3$) equals the predicted $1,2,2,2,2,3$ and $1$ ($\gamma=2$) in seven test cases  \\
\end{longtable}
\endgroup

\begingroup\normalsize\renewcommand{\arraystretch}{1.05}
\begin{longtable}{@{}>{\raggedright\arraybackslash}p{1.05cm}>{\raggedright\arraybackslash}p{6.9cm}>{\raggedright\arraybackslash}p{7.1cm}@{}}
\caption{Numerical verification.}\label{tab:vernum}\\
\toprule
Id & Statement & Result \\
\midrule
\endfirsthead
\multicolumn{3}{@{}l}{\small\emph{Table \thetable\ (continued)}}\\
\toprule
Id & Statement & Result \\
\midrule
\endhead
\bottomrule
\endfoot
$\mathrm{N}1$ & series vs ODE, $\lambda=0.3$ & $2\times10^{-12}$ at $m=0.05$; $6\times10^{-8}$ at $m=1$; start-point independent  \\
$\mathrm{N}2$ & $M$ at $\lambda=0$ vs $\Edd$ (Lane-Emden) & equal to six digits  \\
$\mathrm{N}3$ & profile identity \eqref{eq:profid} & relative residual $\le1.1\times10^{-6}$  \\
$\mathrm{N}4$ & nonlinear solver vs exact modes & order $1.7$-$2.0$ for $N\le160$ (Table \ref{tab:conv})  \\
$\mathrm{N}5$ & surface Puiseux law \eqref{eq:puiseux} & ratio $\to1$, deviations $\propto(M-m)^{1/4}$ (Table \ref{tab:surface})  \\
$\mathrm{N}6$ & dust crossing time, $\alpha=0.3$ & predicted $1.008075$, direct integration $1.008353$ (sampling $3\times10^{-4}$)  \\
$\mathrm{N}7$ & self-similar series (Table \ref{tab:ssseries}) & start-point independence $10^{-13}$; errors as tabulated  \\
$\mathrm{N}8$ & free-surface homologous collapse (Table \ref{tab:surfconv}) & mass-weighted $L^2$ order $1.9$-$2.0$; energy drift $\le2.5\times10^{-13}$  \\
\end{longtable}
\endgroup

\bibliographystyle{unsrt}
\bibliography{bibliog}

\end{document}